\documentclass{aa}
\usepackage{graphicx}
\usepackage{natbib}
\usepackage{psfig}
\bibpunct{(}{)}{;}{a}{}{,}
\newcommand{\be}{\begin{equation}}
\newcommand{\ee}{\end{equation}}

\newcommand{\myarcsec}{\hbox{$.\!\!^{\prime\prime}$}}

\newcommand{\myarcsecnodot}{\hbox{$\;\!\!^{\prime\prime}\;$}}
\newcommand{\myarcminnodot}{\hbox{$\;\!\!^{\prime}\;$}}
\newcommand{\fat}[1]{\mbox{\boldmath $ #1 $}}
\newcommand{\dd}{\,\textrm{d}}

\newcommand{\sectionref}[1]{Sect. \ref{#1}}
\newcommand{\appendixref}[1]{App. \ref{#1}}
\newcommand{\figref}[1]{Fig. \ref{#1}}
\newcommand{\eqref}[1]{eq. (\ref{#1})}
\newcommand{\tableref}[1]{Table \ref{#1}}

\begin{document}
  \title{GaBoDS: The Garching-Bonn Deep Survey}
   \subtitle{IV. Methods for the Image reduction of multi-chip 
     Cameras\thanks{Based on observations made with ESO
     Telescopes at the La Silla Observatory}}

   \author{T. Erben\inst{1}
          \and
	  M. Schirmer\inst{1,2}
          \and
	  J.~P. Dietrich\inst{1}
          \and
	  O. Cordes\inst{1}
          \and
	  L. Haberzettl\inst{1,3}
          \and
	  M. Hetterscheidt\inst{1}
          \and
	  H. Hildebrandt\inst{1}
          \and
	  O. Schmithuesen\inst{1,3}
	  \and 
	  P. Schneider\inst{1}
	  \and 
	  P. Simon\inst{1}
	  \and 
	  J.~C. Cuillandre\inst{4}
	  \and 
	  E. Deul\inst{5}
	  \and 
	  R.~N. Hook\inst{6}
	  \and 
	  N. Kaiser\inst{7}
	  \and 
	  M. Radovich\inst{8}
	  \and 
	  C. Benoist\inst{9}
	  \and 
	  M. Nonino\inst{10}
	  \and 
	  L.~F. Olsen\inst{11,9}
	  \and 
	  I. Prandoni\inst{12}
	  \and 
	  R. Wichmann\inst{13}
	  \and 
	  S. Zaggia\inst{10}
	  \and 
	  D. Bomans\inst{3}
	  \and 
	  R.~J. Dettmar\inst{3}
	  \and 
	  J.~M. Miralles\inst{14,1}
          }

   \offprints{T. Erben: terben@astro.uni-bonn.de}

   \institute{$^{1}$Institut f\"ur Astrophysik und Extraterrestrische 
              Forschung (IAEF), Universit\"at Bonn,
              Auf dem H\"ugel 71, D-53121 Bonn, Germany\\
	      $^{2}$Isaac Newton Group of Telescopes, Apartado de correos 321, 
	      38700 Santa Cruz de La Palma, Tenerife, Spain\\
              $^{3}$Astronomisches Institut der Ruhr-Universit\"at-Bochum,
              Universit\"atsstr. 150, D-44780 Bochum, Germany\\
	      $^{4}$Canada-France-Hawaii Telescope Corporation, 
              65-1238 Mamalahoa Highway, Kamuela, HI 96743\\
	      $^{5}$Leiden Observatory, Postbus 9513, NL-2300 RA
              Leiden, The Netherlands\\
	      $^{6}$Space Telescope European Coordinating Facility, 
              European Southern Observatory,
              Karl-Schwarzschild-Strasse 2, D-85748 Garching bei
              M\"unchen, Germany\\
	      $^{7}$Institute for Astronomy, University of Hawaii, 
              2680 Woodlawn Drive, Honolulu, HI 96822\\
              $^{8}$INAF, Osservatorio Astronomico di Capodimonte, 
              via Moiariello 16, 80131 Napoli, Italy\\
              $^{9}$Laboratoire Cassiop\'ee, CNRS, Observatoire de 
              la C\^ote d'Azur, BP4229, 06304 Nice Cedex 4, France\\
              $^{10}$INAF, Osservatorio Astronomico di Trieste, 
              Via G. Tiepolo 11, 34131 Trieste, Italy\\
              $^{11}$Copenhagen University Observatory, Copenhagen
              University, Juliane Maries Vej 30, 2100 Copenhagen, Denmark\\
              $^{12}$Instituto di Radioastronomia, CNR, Via Gobetti 101, 40129, Bologna, Italy\\
              $^{13}$Hamburger Sternwarte, University of Hamburg, 
              Gojenbergsweg 112, 21029 Hamburg, Germany\\
              $^{14}$European Southern Observatory, 
              Karl-Schwarzschild-Strasse 2, D-85748 Garching bei
              M\"unchen, Germany\\
}

   \date{Received June 15, 2004; accepted June 16, 2004}

   \abstract{ We present our image processing system for the reduction
   of optical imaging data from multi-chip cameras. In the framework
   of the Garching Bonn Deep Survey (GaBoDS; \citeauthor{ses03}
   \citeyear{ses03}) consisting of about 20 square degrees of
   high-quality data from WFI@MPG/ESO 2.2m, our group developed an
   imaging pipeline for the homogeneous and efficient processing of
   this large data set. Having weak gravitational lensing as the main
   science driver, our algorithms are optimised to produce deep
   co-added mosaics from individual exposures obtained from empty
   field observations. However, the modular design of our pipeline
   allows an easy adaption to different scientific applications.
   Our system has already been ported to a large variety of
   optical instruments and its products have been used in various
   scientific contexts.  In this paper we give a thorough description
   of the algorithms used and a careful evaluation of the accuracies
   reached. This concerns the removal of the instrumental signature,
   the astrometric alignment, photometric calibration and the
   characterisation of final co-added mosaics. In addition we give a more
   general overview on the image reduction process and comment on
   observing strategies where they have significant influence on the
   data quality.
     }

   \titlerunning{Methods for WFI data reduction}
   \authorrunning{T. Erben et al.}

   \maketitle

\section{Introduction}
\label{sec:introduction}
\begin{figure*}[ht]
  \centerline{\includegraphics[width=0.7\hsize]{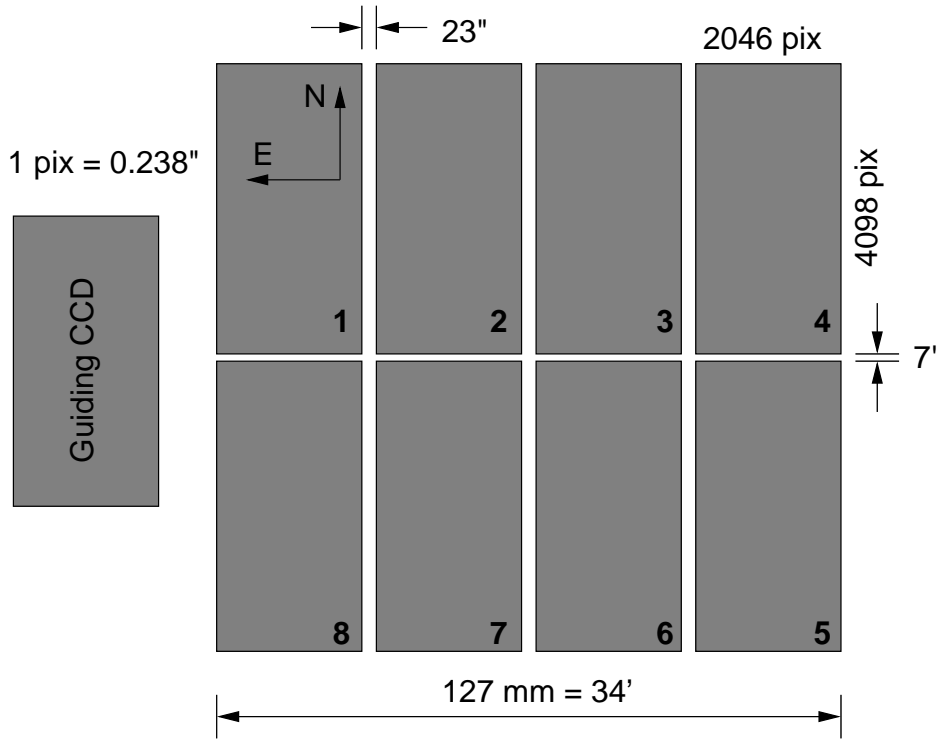}}
  \caption{\label{fig:WFIAREA}\small{Shown is the CCD layout of
  WFI@MPG/ESO 2.2m (not to scale). The mosaic design of CCDs to obtain a larger
  field-of-view generally confronts the user with new issues compared
  to the treatment of single-chip cameras.  The possible overlap of
  sources lying on different detectors on dithered exposures
  significantly complicates the astrometric and photometric
  calibration and the subsequent co-addition process.  First, the
  large field-of-view leads to notable optical distortions from the
  field centre to the edges (see also \figref{fig:wfi_distort}).  These
  distortions have to be known accurately for each CCD to create a
  correctly stacked mosaic. Second, for homogeneous photometry we need
  to take into account different quantum efficiencies and hence different photometric
  zero points of individual chips. Finally, dithering leads to
  non-uniform noise properties in the co-added image (due to the gaps, sensitivity
  and gain differences; see also \figref{fig:weight}). The
  knowledge of the relative noise properties between individual pixels
  in the co-added mosaic is important in the subsequent source catalogue
  creation process (see \figref{fig:weightcat}).  }}
\end{figure*}

During the last decades, optical Wide-Field Imaging has become one of
the most important tools in observational astronomy.  The advances in
this field are closely linked to the development of highly sensitive
detectors based on charge coupled devices (CCDs).  These detectors
have grown in size from a few hundred pixels on a side to the currently largest
arrays of about $4000\times 4000$ pixels.  The need for ever larger
fields-of-view and deeper images, and the technical constraints in the
manufacturing processes of larger CCDs led to the development of
multi-chip cameras in the mid 90's.  Hereafter, we refer to them
simply as \textit{Wide Field Imagers}, or WFIs. The production of such
a mosaic detector array with well aligned CCDs in three dimensions is
very difficult. The misalignments of individual chips, and the
multi-chip nature by itself, are the main reasons that the data
reduction process is significantly more complicated compared to the
treatment of single-chip cameras. Although a multi-chip CCD camera
seems to be a simple collection of $N$ single-chip cameras (see
\figref{fig:WFIAREA}) at first glance, it is, in general, not possible
to treat all images belonging to a certain detector independently
during the complete reduction process.\footnote{This is still possible
for certain scientific objectives not needing any image co-addition
(as variability studies in single frames for instance) or if data are
obtained in a very compact dither pattern so that different chips
never overlap.  As we will show in the course of the paper, compact
dither patterns have in general severe drawbacks on the final image
quality.} To obtain an accurately aligned co-added image on the whole
WFI scale we have to properly account for the mosaic structure with
its gaps, not perfectly aligned chips, and a large field-of-view that
needs to be treated as a curved surface instead of a flat plane.  We
have to apply sophisticated techniques going considerably beyond the
work that has to be done for a single-chip camera with a small
field-of-view.  In addition, a homogeneous photometric calibration and
the characterisation of the noise properties in a co-added mosaic
poses new challenges on calibration techniques.  Aside from the
technical issues related to the mosaic structure of new cameras, the
increasing quality of new optical instruments with their high
sensitivity and throughput demand the use of state-of-the-art
algorithms to optimally exploit the obtained data scientifically.

Up to now, no {\sl standard procedure} to process and treat
WFI data is established among observers. The main reason is that WFI
capabilities have only been available to a broader community for about
half a decade and we are still in the process of fully understanding the
properties of the new instruments and developing the necessary tools
for the data handling, and also for analysing reduced
images. Another issue is the high data flow and the demands on computer
hardware connected already to small and medium-sized WFI projects.
Only for about two years high-end, and for the occasional
observer affordable, desktop PCs are equipped adequately.

That these issues are non-trivial is illustrated by the relatively
small number of publications based on WFI projects compared to the
amount of data acquired. For example, after the WFI at the 2.2m
MPG/ESO telescope (hereafter WFI@2.2m) became operational in January
1999, only 35 refereed papers based on its data appeared until the end
of 2002.  Meanwhile, the rate has risen\footnote{The {\sl ESO
Telescope Bibliography} turns up a total of 107 refereed papers until
December 2004.}, but is still significantly behind those of other ESO
instruments which are under similar pressure by proposers as WFI@2.2m.

In this paper we present the methods and algorithms we use to process
WFI data. The presentation is organised as follows: In
\sectionref{sec:concepts} we introduce the basic concepts of the
GaBoDS pipeline and the philosophy behind the choices we made. We
discuss the advantages and the disadvantages that arise thereof.
The pre-reduction process (i.e. the removal of the instrumental
signature from the data), which is in principle identical for
single- and multi-chip cameras, is described in
\sectionref{sec:prereduction}.  Details of the astrometric and
photometric calibration of the data are presented in
\sectionref{sec:astromphotom}. An explanation of our adopted scheme to
deal with the inhomogeneous noise properties in co-added mosaic data
is given in \sectionref{sec:weightingflagging}, followed by the image
co-addition methods in \sectionref{sec:co-addition}. We perform
quality control checks on co-added data in
\sectionref{sec:co-addquality} 
and draw our conclusions in \sectionref{sec:conclusions}.

No astronomical pipeline will produce the best possible result with data
obtained in arbitrary conditions and strategies. Hence, besides a pure
description of algorithms, our presentation also contains guidelines
how WFI data should be obtained to achieve good results.

We assume familiarity with data processing of optical imaging data. In
this publication we focus on describing the algorithms used for all
necessary processing steps.  Where the reduction differs significantly
from well established algorithms for single-chip cameras (this mostly
concerns the astrometric alignment and the photometric calibration) we
give a scientific evaluation and a thorough estimation of the
accuracies reached.

We note that we mainly work on data from WFI@2.2m and most of the
examples and figures in this paper refer to data from this
instrument. The reader has to be aware that the quoted results, the
accuracies and the overall usability of our algorithms can vary
significantly when being applied to data sets from other cameras.  At
critical points we come back to this issue in the text.
%
\section{Pipeline characteristics}
\label{sec:concepts}
\subsection{Scientific motivation}
While all optical data need the same treatment to remove the
instrumental signature (bias correction, flat-fielding, fringe
correction etc; see \sectionref{sec:prereduction}), the subsequent
treatment of the images strongly depends on the scientific
objectives and on the kind of data at hand. Our primary scientific
interests are weak gravitational lensing studies 
(see e.g. \citeauthor{bas01} \citeyear{bas01} for an extended review)
in deep blank-field 
surveys. These studies mainly depend on shape measurements 
of faint galaxies. To ensure that the light distribution is deformed
as little as possible
by the Earth's atmosphere and the optics of the telescope,
weak lensing data are typically obtained under superb seeing
conditions at telescopes with state-of-the-art optical equipment.  
For a proper treatment of those data, the main requirements on a 
data processing pipeline are the following:
\begin{itemize}
\item We need to align very accurately galaxy images of subsequent 
exposures that are finally stacked. This involves an accurate
mapping of possible distortions (see \figref{fig:wfi_distort}).
\item We need the highest possible resolution, and hence a
co-addition on a sub-pixel basis (see \sectionref{sec:co-addition}).
Together with the previous step, this is crucial for an
accurate measurement of object shapes in the subsequent lensing
analysis.
\item We need to accurately map the noise properties in our
finally co-added images (see \figref{fig:weightcat}). This knowledge
enables us to use as many faint galaxies as possible and to push the
object detection limit. This also requires that the sky-background in
the co-added image is as flat as possible.
\end{itemize}
The algorithms we use are chosen to go from raw images to
a final co-added mosaic with the objectives described above. The
responsibility of the pipeline ends with the co-addition step.
We are fully aware that the chosen
procedures and algorithms may not be optimal for projects
having different scientific objectives such as accurate photometry, the 
investigation of crowded fields or the study of large scale, low 
surface brightness objects for instance.
However, the modular design of our pipeline (see below) allows 
an easy exchange of algorithms or the integration of new methods
necessary for different applications. 
The main characteristics of our current pipeline are summarised in the
following:
\begin{itemize}
\item Ability to process exposures from a multi-chip camera with
$N$ CCDs. To date we have successfully processed data from: 
WFI@2.2m, CFH12K@CFHT, FORS1/2@VLT, 
WFI@AAO, MOSAIC-I/II@CTIO/KPNO, SUPRIMECAM@Subaru, WFC@INT, WFC@WHT,
and several single-chip cameras (e.g. BUSCA@2.2m Calar Alto).
\item Ability to handle mosaic data taken with arbitrary dither patterns.
\item Precise image alignments without prior knowledge about
distortions. The astrometric calibration is performed with the
data itself.
\item Absolute astrometric calibration of the final images to sky coordinates.
\item An appropriate co-addition of data obtained under different
photometric conditions.
\item Image defects on all exposures are identified and marked before
the co-addition process.
\item Creation of weight maps taking into account different noise/gain properties
and image defects for the co-addition.
\item Accurate co-addition on sub-pixel level/ability to rescale data to
an ``arbitrary'' scale (ability to easily combine data from different 
cameras/telescopes).
\end{itemize}
We are currently extending our algorithms to near-IR cameras which
will be described in a forthcoming publication (Schirmer et al. in prep.).
\subsection{Implementation details}
\label{subsec:pipecharacteristics} 
As mentioned above, many tools for WFI data processing are currently
under active development and different groups have already released
excellent software packages for specific tasks. Hence, we built our
own efforts on already publicly available software modules wherever
possible. Many of the algorithms used are very similar to those
developed for the EIS (ESO Imaging Survey) Wide survey in 1997-1999
\citep[see][]{nbc99}. The main pillars of our pipeline are the following
software modules:
\begin{itemize}
\item {\bf The LDAC software\footnote{available at ftp://ftp.strw.leidenuniv.nl/pub/ldac/software/}:}
The LDAC (Leiden Data Analysis Centre) software was the backbone of 
the first EIS pipeline. It provides a binary catalogue format (in the
form of system-independent binary FITS tables)
including a large amount of tools for their handling. Moreover, this module
contains software for the astrometric and photometric calibration of 
mosaic data.
\item {\bf The EIS Drizzle\footnote{available via ESO's SCISOFT CD
(see http://www.eso.org/scisoft/)}:} A specific version of the IRAF
package {\tt drizzle} \citep[][]{frh02} was developed
for EIS. It directly uses the astrometric calibration provided by the
LDAC astrom part and performs a weighted linear co-addition of the
imaging data.
\item {\bf TERAPIX software\footnote{available at http://terapix.iap.fr/soft/}:}
\citep[][]{bmr02}
{\tt SExtractor} is used to obtain object catalogues for the astrometric calibration.
Moreover it produces a cosmic ray mask in connection with {\tt
  EYE}\footnote{{\tt EYE} (Enhance
Your Extraction) allows the user to generate image filters for the detection
of small-scale features in astronomical images by machine learning (neural networks).
These filters are loaded into {\tt SExtractor} that performs the actual detection. We use
such filters to detect cosmic rays in our images.} in addition
to smoothed and sky-subtracted images at different parts of the pipeline.
{\tt SWarp}, the TERAPIX software module for resampling and co-adding
FITS images, is used alternatively to {\tt EIS Drizzle} for the final image 
co-addition. 
\item {\bf FLIPS\footnote{see http://www.cfht.hawaii.edu/$\sim$jcc/Flips/flips.html}:}
\citep[][]{mac04}
{\tt FLIPS} is one of the modules for data pre-reduction. It is optimised to
perform operations on large format CCDs with minimal memory
requirements at the cost of I/O performance. 
\item {\bf Eclipse and qfits tools\footnote{available at http://www.eso.org/projects/aot/eclipse/}:}
\citep[][]{dev01}
We use several stand-alone FITS header tools from the {\tt Eclipse} package to
update/query our image header. Moreover, tools based on the {\tt
  qfits} library offer an alternative  to {\tt FLIPS} for image pre-reduction.
\item {\bf Astrometrix\footnote{available at http://www.na.astro.it/$\sim$radovich/wifix.htm}:}
{\tt Astrometrix} (developed at TERAPIX) is another module for obtaining astrometric calibration.
\item {\bf IMCAT utilities\footnote{available at http://www.ifa.hawaii.edu/$\sim$kaiser/imcat/}:}
From the {\tt IMCAT} tools we extensively use the image calculator {\tt ic}.
\end{itemize}
These tools have been adapted for our purposes if necessary and
wrapped by UNIX/bash shell scripts to form our pipeline. Our main
effort is to provide the necessary interfaces for the communication
between the individual modules and to add instrument and science
specific modules for our purposes (as a thorough quantification of PSF
properties for instance).  With our implementation approach we can
make use of already well-tested and maintained software packages.  The
description above also shows that we can easily exchange modules as
soon as new algorithms or better implementations for a certain task
become available. To further enhance the modularity of the pipeline,
we split up the whole reduction process into small tasks (a complete
reduction process for WFI@2.2m data typically involves the call to
20-30 different scripts but {\sl superscripts} collecting several
tasks can easily be written). This ensures a very high flexibility of
the system and that potential users can easily adapt it to their needs
or to specific instruments.

Nearly the whole system is implemented in ANSI C and UNIX bash shell
scripts ({\tt EIS Drizzle} is written in FORTRAN77 and embedded into
IRAF, {\tt Astrometrix} is developed under Perl+PDL and parts of our
photometric calibration module are implemented in Python). Thus we
have full control over the source codes and can port the pipeline to
different UNIX flavours.  \tableref{table:testedmachines} lists the
systems on which we successfully used it so far.

The main disadvantage of building up a pipeline from many different
software modules instead of developing a homogeneous system from
scratch is that it becomes very difficult to automatically control the
data flow or to construct a sophisticated error handling and data
integrity checking system.  Furthermore, our pipeline so far offers
only very limited possibilities to store the history of the reduction
process or to administrate raw and processed image products. Also,
formal speed estimates for the throughput of our processing system are
low compared to homogeneous systems (see \tableref{table:cputime}).
Thus, the usability for large, long-term projects of the system is
limited at this stage.

With the compactness, the flexibility and the usability of our system
for the occasional user\footnote{The usage of our pipeline can easily
be learned within a few days by users having good experience in the
reduction of single-chip cameras, as was experienced by several guests
who made use of our visitor programme.} we regard our tools
complementary to the survey systems developed by other groups such as
TERAPIX\footnote{http://terapix.iap.fr/} or
ASTRO-WISE\footnote{http://www.astro-wise.org/}. We note that the EIS
team recently released a WFI pipeline
\citep[][\citeyear{van04}]{van02} with a functionality similar to
ours\footnote{http://www.eso.org/eis/}.  An alternative, and widely
used reduction package for CCD mosaics is the {\tt mscred} environment
within {\tt IRAF}. It is described in \citet{val02}.
\begin{table}
\begin{center}
\caption{Listed are the different UNIX architectures on which we used
our pipeline so far. Besides the system we list the compilers and the
co-addition module used. We compared the final co-added images from
different systems with respect to flux and shape measurements from
extracted objects. The individual reductions do not show significant
differences.}
\begin{tabular}{cccc}
\hline
\hline
System & compiler & co-addition \\
\hline
Linux/i386 & gcc-2.95.3   & {\tt SWarp}, {\tt EIS Drizzle}  \\
Linux/alpha & gcc-2.96    & {\tt SWarp}          \\
Linux/AMD64 & gcc-3.3.4    & {\tt SWarp}          \\
SunOS      & gcc-2.95.2   & {\tt SWarp}, {\tt EIS Drizzle}  \\
IBM AIX    & gcc-2.95.2   & {\tt SWarp}           \\
HP-UX      & gcc-2.7.2.2  & {\tt SWarp}           \\
OSF1       & gcc-2.7-97r1 & {\tt SWarp}           \\
IRIX64     & gcc-3.0.4    & {\tt SWarp}           \\
\hline
\end{tabular}
\label{table:testedmachines}
\end{center}
\end{table}
\subsection{Parallel processing}
A great deal of the reduction processes is performed independently on
individual chips or on a set of images belonging to the same CCD. Only
very few steps require information or imaging data from the complete
mosaic. This mainly concerns the estimation of a global astrometric
and photometric solution for the entire mosaic (which is done on
catalogues) and the final image co-addition.  Hence, most processing
tasks can naturally be performed in parallel if the mosaic data is
split up on a detector basis and if a multi-processor architecture is
at hand. Our pipeline is designed to perform this kind of
parallelisation on a single multi-processor machine as well as on
machines connected over a network (such as a Beowulf Cluster for
instance). This parallelisation scheme of performing the same task
simultaneously on different machine processors rather than using
explicit parallel code for the individual tasks has two main
advantages:
\begin{itemize}
\item No new software has to be developed for the parallel
processing. We developed a {\sl parallel manager} that launches
and surveys script execution on the individual cluster nodes. The
parallel manager is also used in standard, single-node processing 
so that the pipeline operation is unified in single- and
multi-processor mode. 
\item As the operations on different chips are completely independent
and no communication is necessary between the tasks, no message
passing/synchronisation or data exchange schemes have to be
implemented. We only synchronise the processing after every processing
step, waiting until a certain job is done on all chips before the next
task is launched.
\end{itemize}
The major disadvantage in our scheme not allowing a parallelisation in
the processing of a given chip is that resources are not used
optimally if more nodes/processors than chips are available (nodes are
simply staying idle for the complete process) or if the number of
nodes is not an integer multiple of chips (some nodes have to work on
more chips than others, basically doubling processing time).
Nevertheless, we think that the advantages heavily outweigh this
disadvantage for our purposes. Moreover, most multi-chip cameras so
far have $2^N$ chips matching typical architectures of computer
clusters. In \tableref{table:cputime} we compare processing times from
single- and multi-node processing of an observing run.

\begin{table}
\caption{\label{table:cputime} Processing times for a WFI@2.2m set
consisting of 10 Bias, 12 Dark, 12 Skyflat and 11 Science frames.  The
third column lists the percentage of the processing time with respect
to the single-node case. The number in parentheses gives the
percentage if the whole process could be parallelised, i.e. 100\%
divided by the number of nodes. The fourth column lists the formal
number of Science frame pixels processed in one second. The times
include the full processing from raw data as they arrive from the ESO
archive to the final co-added images [it does not include the time for
eye-balling and masking of images (see \sectionref{sec:manualpass})].
We note that the parts that cannot be parallelised (an estimate for
the global astrometric alignment with {\tt Astrometrix}, the global
photometric solution and the final co-addition with {\tt SWarp}) and
have to be done on a single processor add up to 24 min in our example.
This becomes a very significant fraction of the total processing time
in the 4 and 8 node cases. Each node is equipped with an Athlon XP
2800+ processor and 1 GB of main memory. For a comparison, for the
current EIS pipeline a throughput of 0.4-0.5 Mpix/s on a
double-processor machine is quoted \citep[][]{van02}.}
\begin{center}
\begin{tabular}{cccc}
\hline
\hline
used nodes & time [min] & time [\%] & Kpix/s \\
\hline
1 & 240 & 100   (100) & 50 \\
2 & 131 & 54.0   (50) & 90 \\
4 &  77 & 32.0   (25) & 152 \\
8 &  52 & 21.7 (12.5) & 226 \\
\hline
\end{tabular}
\end{center}
\end{table}

\subsection{Terminology}
\label{subsec:terminology}
We use the following terminology throughout the paper:
\begin{itemize}
\item{CCD, chip -- one of the detectors in a WFI}
\item{exposure -- a single WFI shot through the telescope}
\item{image -- the part of the exposure that belongs to a particular CCD}
\item{frame -- this term is used as a synonym for an image.}
\item{BIAS -- an exposure/image with zero seconds exposure time}
\item{DARK -- an exposure/image with non-zero exposure time, keeping the
  shutter closed}
\item{FLAT -- an exposure/image of a uniformly illuminated area; this can
  be the telescope dome giving DOMEFLATs or sky observations during
  evening and/or morning twilight providing SKYFLATs}
\item{SCIENCE -- an exposure/image of the actual target, not a calibration image}
\item{SUPERFLAT -- properly stacked SCIENCE data to extract large
  scale illuminations or fringe patterns.}
\item{STANDARD -- an exposure/image of photometric standard stars}
\item{other terms written in CAPITAL letters denote additional 
  images or calibration frames. Their meaning will be clear within the
  context.}
\item{mosaic -- this term is used as a synonym for exposure or for
the final co-added image.}
\item{dithering -- offsetting the telescope between the exposures}
\item{overlap -- images from different CCDs and exposures overlap if
  the dithering between the exposures was large enough}
\item{stack -- a set of $n$ images belonging to the same chip that
have been combined pixel by pixel.}
\item{Names of software packages are written in the {\tt TypeWriter} font.}
\end{itemize}
\section{Pre-reduction ({\sl Run} processing)}
\label{sec:prereduction}
In the following we describe our algorithms for the pre-reduction of
optical data, i.e. the removal of the instrumental signature.  The
first issue is the compilation of data for this step.  For most of the
effects to be corrected for (instrument bias, bad CCD pixels,
flat-field properties) we can safely assume stability of the instrument
and the CCD characteristics over several days or even a few
weeks. Hence, in many cases we can collect data from a complete {\sl
observing run} and benefit from having many images for the necessary
corrections from which most are of statistical nature. As described
below, the matter can become more complex when dealing with strong
fringes in red passbands. Here, the time scales from which SCIENCE
data can safely be combined in the process is much shorter, sometimes
only a couple of hours. The issue is further discussed in
\sectionref{sub:guidelinesprered}.  In any case we say that we
perform the pre-reduction process on a {\sl Run basis} regardless of
how long this period actually is. The pre-reduction is done
independently on each CCD of a mosaic camera. Only in one step, the
sky-background equalisation (see \sectionref{subsec:gainequalisation}), the
action to be performed on a CCD depends on properties from the rest of
the mosaic. Hence, unless stated otherwise, each step described below
has to be performed on a detector basis. For this part of the pipeline
we can use two different software packages, one based on {\tt FLIPS}, the
other on {\tt Eclipse}. {\tt FLIPS} is very I/O
intensive but has very low memory requirements, while {\tt Eclipse} reduces
the necessary I/O to a minimum and operates very efficiently on
imaging data by keeping it in virtual memory. Depending on the size of
the data set and the computer equipment at hand, one or the other is
preferable. In the following we focus on the description of the
{\tt Eclipse} package and we mention {\tt FLIPS} where its functionality differs
significantly from {\tt Eclipse}.
\subsection{Handling the data and the FITS headers}
\label{subsec:datahandling}
The variety of ways in which header information and raw data from WFIs
are stored in FITS files are as large as the number of instruments,
i.e. so far there is no established standard on the FITS format of CCD
mosaics. In order to cope with the different formats and to unify the
treatment, we perform the following tasks on raw data:
\begin{enumerate}
\item If the data are stored in Multiple Extension Fits (MEF) files, we
extract all the images from them. All subsequent pipeline processing
is done on individual images and also our pipeline parallelisation is
based on the simultaneous execution of the same task on different
images.
\item We substitute the original image headers from the chips by a 
new one containing only a minimum set of keywords necessary for a 
successful pipeline processing (see \appendixref{sec:imageheader}). 
Especially the astrometric calibration depends on correct header entries.
\item If necessary we flip and/or rotate individual CCDs to bring all
images of the mosaic to the same orientation with respect to the
sky. Only rotations by 90 degrees, that do not require pixel
resampling, are performed.
\end{enumerate}
All these tasks are performed by a {\tt qfits}-based utility.
\subsection{Modes, medians, and the stacking of images}
\begin{figure}[ht]
  \includegraphics[width=1.0\hsize]{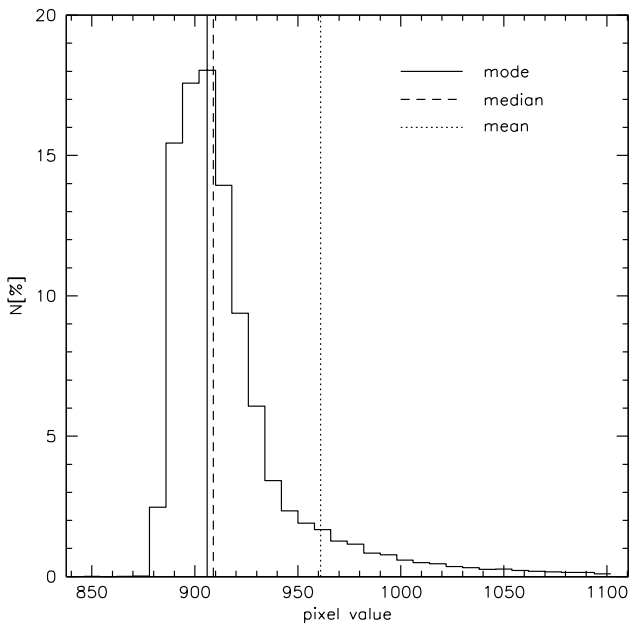}
  \caption{\label{fig:pixdist}\small{Shown is a typical pixel value
  distribution in an astronomical image. The clear peak denotes the
  brightness of the night sky. Due to the presence of objects the
  distribution is strongly skewed towards high values. The most
  accurate estimate for the sky comes from analysing the histogram and
  determining the peak (the mode of the distribution) directly. The
  median (for a set of $N$ points the median element is defined so
  that there are as many greater as smaller values than the median in
  the set) also gives a very robust (and for our purposes sufficient
  though biased) estimate for it. Due to the long tail of the
  distribution, a straight mean is completely useless as an estimate
  for the sky value. As the values to the left of the mode form half
  of a Gaussian distribution, an estimate for the sky variance can be
  obtained from that part.}}
\end{figure}
In most pre-reduction steps we have to perform robust statistical
estimates on a set of pixels following Poisson statistics.
At some points we have to estimate the sky-background and its
variance in SCIENCE data. A typical distribution of
pixel values on a SCIENCE image is shown in \figref{fig:pixdist}.
The second common operation is to statistically stack several
input images to produce a noise-reduced calibration frame. This
master image is then used for subsequent calculations on
SCIENCE images. In the following we define several 
{\sl operators} acting on images:
\begin{itemize}
\item {\sl value=median(image):} \\
The median of an image is estimated. This is done by
collecting a representative sample from the CCDs pixels (we
take about 5\% of the detector pixels in a region around the 
CCD centre),
sorting the obtained pixel array and returning its middle element
(in case the array has an even number of elements we return
the mean of the two middle elements).
\item {\sl value=mode(image):} \\ 
The mode of an image is
estimated. We consider the same representative pixel sub-sample as for
the median estimation, build a smoothed histogram and return the peak
value.
\item {\sl image=rescale(images):}\\
This operator rescales a set of images so that they have
the same median after the process. The resulting median is chosen
to be the mean of the medians (we write {\sl meanmed} for it)
from the input images. Hence on each input image we perform 
the operation
{\sl image $\rightarrow$ image * meanmed / median(image)}.
The operator is usually applied to a set of images before
they are stacked (as SKYFLATs or SCIENCE images). The median
of the stacked image is then by construction equal to 
{\sl meanmed}.
\item {\sl image=stack(images):} \\
A set of input images is stacked to produce a noise-reduced
{\sl master image}. The following procedure is performed
independently on each pixel position. We collect all pixel
values from the input images into an array, sort it and
reject several low and high values (typically we reject the
three lowest and three highest values if we have 15 input
images or more). From the rest we estimate the median that
goes into the master image. Here {\tt FLIPS} uses a more 
sophisticated algorithm. On the remaining array it
first performs an iterative sigma clipping. 
It estimates mean and sigma, rejects low and high values 
(typically pixels lying more than 3.5 sigma below and above the 
mean) and repeats the procedure until no more pixels 
are rejected. From the rest it returns the median
for the master image.
\end{itemize}
We will use additional pseudo operators in the following whose
meaning and behaviour will be clear within the context. All
calculations with these operators are written in {\sl slanted}
notation.
\subsection{A first quality check, Overscan correction, master BIASes and DARKs}
\label{subsec:overscanbiasdark}
Before any exposure enters the reduction process, we estimate its
mode. If this estimate lies outside predefined boundaries, the
exposure will not be processed. This rejects most of the
completely faulty images (such as saturated SKYFLATs) at the very
beginning.  After this initial quality check, the first step in the
reduction process of each image is the correction of an overall
estimate for the BIAS value by considering pixels in not illuminated
parts of each CCD (the overscan region).  This first-order BIAS
correction is done by collecting for each line all pixels in the
overscan region, rejecting the lowest and highest values (usually the
overscan regions contains about 40 columns and we reject the 5 lowest
and 5 highest values) and estimating a straight mean from the
rest. This mean is subtracted from the corresponding line. After this
correction, the overscan regions are trimmed off the images.  For
correcting spatially varying BIAS patterns in the FLAT or SCIENCE
images, a {\sl master BIAS} is created for each CCD from several
individual BIAS exposures:
\begin{enumerate}
\item Each BIAS exposure is overscan corrected and trimmed.
\item The master BIAS is formed by {\sl masBIAS=stack(BIAS frames)}
\end{enumerate} 

In the same way, a master DARK ({\sl masDARK}) frame is created.  So
far, we do not correct FLAT or SCIENCE frames for a possible dark
current but we use the masDARK for the identification of bad pixels,
rows and columns (see \sectionref{sec:weightingflagging}).
\begin{figure*}[ht]
  \includegraphics[height=1.0\hsize,angle=-90]{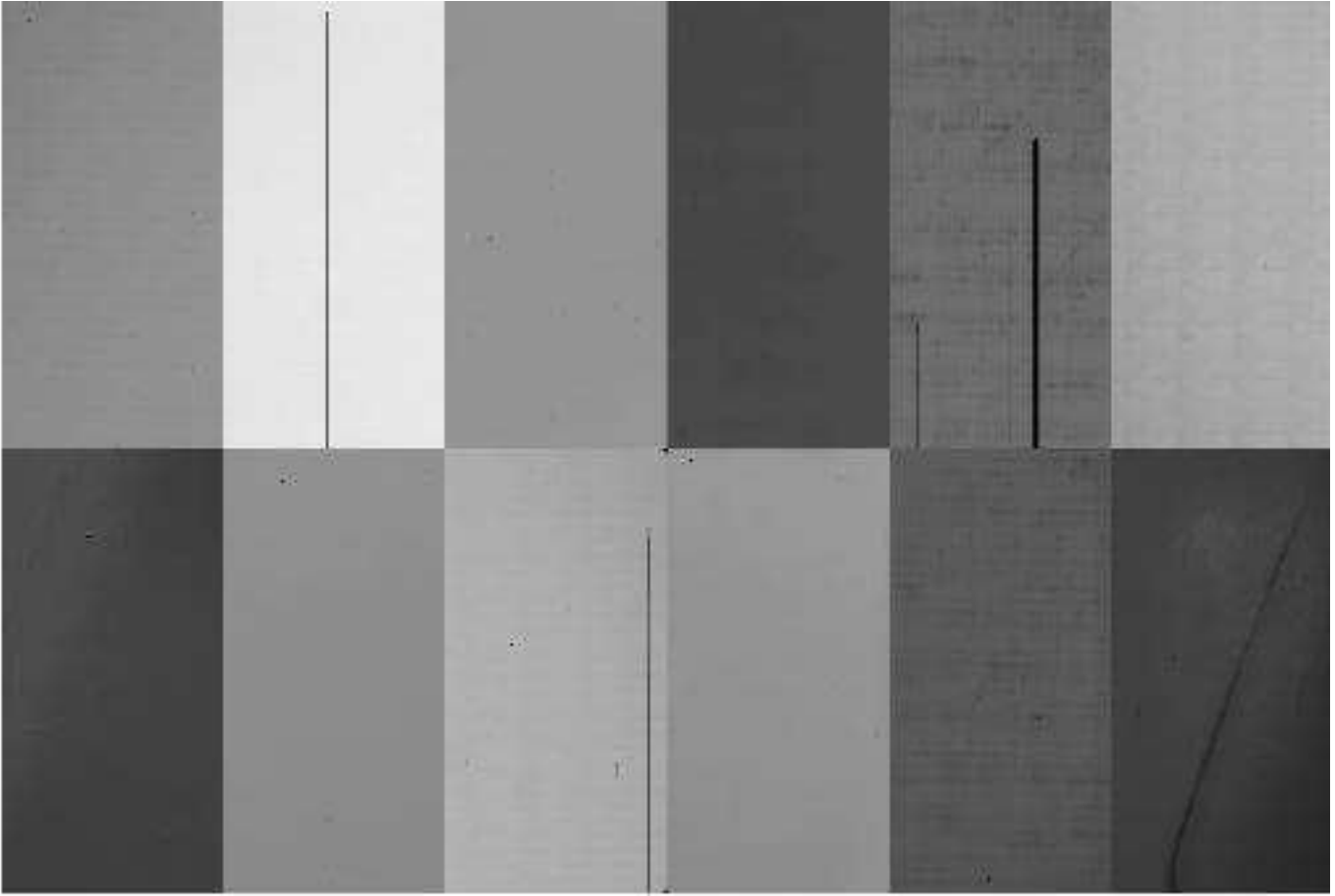}
  \caption{\label{fig:flat}\small{A skyflat from the CFH12K@CFHT
  camera \citep[][]{cls00}. We
  note the significant intrinsic sensitivity and gain differences between the
  chips. The counts of the second left detector in the top row are about
  a factor 1.3 higher than that of the two chips at the corners in the
  bottom row. Within a CCD, large-scale gradients are up to $8\%$.
  Especially the four chips on the right also show small-scale
  sensitivity variations across the whole CCD.}}
\end{figure*}
\subsection{Methods for flat-fielding}
\label{subsec:howtoflat}
The illumination of a CCD and its sensitivity is in general not homogeneous
over the detector area. Illumination usually varies on large 
scales, but sensitivity can change significantly from pixel to
pixel (see \figref{fig:flat}). In 
order to overcome these effects and to achieve a homogeneous surface
brightness, SCIENCE images have to be flat-fielded. The flat-field
pattern typically varies with the observation configuration and hence
calibration images have to be obtained for each filter to be used.
There are three major
possibilities to obtain an estimate for the flat-field-pattern:
\begin{enumerate}
\item {\bf DOMEFLATs:} \\ One can obtain flat images by pointing 
the telescope to an illuminated, blank surface in the dome. These
DOMEFLATs can be taken during day time, the images do not contain
any astronomical sources, we can take as many exposures as we wish and
we can adjust the count rate in the frames. This ensures FLATs with a
very high signal-to-noise. However, especially for WFIs, the uniform
illumination of a large enough area is very difficult to achieve. In
addition, lamps emit light at a maximum temperature of around 3000 K
so that the spectral distribution is very red. In red filters this can
lead to an excess amount of fringing as compared to SKYFLATs.  For
these reasons, the flat calibration of SCIENCE frames with DOMEFLATs
is usually inferior to the two methods described below (especially at
the correction of large-scale illumination variations). We only use
them if SKYFLATs are not available.
\item {\bf SKYFLATs:} \\ One can also obtain flat images by observing the
sky with short exposure times during the evening and morning
twilight. These flats resemble the night sky during SCIENCE
observations better than dome flats, although for WFIs covering an
area of $30\arcmin\times 30\arcmin$ or more we expect variations of
the twilight sky over their field-of-view. At many telescopes, an
automatic setup to obtain exposures with a desired count-rate are
available and hence these flats also have high $S/N$.  See also
\citet{tyg93} for a guide to obtain twilight flats if an automatic
procedure is not available. However, only a
very limited number can be obtained during the twilight phase. This
holds especially if several filters have to be calibrated and/or if
the readout time of the detector is rather long. Using an exposure
time too short in bright twilight can lead to a non-uniform
illumination due to shutter effects; see
\sectionref{sub:guidelinesprered}. However, CCDs are usually very
stable over a few days or weeks so that a larger number of SKYFLATs
taken during several nights can be combined into good master flat
images.
\item {\bf SUPERFLATs:}\\ \label{item:superflat} In addition, one can
try to extract the flat-field pattern from the SCIENCE observations
itself. If a sufficient number of SCIENCE exposures is at hand
(usually more than a dozen), and if the dither pattern was
significantly larger than the largest object in the field, each pixel
of the camera will see the sky-background several times. Hence, a
proper combination of these exposures yields a master FLAT that
closely represents the night sky during observations. We refer to such
a FLAT as a SUPERFLAT.  A straightforward application of this method
is often hampered by several factors: During phases of grey and bright
time, the night sky shows large gradients and variable reflections in
the dome and telescope can occur. Thus, a careful selection of images
that go into the SUPERFLAT has to be done. In medium/narrow band
filters and in the ultra violet, the counts and hence the $S/N$ of
SUPERFLATs in these bands are typically low.  Furthermore, the
technique is very difficult to apply in programmes where the imaged
target has a size comparable to the field-of-view.  In this case also
large dither patterns cannot assure sufficient sky coverage on the
complete mosaic.
\end{enumerate}
As the superflat technique cannot be applied in general we use
the following two-stage flat-fielding process:
\begin{enumerate}
\item In a first step, the SCIENCE observations are flat-fielded
with SKYFLATs (if those are not at hand we use DOMEFLATs instead).
This typically corrects the small-scale sensitivity variations and
leaves large-scale gradients around the $3\%$ level on the scale
of a chip (see \figref{fig:reduced}).
\item If the data permit it, a SUPERFLAT is created out of the 
flat-fielded SCIENCE observations, smoothed  with a large kernel to extract remaining
large-scale variations and is applied to the individual images. For our
empty field observations at the WFI@2.2m this leaves typical large-scale
variations around the $1\%$ level in $U, B, V, R$ broad band
observations (in $I$ the presence of strong fringing often leads
to significantly higher residuals). 
\end{enumerate}
This two-stage flat-fielding process is very similar to the
method adopted by \citet{ars01}.
\subsection{The creation of master DOME-/SKYFLATs}
\label{subsec:flat}
The master FLAT (SKYFLAT or DOMEFLAT) 
for each CCD is created as follows:
\begin{enumerate}
\item All individual FLAT exposures are overscan-corrected and
trimmed.
\item The masBIAS is subtracted from all images.
\item The FLATs are rescaled to a common median:
{\sl rescFLAT=rescale(FLAT)}.
\item We form the master FLAT by {\sl masFLAT=stack(rescFLAT frames)}
\end{enumerate}
\subsection{Sky-background equalisation}
\label{subsec:gainequalisation}
Within a mosaic, CCDs have different quantum efficiencies, varying
intrinsic gains and hence
different photometric zero points (see \figref{fig:flat}). For the
later photometric calibration it is desirable to equalise the
photometric zero points in
all detectors, which is achieved by scaling all CCDs to the same
sky-background.
We rescale all chips to the median of the CCD with the highest count-rate
during flat-fielding. If possible we perform this step within the
SUPERFLAT correction as its median estimation is more robust than with
the SKYFLATs.  This is because SKYFLATs show larger variations in
brightness than the SUPERFLATs which are calculated from already
flat-fielded data. We estimate that photometric zero points of the
mosaic agree with an rms scatter of about 0.01-0.03 mag after sky-background
equalisation (see \sectionref{subsec:setcharacteristics}).
\subsection{SCIENCE image processing, the creation and application of the SUPERFLAT}
\label{subsec:science}
\begin{figure*}[ht]
  \includegraphics[height=1.0\hsize,angle=-90]{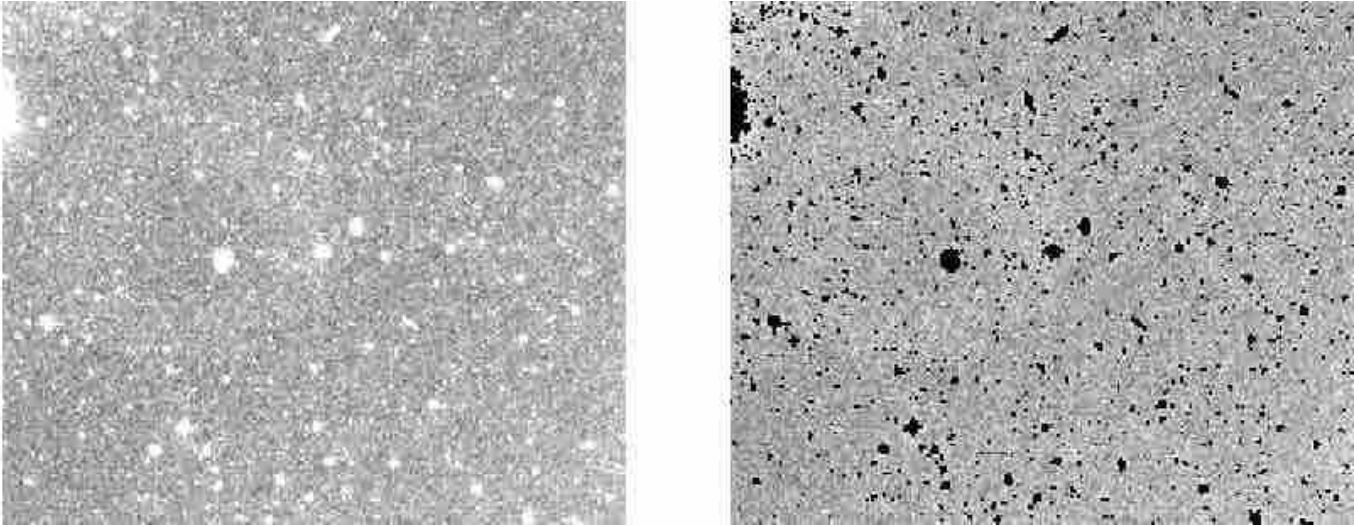}
  \caption{\label{fig:sub}\small{For the creation of SUPERFLATs we
  mask pixels belonging to astronomical sources in the SCIENCE
  frames. The left panel shows a SCIENCE image before and the right
  panel after objects have been detected by {\tt SExtractor} and flagged. We
  mask all structures having 50 contiguous pixels with 1.0 sigma above
  the sky-background.  This helps to pick up also pixels of extended
  halos around bright stars.  For images containing strong fringing we
  change the parameters to 7 pixels above 5 sigma as otherwise fringes
  would be detected and masked.}}
\end{figure*}

\begin{figure*}[ht]
  \includegraphics[height=1.25\hsize,angle=-90]{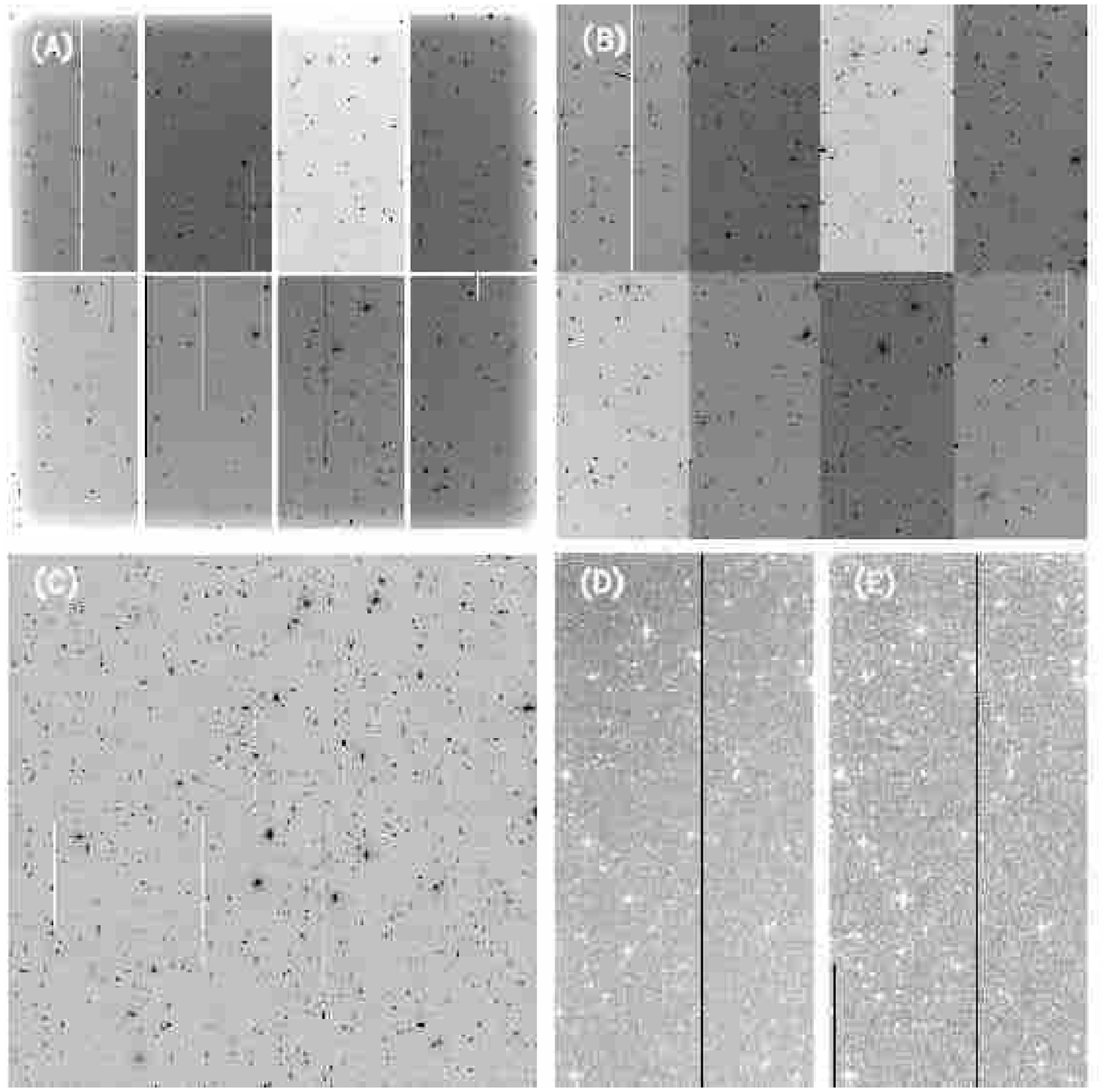}
  \caption{\label{fig:reduced}\small{The pre-reduction steps on a
  $V$-band exposure from WFI@2.2m: Panel (A) shows the raw
  image, panel (B) the result from applying masBIAS, trimming
  the CCDs and flat-fielding with the masFLAT. The high $S/N$ masFLAT
  takes out small-scale variations but leaves large-scale residuals of
  up to 3\% on the scale of the CCD [panel (D)].  
  These and the differences in the sky count-rate are removed after the
  application of ILLUMINATION giving a flatness of 1\% over the entire
  mosaic in most of the cases as seen in panels (C) and (E).}}
\end{figure*}

The individual steps in the SCIENCE frame processing are:
\begin{enumerate}
\item The images are overscan-corrected and trimmed. Afterwards 
the masBIAS is subtracted and the frames are divided by masFLAT.
We write {\sl flatSCIENCE=(SCIENCE$-$masBIAS)/masFLAT}.
In the case where no SUPERFLAT correction is applied to the data,
masFLAT is rescaled at this step to equalise the sky-background between
individual detectors (see \sectionref{subsec:gainequalisation}).
\item For the creation of the SUPERFLAT we first remove
astronomical sources from the flatSCIENCE images. To this end
we run {\tt SExtractor} \citep{bea96} on them and create a new set of 
images where pixels belonging to objects (i.e. above certain detection
thresholds) are flagged 
({\sl objSCIENCE=flatSCIENCE$-$OBJECTS}).
See also \figref{fig:sub}. The flagged pixels are not taken into account
in the subsequent processing.
\item The objSCIENCE images are rescaled to a common median
{\sl rescobjSCIENCE=rescale(objSCIENCE)}.
\item We calculate the SUPERFLAT 
[{\sl SUPERFLAT=stack(rescobjSCIENCE frames)}].
If all input pixels of a given position are flagged (i.e. all the 
images had an object at that position) we assign the {\sl meanmed} 
value from the objSCIENCE frames to the SUPERFLAT.
\item The SUPERFLAT is heavily smoothed by creating a {\tt SExtractor}
BACKGROUND check-image with a background mesh of 512 pixels (see
\citeauthor{ber03} \citeyear{ber03} on the SExtractor sky-background
estimation). This image, the illumination correction image, forms the
basis for removing large-scale flat-field variations ({\sl
ILLUMINATION=smooth(SUPERFLAT)}).
\item The flatSCIENCE images are divided by the ILLUMINATION image
which has been rescaled to equalise the sky-background of the different
detectors (see \sectionref{subsec:gainequalisation}). 
We write {\sl illumSCIENCE=flatSCIENCE/rescaled(ILLUMINATION)}.
\end{enumerate}
For blue passbands not showing any fringing, the pre-reduction ends here
and the illumSCIENCE images are used in the subsequent processing.
See \figref{fig:reduced} for an example of a pre-reduced $V$-band
exposure. 
\subsection{Fringing correction in red passbands}
\label{subsec:fringingcorrection}
\begin{figure*}[ht]
  \includegraphics[width=0.97\hsize,angle=-90]{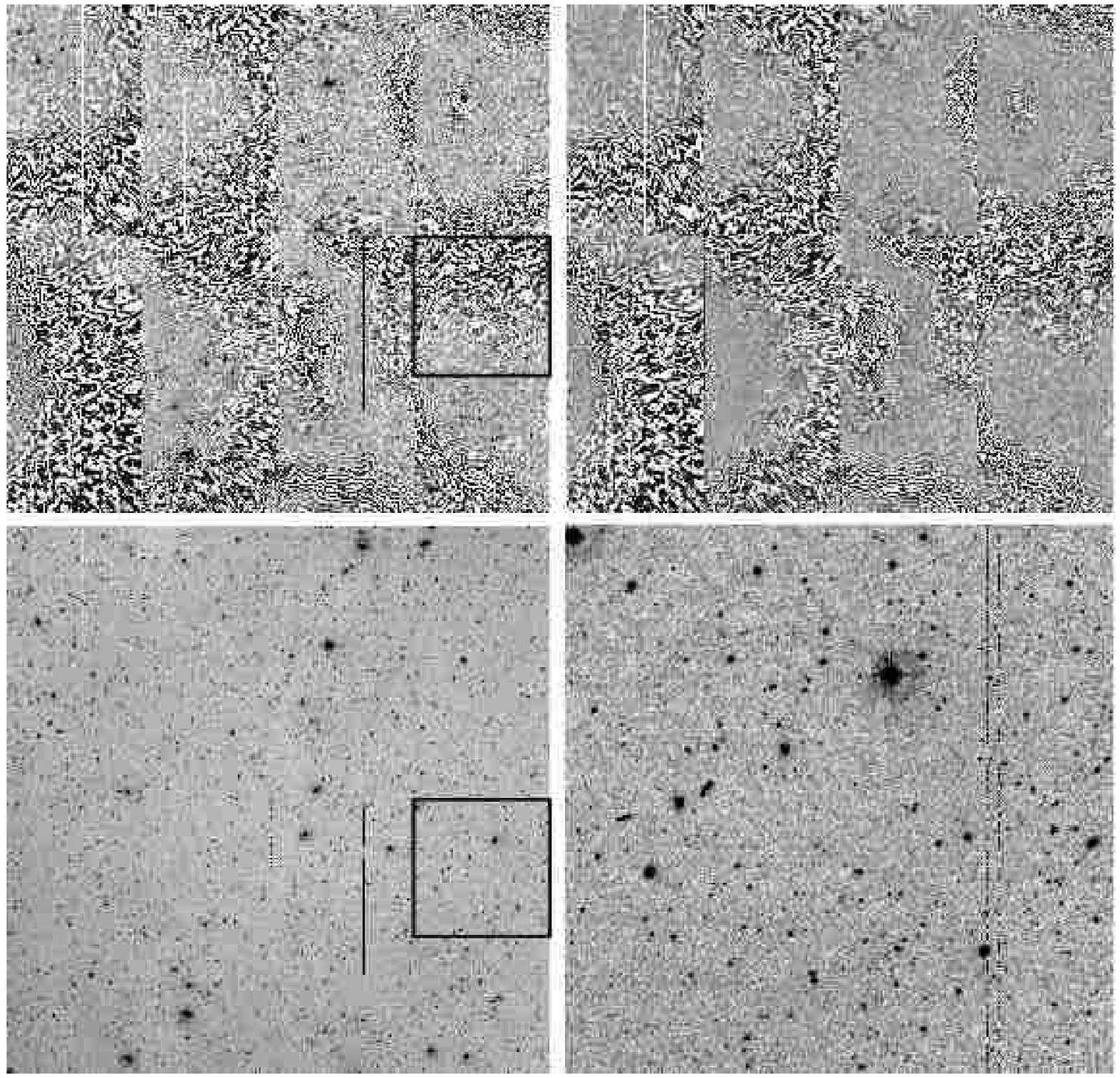}
  \caption{\label{fig:fringes}\small{Fringing correction in WFI@2.2m
  $I$-band data: For the shown correction, fifteen 300 s exposures
  have been obtained within 1.5 hours during stable, photometric
  conditions with a large dither pattern (3 arcmin in Ra and Dec)
  on a blank-field target. The upper left panel shows a SCIENCE
  frame before fringing correction, the upper right panel the
  extracted fringe pattern from the SUPERFLAT and the lower left 
  panel the SCIENCE
  image after the correction. The lower right panel shows a zoom
  of a representative region (marked in the left two panels) after the correction. 
  The residual fringes have an amplitude of about 0.1\% of the night sky.}}
\end{figure*}

Fringing is observed as an additional, additive instrumental signature
in red passbands. It is most prominent on cameras that use thinned
CCDs and hence are optimised for observations in blue
passbands. Fringes show up as spatially quickly varying, wave-like
structures on the CCDs (see \figref{fig:fringes}). The {\sl geometry}
of these patterns usually does not change with time since this
interference effect is created in the substrate of the CCD
itself. WFI@2.2m shows fringes with an amplitude of about $1\%$ as
compared to the sky-background in broad-band $R$.  In the $I$- and
$Z$-bands, fringing becomes much stronger and reaches up to about
$10\%$ of the night sky. Unfortunately, contrary to the geometry of
the fringing, its {\sl amplitude} is not stable in consecutive SCIENCE
exposures since it strongly depends on the night sky conditions (as
brightness, cloud coverage), the position on the sky and the airmass.
If a reasonably good SUPERFLAT can be constructed from a sufficient
number of SCIENCE frames obtained under stable sky conditions, a
possible way to correct for fringes is the following:
\begin{enumerate}
\item Besides the large-scale sky variations not corrected for
by SKYFLATs, the SUPERFLAT contains the fringe pattern as an additive
component. Hence, the fringe pattern can be isolated from the
SUPERFLAT by {\sl FRINGE=SUPERFLAT$-$ILLUMINATION}. 
\item Individual SCIENCE frames are corrected for fringing by
{\sl fringeSCIENCE=illumSCIENCE$-$f $\times$ FRINGE}. We assume
that the fringe amplitude directly scales with sky brightness and
{\sl f} is calculated by 
{\sl f=median(illumSCIENCE)/median(ILLUMINATION)}. 
\end{enumerate}
In the case of WFI@2.2m, this method removes very efficiently
the low-level fringes in the $R$-band. Fringing is usually no
longer visible and we estimate possible residuals well below
$0.1\%$. In the case of $I$- and $Z$-bands,
fringes can be suppressed to a level of about $0.1\%$ of the sky
amplitude if a very good SUPERFLAT can be constructed (see
\figref{fig:fringes}). If this is not the case, our approach may
perform very poorly in reducing the fringe amplitude and we cannot
propose a pipeline solution to the problem at this stage.
\subsection{Guidelines for constructing calibration images}
\label{sub:guidelinesprered}
The success of the image pre-reduction heavily depends on the 
quality of calibration data at hand. In the creation of flatSCIENCE,
we propagate the noise in masBIAS and masFLAT to our SCIENCE frames.
As the noise in these calibration frames is of statistical
nature we can diminish it by using as many images in the stacking
process as possible. For the successful creation of a SUPERFLAT,
and hence for the later quality of large-scale illumination and
fringing correction, not only the number of images is
important but it is essential that each pixel in the mosaic
sees blank sky several times during SCIENCE observations. This
suggests that the best observing strategy to achieve this goal
is a large dither pattern between consecutive exposures (ideally 
it should be wider than the largest object in the field). In
the case that the target occupies a significant fraction of the
mosaic (as big galaxies or globular clusters for instance) the
best strategy is to observe a neighbouring blank field
for a short while if a SUPERFLAT and/or fringe correction is important.
In the following we give some additional guidelines leading to 
good results in most of our own reductions:
\begin{itemize}
\item For the stacking process for masBIAS (masDARK) and masFLAT,
about 15-20 individual images of each type should be acquired. As the
noise in the final images is inversely proportional to the square root
of the number of input images this is a good compromise between the
number of exposures that can be taken during a typical observing run (in the
case of SKYFLATs) and the noise reduction.
\item The observer should find out the minimum exposure time required for the 
FLATs in order to suppress shutter effects. This can be done by taking
DOMEFLATs or SKYFLATs with varying exposure time. One then normalises the flat
fields and divides them. Any systematic residuals are then due to shutter
effects, and one can determine the minimum usable exposure time. 
This is
illustrated in Fig. \ref{whtshutter} for the prime focus imager of the 4.2m 
William Herschel Telescope (WHT) at La Palma, which has a comparatively slow 
iris like shutter. Thus, images with short exposure times receive more light 
in the centre than in the corners. In any case, a SUPERFLAT of the data would 
recover any such shutter-related inhomogeneities in the FLAT, but we recommend to
get everything correct from the beginning if possible.

For newer or future WFIs such as OmegaCAM, shutters are constructed 
in which two blades move with constant speed from one side to the other
exposing and covering each pixel of the detector array for exactly the same 
time \citep[see e.g.][]{rkm04}.
\begin{figure}[ht]
  \includegraphics[width=1.0\hsize]{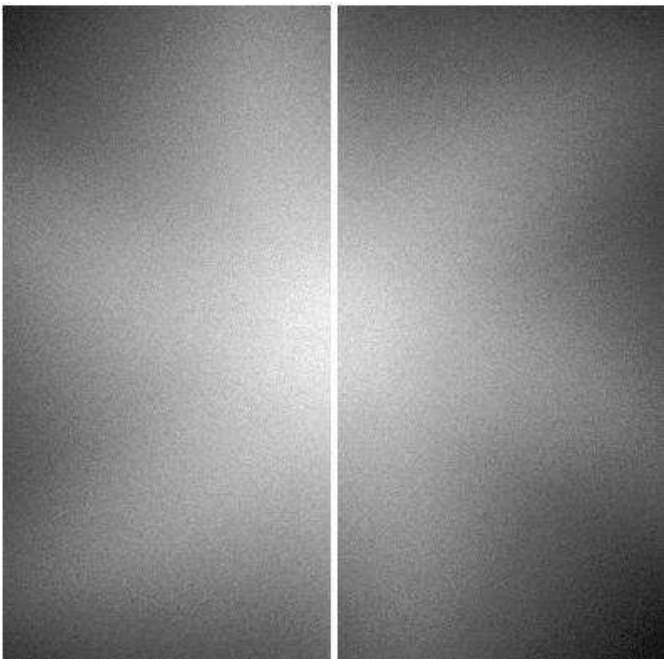}
  \caption{\label{whtshutter} Ratio of two normalised sky flats taken with the
  prime focus imager of the WHT (2 CCDs). The images were exposed for 0.3 and 1.0
  seconds, respectively. One can see the way the iris shutter exposed
  and covered the detector again. The illumination difference between the 
  brightest and faintest part in this representation is 20\%. For the WHT
  prime focus imager it is recommended to expose for at least 2.0 seconds.}
\end{figure}
\item As mentioned above, the creation of SUPERFLATs is not trivial in
general. This is primarily an issue of which SCIENCE images to
include. In the case of blue passbands, where only the smoothed
ILLUMINATION image is used to correct for large-scale illumination
gradients, it is safe to construct it from all SCIENCE frames of
several, consecutive nights. Blue passbands are typically observed
during periods of dark time, and hence the sky conditions are
sufficiently stable. Also, with the moon down we observed no
significant dependence of large-scale sky variations on telescope
pointing. Hence, a robust SUPERFLAT can be obtained with a sufficient
number of images during an observing run. This is no longer the case
if we need to correct for strong fringing. We observe that our
assumption of a sole dependence of the fringe amplitude on sky
brightness typically fails when constructing the SUPERFLAT from
SCIENCE frames of different sky positions. As the appearance of the
overall pattern is very stable, this means either that the fringe
amplitude is no longer a linear function of sky brightness or that the
sky-background varies on scales significantly smaller than a CCD so
that the scaling factor becomes dependent on CCD position.  In
addition, a similar behaviour is observed as a function of time,
depending on the excitation of the OH$^-$ night-sky emission lines.
For our blank-field observations we obtain good results in the
fringing correction if we observe the same target between 10 and 15
times with a large dither pattern within an hour (at WFI@2.2m with an
overhead of about 2 minutes per image this can be achieved with 300 s 
exposures).
\item If a very good SUPERFLAT could be constructed, flat-fielding
results from our proposed method and from a pure SUPERFLAT application
(i. e. using flat-fielding method \ref{item:superflat} in
\sectionref{subsec:howtoflat}) should be compared if flatness in
SCIENCE images is crucial. Especially if only a small number of
individual frames went into the construction of masFLAT, the direct
SUPERFLAT approach often gives better results.
\item Offsetting the telescope between the exposures is fundamental for
high-quality, high-$S/N$ mosaics. The \textit{dither box}, i.e. the box 
that encloses all dither offsets, should be clearly larger than the gaps
between the CCDs and the objects in the field. This has several advantages as 
compared to the \textit{staring} mode (no offsets at all) or to the 
application of only small offsets.

First, the CCDs in a multi-chip camera are fully independent from each
other. They see the sky through different sections of the filter, and they 
have their individual flat-fields, gains and read-out noise. Choosing a wide
dither pattern is the easiest way to establish an accurate global photometric and
astrometric solution for the entire mosaic, based on enough overlap objects.

Second, the wide dither pattern allows for a significantly better 
superflattening of the data, since the objects do not fall on top of themselves 
in the stacks and thus for every pixel a good estimate of the background can 
be obtained. Besides, remaining very low-amplitude patterns in the sky 
background caused by improper flat-fields etc. do not add in the mosaic, but 
are averaged out. Thus, a wide dither pattern will lead to an improved sky
background from which the $S/N$ will benefit; see also Fig. 
\ref{fig:goodbadflat}.
\begin{figure*}[ht]
  \centerline{\includegraphics[width=0.47\hsize]{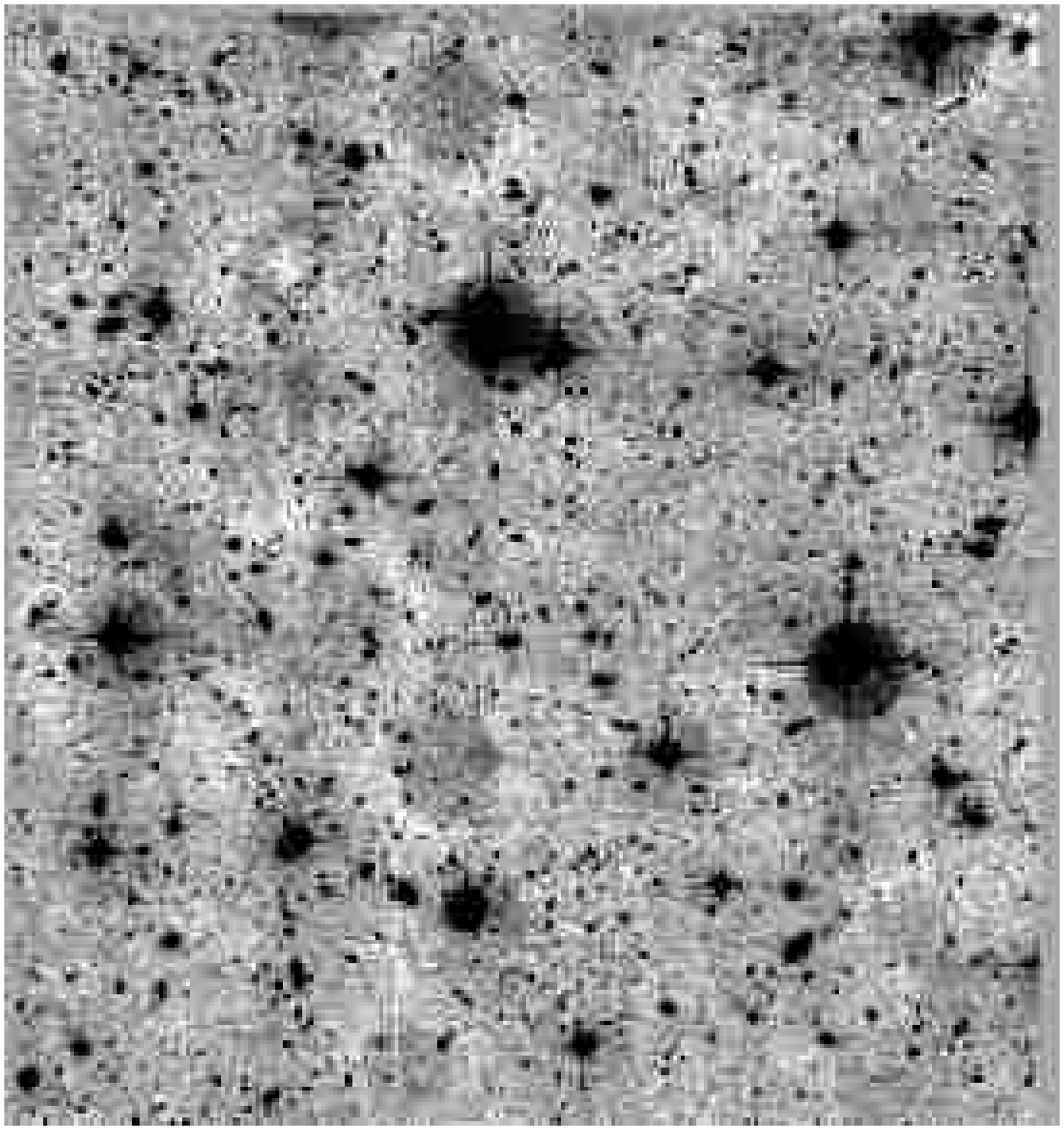}
  \includegraphics[width=0.47\hsize]{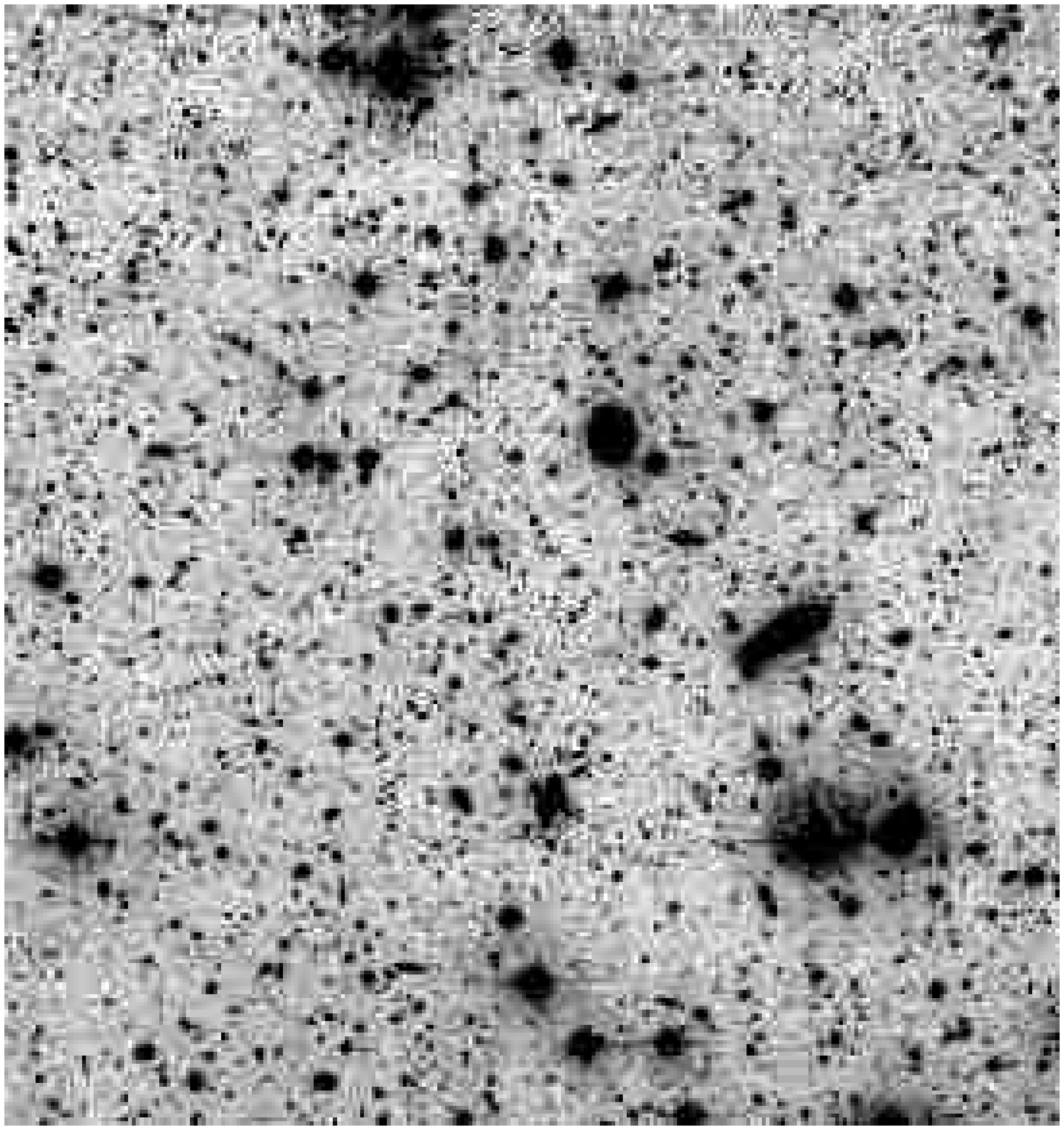}}
  \caption{\label{fig:goodbadflat}Sky-background variations in a mosaic
  obtained from observations of several fields in staring mode (left), 
  and for one obtained with a wide dither
  pattern (right). The contrast scaling for both mosaics is the same. Thus,
  science applications that require a very uniform background, e.g. the search
  for low surface brightness galaxies, profit in general substantially
  from large dither patterns.}
\end{figure*}

Third, the object $S/N$ in a mosaic with a (wide) dither pattern is
significantly superior to one obtained in staring mode. The reason for this is 
that the master BIAS and the master FLAT are not noise-free, since they are
created from a finite number of images. An identical copy of this 
noise is then created in each SCIENCE image during the pre-processing. If no 
dither offsets are applied, then the calibration noise in the $N$ SCIENCE 
images is stacked on top of itself during the co-addition, and thus increases 
in the same way as the flux of the objects ($\propto N$) instead of averaging 
out ($\propto \sqrt{N}$). This is enforced by the use of an auto-guider, which
keeps the sources in subsequent images exactly at the same pixel position.
Depending on the ratio of the calibration noise and the noise in the 
uncalibrated images themselves, the effective exposure time of the mosaic can 
be very significantly reduced. This is especially true for observations gained 
under excellent (dark) conditions, since the sky noise in the SCIENCE images 
is then reduced, and the noise from the calibration frames becomes more 
dominant. One should therefore always aim at a sufficient number of 
calibration exposures, \textit{and} go for larger dither offsets. This holds 
for single-chip cameras, too.
\end{itemize} 
\subsection{Image Eyeballing and Manual Masking}
\label{sec:manualpass}
\begin{figure*}[ht]
\begin{center}
  \begin{minipage}[t]{0.45\textwidth}
    \psfig{figure=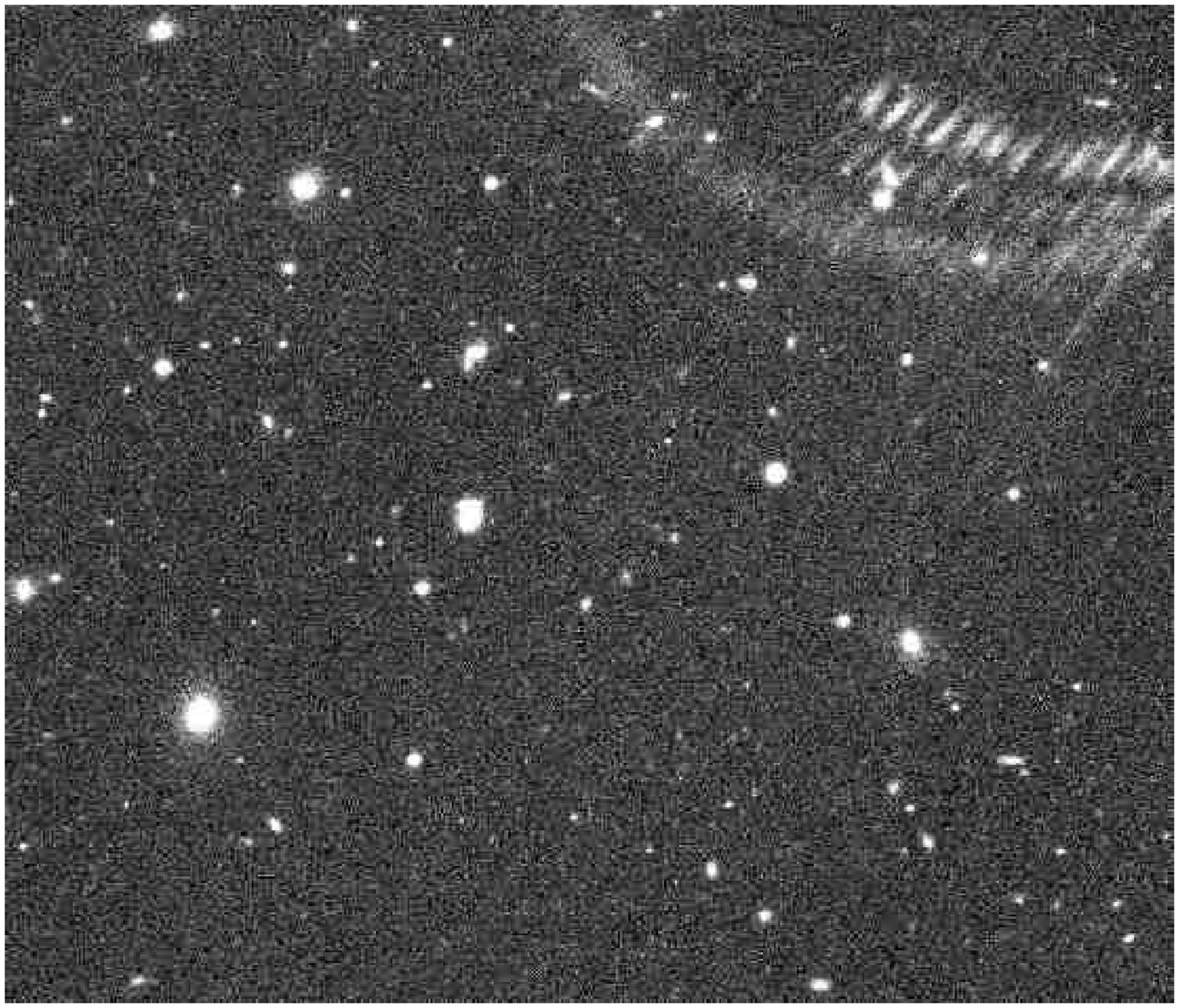,width=\hsize,angle=0}
  \end{minipage}
  \begin{minipage}[t]{0.45\textwidth}
    \psfig{figure=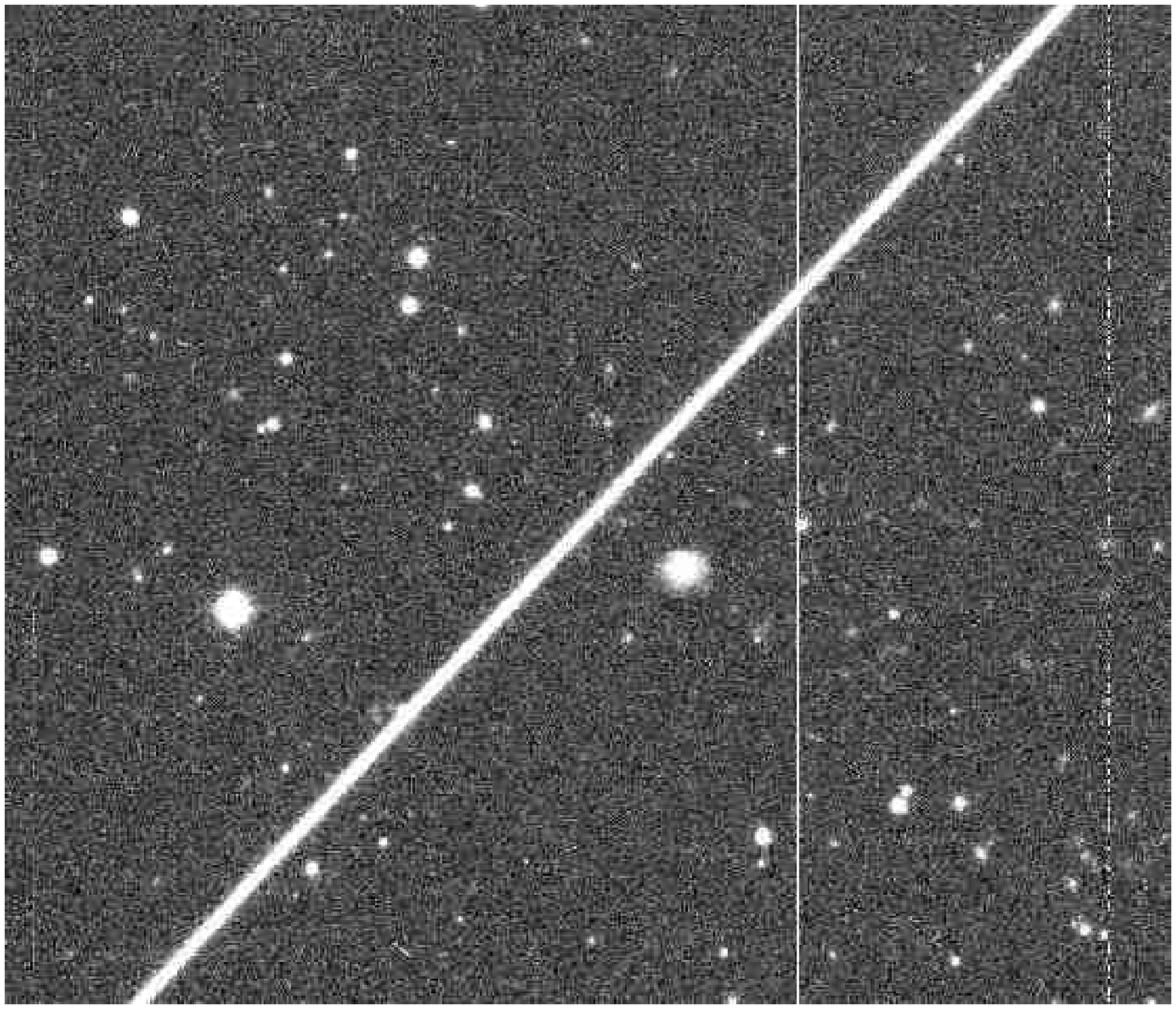,width=\hsize,angle=0}
  \end{minipage}
\end{center}
\begin{center}
  \begin{minipage}[t]{0.45\textwidth}
    \psfig{figure=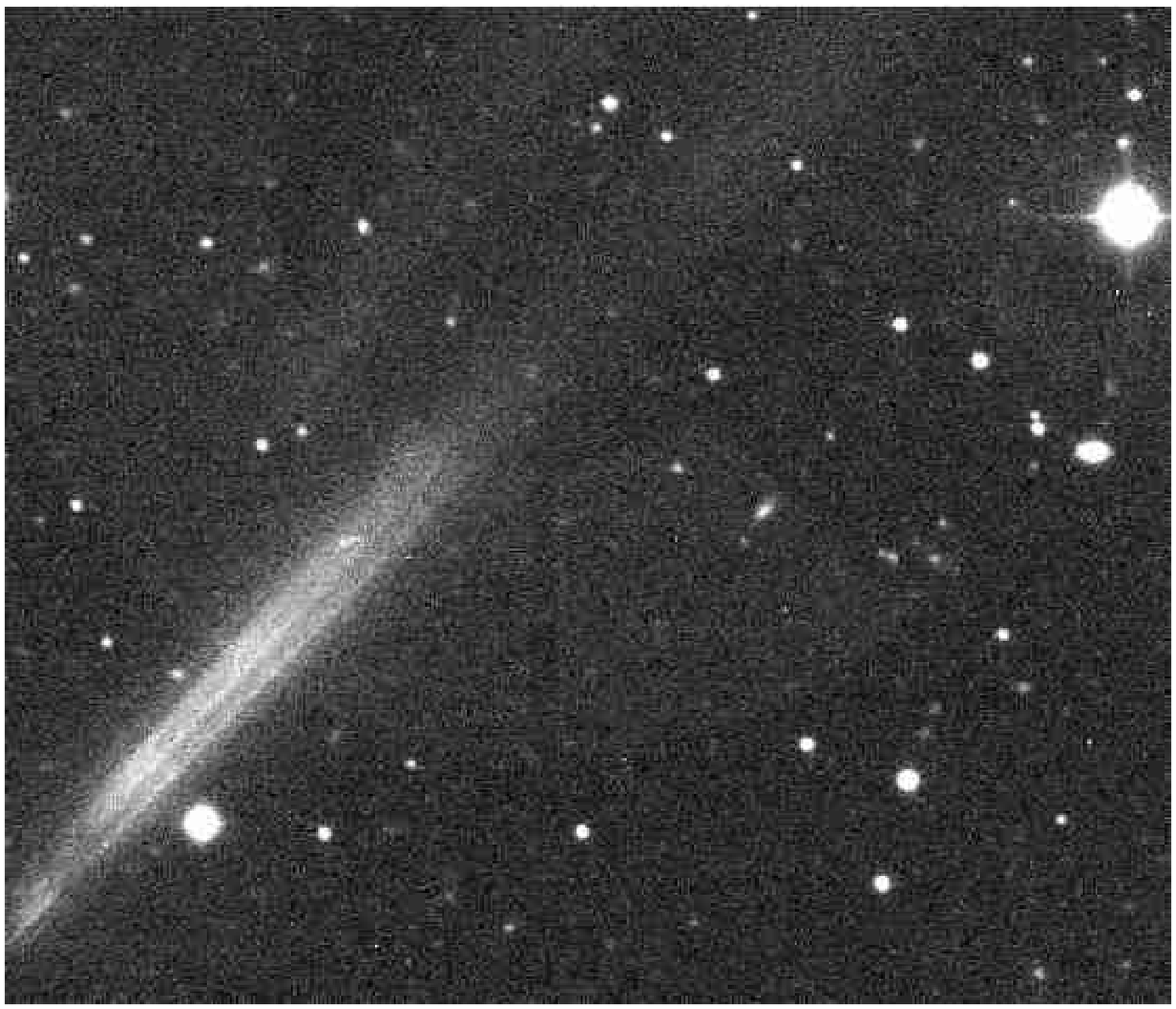,width=\hsize,angle=0}
  \end{minipage}
  \begin{minipage}[t]{0.45\textwidth}
    \psfig{figure=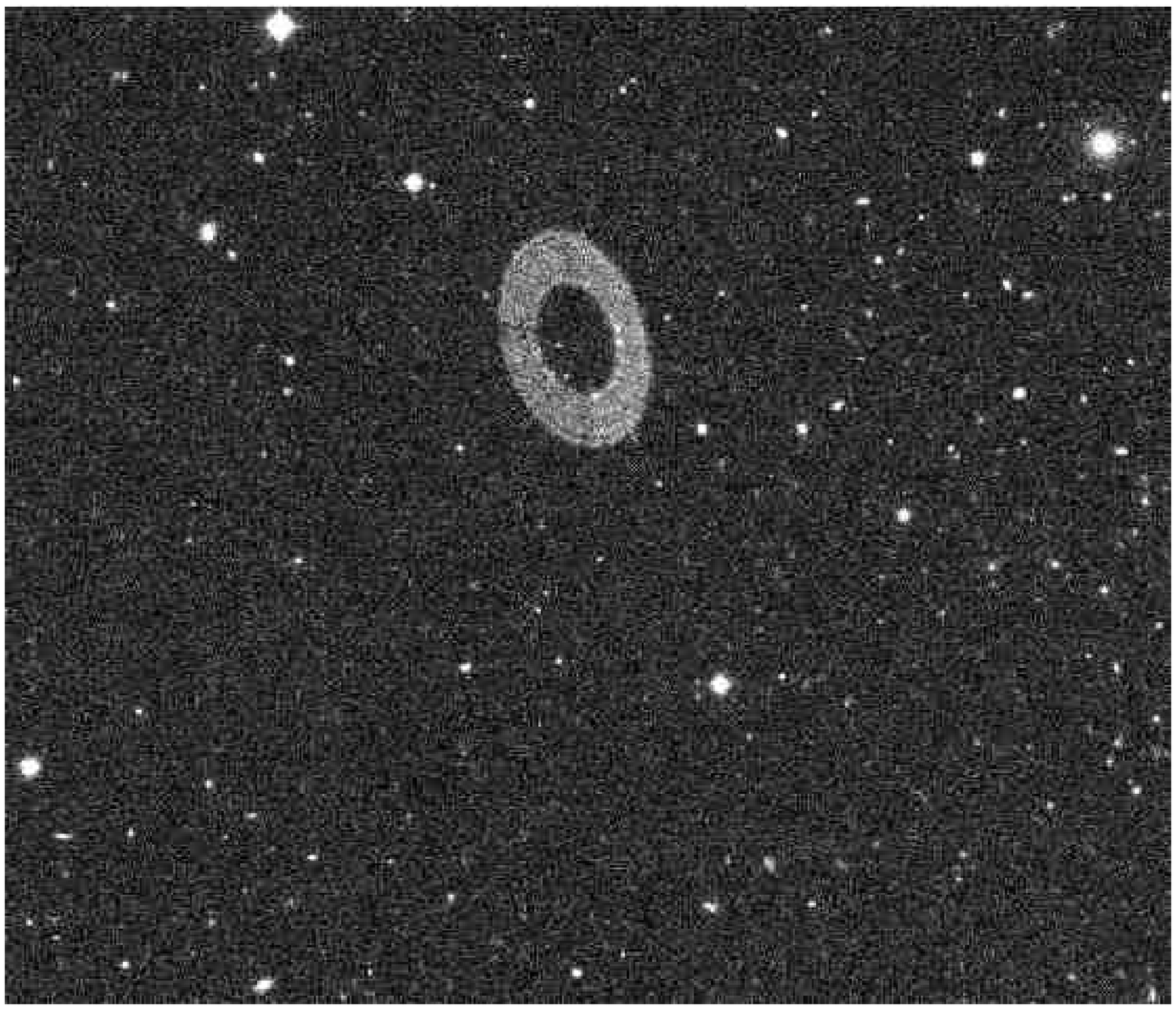,width=\hsize,angle=0}
  \end{minipage}
\end{center}
\caption{Examples of large-scale defects: The upper right panel shows
a satellite track, the other three panels various forms of reflections
from bright stars inside and outside a WFI@2.2m field. All such
features are typically unique to individual frames and do not appear
in subsequent, sufficiently dithered exposures. Currently we have to
register and to mask these defects by hand to prevent that they enter
the final image co-addition (see \sectionref{sec:weightingflagging}).
Masks are built using the {\sl polygon} option of the {\tt ds9}
visualisation tool.}
\label{fig:defects}
\end{figure*}

After the pre-reduction we visually inspect all SCIENCE images to identify
potential problems needing manual intervention:
\begin{itemize}
\item Faulty exposures (e.g. data taken out of focus) 
having passed the initial quality check
(\sectionref{subsec:overscanbiasdark}) are removed at this stage.
\item Problems of the pre-reduction process are identified and
necessary steps are iterated (this concerns
mostly the selection of SCIENCE frames entering the SUPERFLAT).
\item The most important aspect of this manual pass through all
SCIENCE data is the identification of large-scale image defects that
must not enter the final co-addition process and which have to be
manually masked at this stage. This will be described in more detail
in \sectionref{sec:weightingflagging}.  See also \figref{fig:defects}.
\end{itemize} 
\subsection{Standard star processing and absolute photometric calibration}
\label{sec:standards}
Besides the SCIENCE images we have to process STANDARD frames
to obtain an absolute photometric calibration.
A fully automated absolute photometric calibration up to the final
co-added image is not yet implemented in our system. As a first step
in this direction, the pipeline allows the
photometric calibration of individual nights of a {\sl Run}.
In this section we describe how STANDARD exposures are pre-reduced
and analysed to extract photometric parameters. If not stated
otherwise, all magnitudes in this paper are given in the {\sl Vega} system.

STANDARD exposures are reduced in a fashion similar to SCIENCE
exposures. They are first overscan corrected, then the masBIAS is
subtracted. Flat-fielding is done with the masFLAT generated in the
pre-reduction of the SCIENCE exposures and the SUPERFLAT image created
from the SCIENCE exposures is applied to the pre-reduced STANDARD
exposures. Using the SUPERFLATs created from the SCIENCE images is
important to ensure photometric homogeneity between SCIENCE and
STANDARD exposures.

If necessary the FRINGE maps created from the SCIENCE images are
rescaled and subtracted from the superflatted STANDARD exposures. This
step is particularly difficult and often leads to non-satisfactory
results in the $I$-band. The reason is that STANDARD exposures are
usually taken at different times during the night (sometimes even in
twilight), at different sky positions and various airmasses. This
often leads to fringing patterns with intensities very different from
those present in the SCIENCE exposures.

To generate object catalogues that can be used for photometric
calibration bad pixel masks for the STANDARD images are created.
They mark pixels not suitable
for photometric measurements, such as bad pixels or cosmic rays. 
The creation of these masks (FLAG images henceforth) 
is described in more detail in \sectionref{sec:weightingflagging}.

Using the FLAGs object catalogues are created with
{\tt SExtractor} from the reduced STANDARD exposures. To match the
found objects with \citet{lan92} or \citet{ste00}
standard star catalogues an astrometric solution for the STANDARD images has
to be derived first. This is done with the LDAC astrom tools to match
objects found in the STANDARD images to the USNO-A2 catalogue. The
number of objects found in the STANDARD exposures is sufficiently high
to derive a second-order astrometric solution separately for each
image. This is enough to reliably
merge standard star catalogues with our catalogues. All fluxes in our
STANDARD fields are normalised to an exposure time of 1 s.

From the database of matched standard stars, the coefficients of 
the equation, 
\begin{eqnarray}
  \label{eq:1}
  \mbox{Mag}-\mbox{Mag}_{\rm inst} = \mathit{ZP} + \mathit{CE} * {\rm airmass} 
  + \mathit{CT} * {\rm CI}\;,
\end{eqnarray}
i.e. the zero point $\mathit{ZP}$, the extinction coefficient $\mathit{CE}$, and the
colour term $\mathit{CT}$ are determined, where $\mbox{Mag}$ is the standard
star's magnitude, $\mbox{Mag}_{\rm inst}$ is the instrumental
magnitude measured on the reduced
standard frame (we use the {\tt SExtractor} MAG\_AUTO parameter for
this purpose), and CI is a colour index, e.g. $(B-V)$. This is done
using a non-linear least-squares Marquardt-Levenberg algorithm with an
iterative $3\sigma$ rejection, which allows rejected points to be used
again if they are compatible with later solutions. As this algorithm
is not guaranteed to converge, the iteration is aborted as soon as one
of the following three criteria is true:
\begin{enumerate}
\item The iteration converged and no new points are rejected.
\item The maximum number of iterations (set to 20) is reached.
\item More than a fixed percentage (set to 50\%) of all points are
  rejected.
\end{enumerate}
In a first step all three coefficients are fit simultaneously. However,
in order to reliably estimate the extinction coefficient standard star
observations must be spread over a range of airmasses. This is
sometimes neglected by observers. To find an acceptable photometric
solution in this case, the user can supply a default value for the
extinction coefficient. The fit is then repeated with the extinction
coefficient fixed at the user supplied value and the zero point and
colour term as free parameters.  

Although wide-field observations of Landolt/Stetson fields typically
cover a wide range of stellar colours, the user can also supply a
default value for the colour term. This is then used for a 1-parameter
fit in which the zero point is the only free parameter.

In an interactive step the user can then choose between the 1-, 2-,
and 3-parameter solution, or reject the night as non-photometric. The
FITS headers of the frames belonging to the same night are then
updated with the selected zero point and extinction coefficient or
left at the default value $-1.0$, indicating that no photometric
calibration for that frame is available. 

We note that we perform the fit simultaneously for all mosaic
chips. As discussed in \sectionref{subsec:setcharacteristics} zero
points of individual chips agree within 0.01-0.03 mag after sky-background
equalisation. We do not take into account possible colour term
variations between individual images that are expected due to slightly
different CCD transmission curves. Notable differences can occur in
the $U$ and $Z$ filters that are cut off by the CCD transmission.

\begin{figure*}[ht]
  \centerline{\includegraphics[width=0.47\hsize]{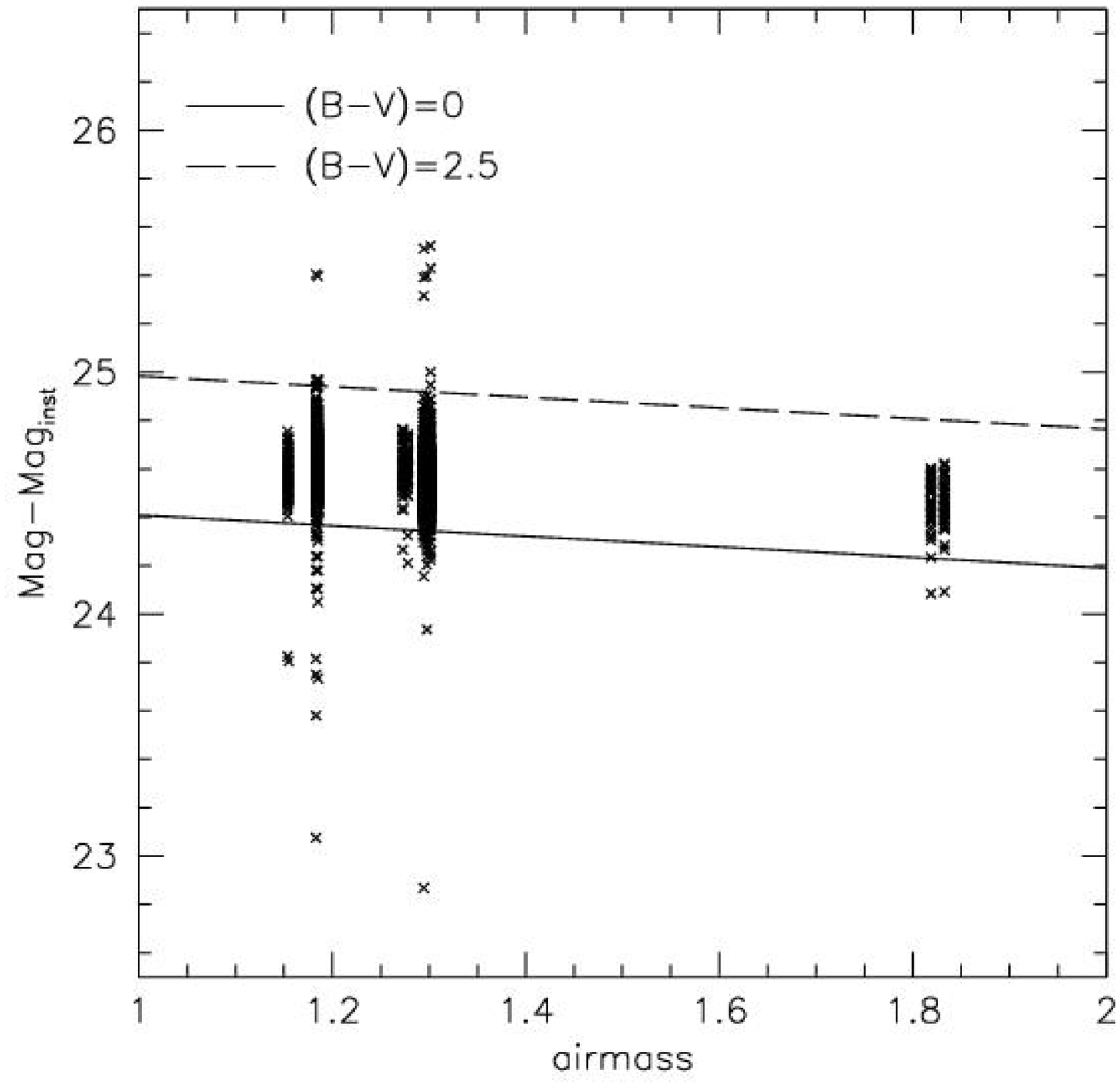}
  \includegraphics[width=0.47\hsize]{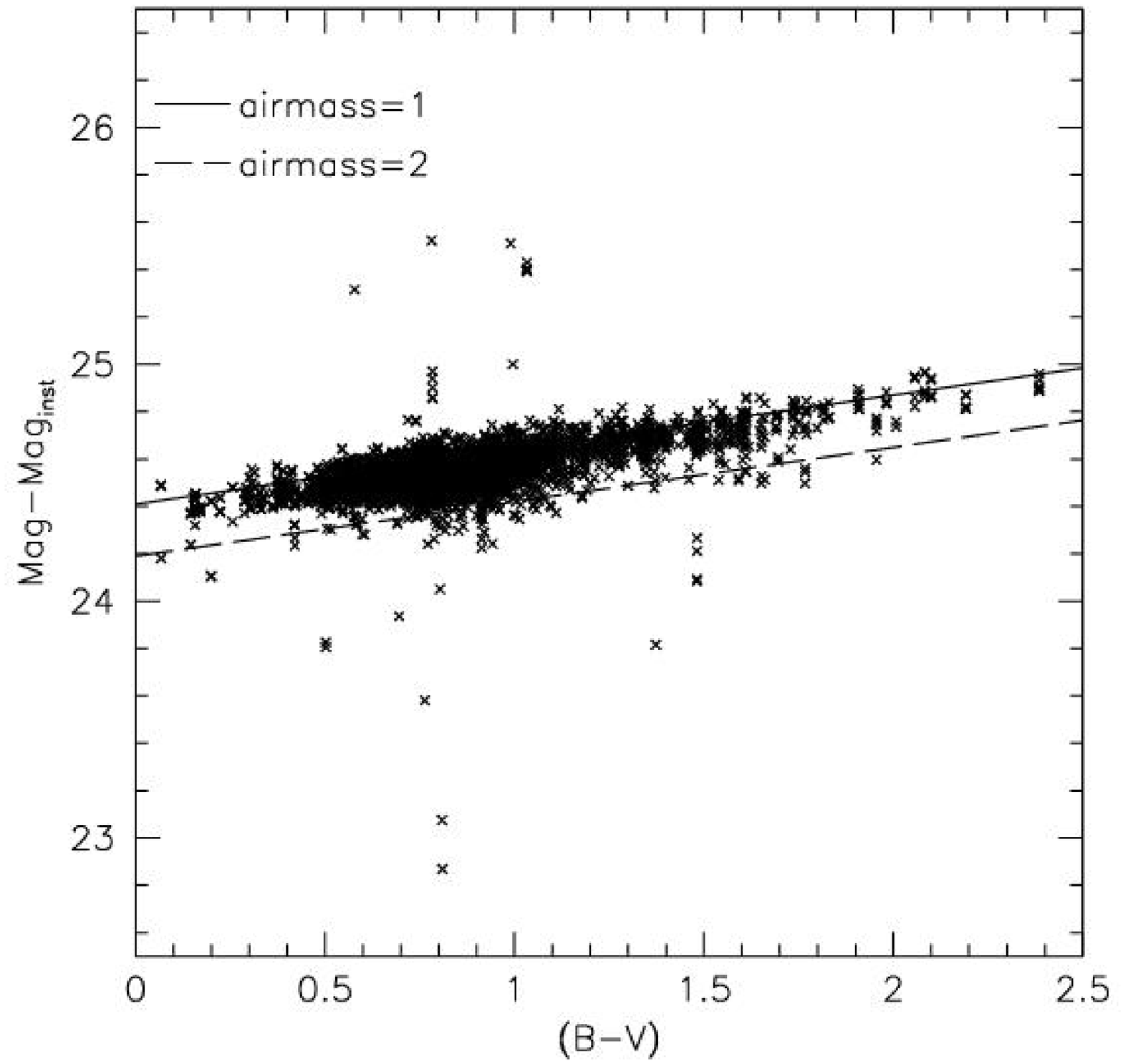}}
  \caption{\label{fig:absphotom}\small{The plots illustrate our 
zero point estimation in broad-band $B$ during a clear
night. Eleven standard field observations with a good airmass
coverage were obtained giving a total of about 4000 individual
standard star measurements (the Stetson standard star catalogue was
used here). Points indicate the measurements, the
lines the result of a three parameter fit of \eqref{eq:1}.}}
\end{figure*}

\figref{fig:absphotom} illustrates our photometric calibration which
is implemented in Python. We compared our own zero point and
extinction measurements for several nights with those recently
obtained by the EIS team. \tableref{tab:zpcomp} shows that the
measurements are in very good agreement.
\begin{table*}
\caption{\label{tab:zpcomp} The table shows comparisons of our own
photometric zero point and extinction measurements with those from the
EIS team.  We independently calibrated several nights in the WFI@2.2m
broad-band $B$ filter (ESO Filter ID: BB\#B/123\_ESO878). The EIS
values were taken from the EIS XMM product logs (see
http://www.eso.org/science/eis/surveys/release\_55000024\_XMM.html) of
the fields listed in column 2. From the comparison we excluded the
measurement of one non-photometric night (30/06/2003).  The 7th column
lists $\Delta ZP$=Bonn\_ZP-EIS\_ZP and the 8th column the number of
parameters fitted in \eqref{eq:1}. Values marked with an asterisk were
not fitted but default values have been used instead. We note that the
results are in very good agreement.}
\begin{center}
\begin{tabular}{cccccccc}
\hline
\hline
Date & Field & EIS\_ZP & EIS ext. & Bonn\_ZP & Bonn ext. & $\Delta ZP$ & \#par (Bonn) \\
\hline
22/06/2003 & BPM16274 & 24.80 & $-$0.199 & 24.80 & $-$0.16 & 0.00  &    3 \\
05/03/2003 & HE1104$-$1805 & 24.86 & $-$0.242 & 24.91 & $-$0.30 & 0.05  &    3 \\
30/03/2003 & NGC4666 & 24.71 & $-$0.184 & 24.75 & $-$0.14 & 0.04  &    3 \\
06/08/2003 & LBQS\_2212-1759 & 24.82 & $-$0.205 & 24.88 & $-$0.22(*) & 0.06  &    2 \\
01/04/2003 & Q1246-057 & 24.64 & $-$0.122 & 24.67 & $-$0.08 & 0.03  &    3 \\
27/01/2003 & SHARC-2 & 24.58 & $-$0.108 & 24.51 & $-$0.04 & $-$0.07 &    3 \\
\hline
\end{tabular}
\end{center}
\end{table*}
\subsection{{\sl Runs} and {\sl Sets}}
As discussed in the beginning of \sectionref{sec:prereduction} the
pre-reduction is done on a {\sl Run} basis usually containing
observations from different patches of the sky.  Before proceeding we
need to split up the SCIENCE exposures of a {\sl Run} into the {\sl Sets}
that have to be co-added later. By {\sl Set} we mean the series of all 
exposures of the same target in a particular filter.
This means that all the following
reduction steps up to the final co-addition have to be done on each {\sl Set}
independently. We note that a {\sl Set} can have been observed in multiple
{\sl Runs} so that all {\sl Runs} containing images of a certain {\sl Set} have to be
processed at this stage.
\section{Astrometric and photometric {\sl Set} calibration}
\label{sec:astromphotom}
\subsection{Astrometry}
\label{astrometric}
After the pre-processing, a global astrometric solution and a global relative
photometric solution is calculated for all SCIENCE images. This is where the 
reduction of WFI data becomes much more complicated than the one for 
single-chip cameras.

\begin{figure}[ht]
  \includegraphics[width=1.0\hsize]{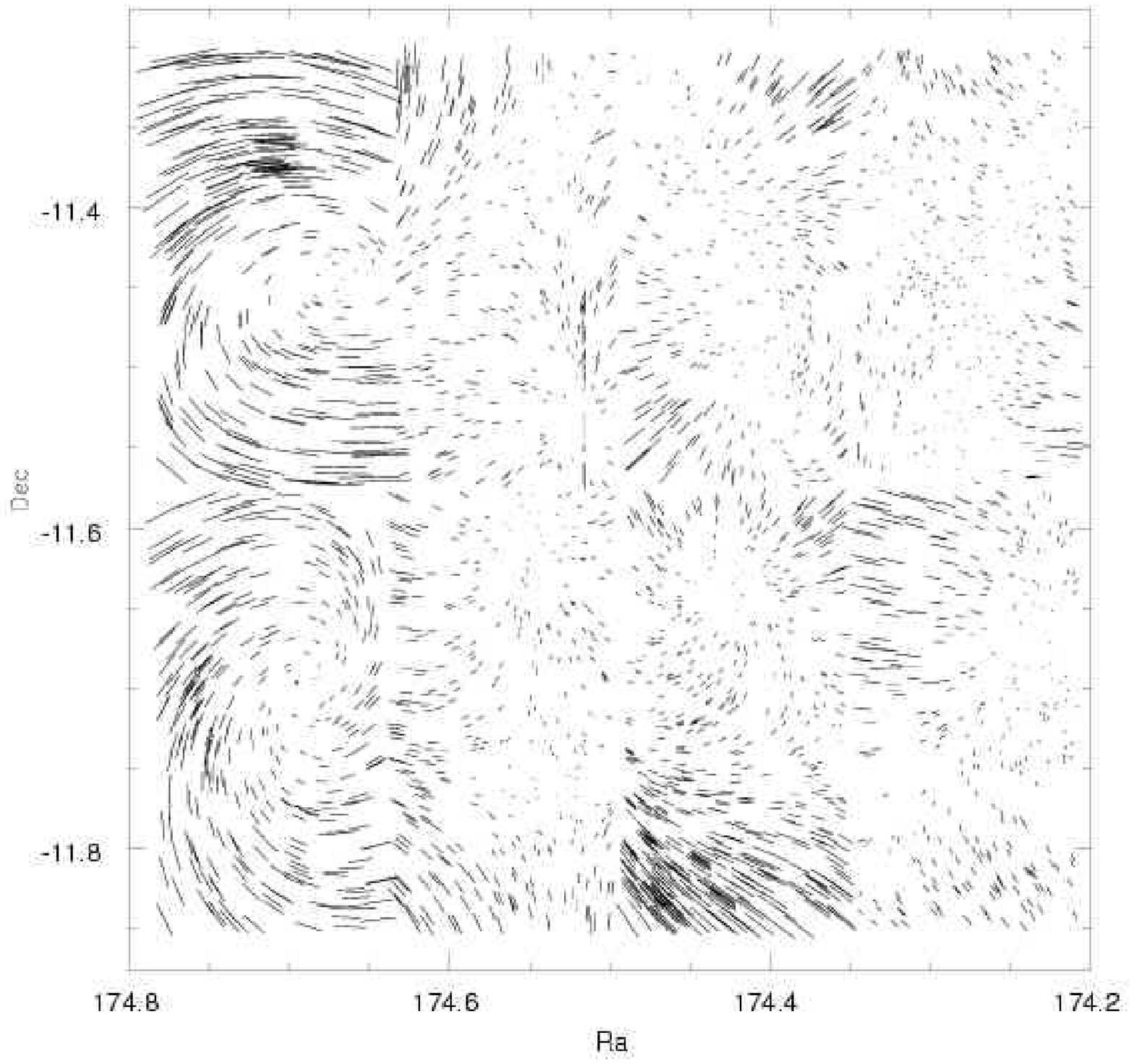}
  \caption{\label{fig:wfi_distort}\small{Difference in
    object position between a single-shift approach and a full
    two-dimensional second-order astrometric solution for the
    WFI@2.2m. In other words, shown are the higher-order terms needed
    for matching the images to the sky. The patterns belonging to the
    left two CCDs are due to a rotation with respect to the
    mosaic. The maximum position difference in the plot is about six
    pixels, still a fairly small value compared to other telescope
    designs. It becomes clear that a single, global distortion
    polynomial for all 8 CCDs does not work. Instead, every image has
    to be treated individually.}}
\end{figure}

In a first step, high $S/N$ objects in each image are detected by {\tt
SExtractor}, and a catalogue of non-saturated stars is
generated. Typically, we use all objects having at least 5 contiguous
pixels with $5\sigma$ above the sky-background in the following
analysis (these numbers may vary according to filter and exposure
time; in the $U$-band for instance, we need to lower these
thresholds in order to have enough sources to compare with a standard
star catalogue).  This usually gives us between 3 and 6 objects per
square arcmin in high-galactic latitude empty field observations.
Based on a comparison with the USNO-A2 \citep[see][]{mcd98}
astrometric reference catalogue (or any other reference catalogue), a
zero-order, single shift astrometric solution is calculated for each
image. For a single-chip camera with a small field-of-view such an
approach is often sufficient, but it no longer holds for multi-chip
cameras with their large field-of-view. In this case, the CCDs can be rotated
with respect to each other, tilted against the focal plane, and in
general cover areas at a distance from the optical axis where field
distortions become large.  \figref{fig:wfi_distort} shows the
difference between a zero order (single shift with respect to a
reference catalogue) and a full astrometric second-order astrometric
solution per image. From this figure it is obvious that the simple
shift-and-add approach will not work for the entire mosaic. The issue
is further complicated by the gaps between the CCDs and large dither
patterns that are used to cover them. Thus, images with very different
distortions overlap. In addition, due to 
the large field-of-view, one must take into account that the spherically 
curved sky is mapped into the flat tangential detector plane. 

In the second step, Mario Radovich's {\tt
Astrometrix}\footnote{http://www.na.astro.it/$\sim$radovich/WIFIX/astrom.ps}
package is used to determine third-order polynomials for the
astrometric solution of each image. This corrects for the
aforementioned effects, and thus allows for the proper mosaicing in
the later co-addition process. For this purpose all high $S/N$ objects
(stars and galaxies) detected in the first step are matched, including
those from the overlap regions. The latter ones are most important in
establishing a global astrometric (and photometric) solution, since
the accuracy of available reference catalogues such as the USNO-A2
with an rms of about $0\myarcsec3$ is insufficient for sub-pixel
registration. Thus the astrometric solution is determined from the
data itself instead of the reference catalogue, which is usually based
on data taken with a much smaller angular resolution than the images
that are processed. The reference catalogue is used only to fix the
solution with respect to absolute sky coordinates within its stated
rms.  A wide dither pattern is required to compute a reliable global
astrometric solution, so that different regions of the CCD are mapped
by overlapping sources and the astrometric solution is properly
constrained.  For a more detailed description of the {\tt Astrometrix}
tool see \citet{rad02} and \citet{mrb03}.
\subsection{Accuracy of our Astrometric calibration}
\label{sec:astromaccuracy}
\begin{figure}[ht]
  \includegraphics[width=1.0\hsize]{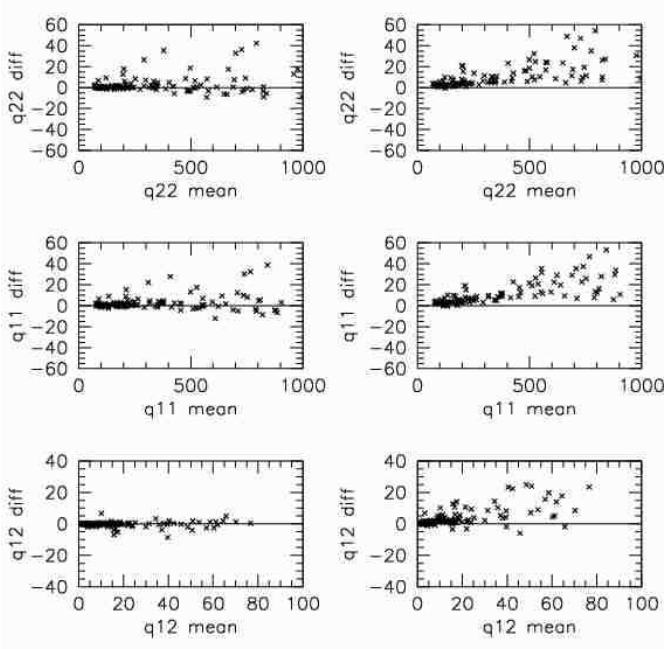}
  \caption{\label{fig:qdiff}\small{Shown are measurements for
  $q_{ij}^{\rm diff}$ from 152 stars lying within one chip of
  30 WFI@2.2m exposures. The quantities are plotted in arbitrary 
  units. The co-addition was performed with {\tt SWarp} and
  the resampled images (according to the astrometric solution
  from {\tt Astrometrix}) have been used for the calculation of
  $q_{ij}^{\rm mean}$ (see \sectionref{sec:co-addition}).
  All stars have at least 10 contiguous pixels
  with 10$\sigma$ above the sky-background in the individual images
  and hence all the measurements involved have high $S/N$.
  The left panels show the result for $q_{ij}^{\rm diff}$, where
  the co-added mosaic was stacked according to the astrometric solution
  provided by {\tt Astrometrix}. We checked that the outliers showing
  a higher value in the co-added image are uniformly distributed over
  the images. For the right panels we shifted all
  individual images with a random offset within an image pixel
  before co-adding them. This mimics a co-addition by a simple
  integer pixel shift and we see a clear, systematic offset for
  all three components.}}
\end{figure}

\begin{figure}[ht]
  \includegraphics[width=1.0\hsize]{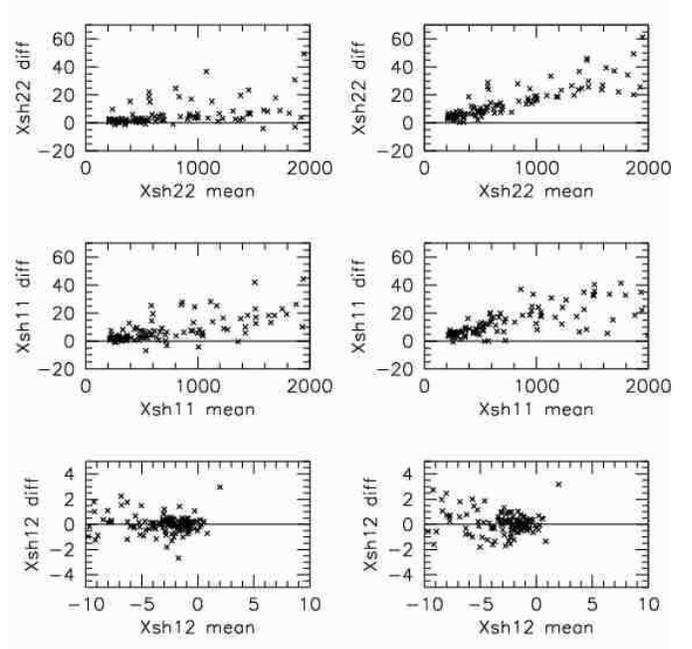}
  \caption{\label{fig:xshdiff}\small{The same as \figref{fig:qdiff}
  for $X^{\rm sh}$ in arbitrary units. The agreement between the
  measurement in the mosaic and the expectation is not as good as for
  $q$ and $X^{\rm sm}$ (see \figref{fig:xsmdiff}) and a systematic
  offset of about $1\%$ is seen in the 11 component.}}
\end{figure}

\begin{figure}[ht]
  \includegraphics[width=1.0\hsize]{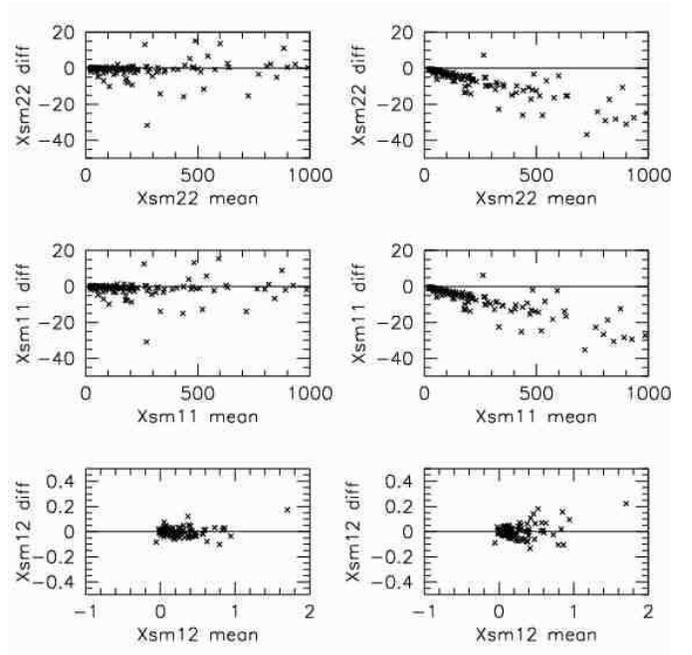}
  \caption{\label{fig:xsmdiff}\small{The same as \figref{fig:qdiff}
  for $X^{\rm sm}$ in arbitrary units. For this quantity, the
  agreement between the values in the mosaic and the predictions is
  excellent for all three components.}}
\end{figure}

\begin{figure}[ht]
  \includegraphics[width=1.0\hsize]{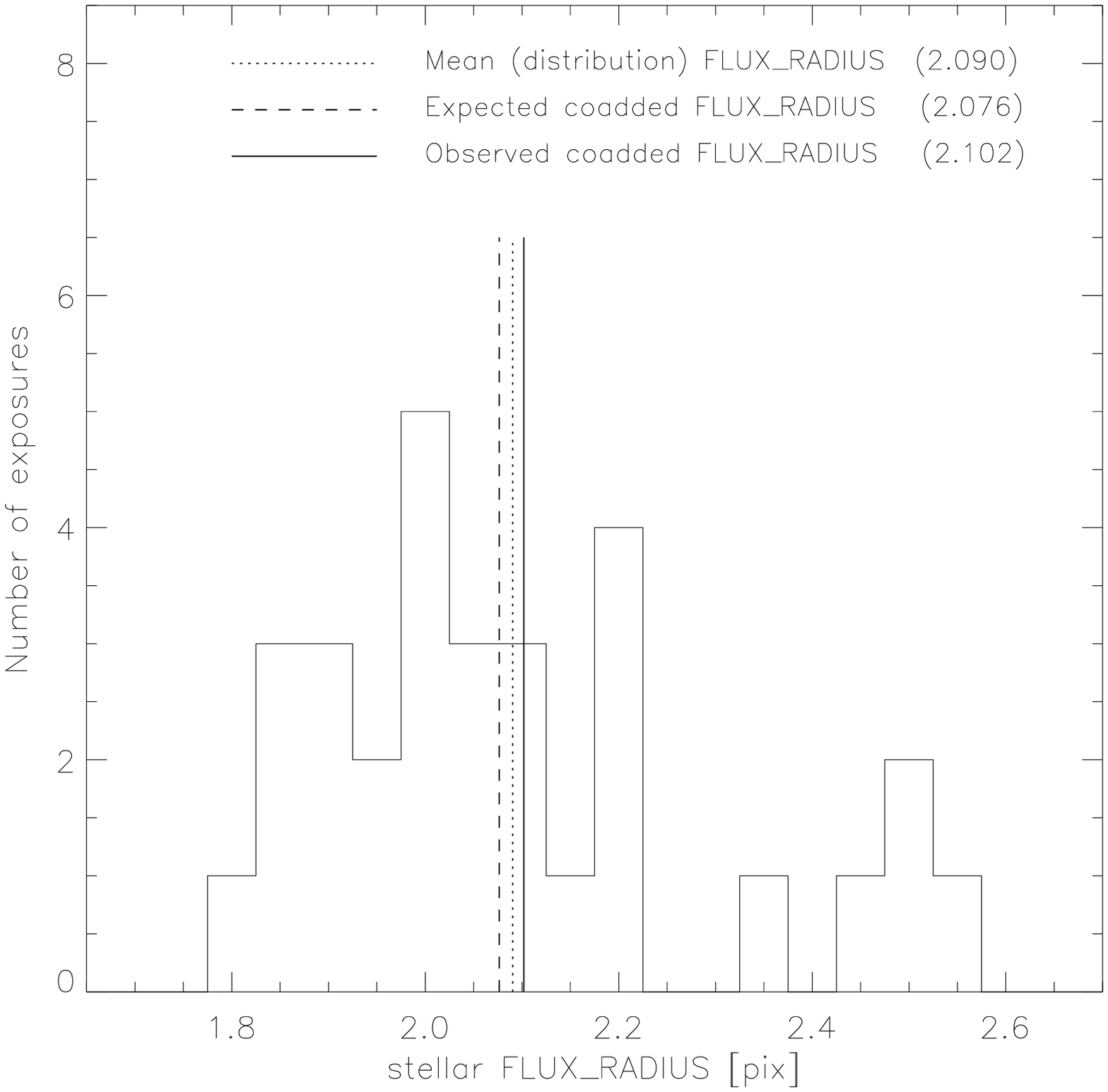}
  \caption{\label{fluxrad}\small{Shown is the image seeing (half-light radius)
      of a set of 30 WFI@2.2m exposures. The mean is indicated by the dotted 
      line. The solid line shows the image seeing in the co-added image, using
      {\tt Astrometrix} for the astrometric solution and {\tt SWarp} for 
      the co-addition. Assuming that the stellar light profile in the
      individual exposures is a Gaussian, one would expect an image seeing as 
      the one indicated by the dashed line, provided that both the astrometric 
      solution and the co-addition method work perfectly well. One can see that 
      we are very close to the optimal value, which is only 
      $0.026\approx 1/40$th of a pixel smaller than the actual measured image
      seeing in the co-added mosaic. One pixel corresponds to
      $0\myarcsec238$. This analysis is based
      on 29 stars that were in common in an arbitrarily selected CCD for all
      30 WFI@2.2m exposures.}}
\end{figure}

The astrometric calibration ensures that we can produce a deep mosaic
out of individual images. The accuracy with which the images
have to be aligned strongly depends on the scientific application in
mind. In most cases, the final goal is the measurement of
moments $M^{mn}$ from the light distribution $I(\theta)$ of objects, 
that is:
$$
M^{mn}=\int {\rm d}^2\theta\,\theta_1^m\theta_2^n W(\theta)
I(\fat\theta); m,n \in {\cal N}_0
$$
where $W(\theta)$ is some weighting function, $\theta=|\fat\theta|$ and 
$(\theta_1, \theta_2)=(0,0)$ is the centre of the object,
which is chosen to be the point where the weighted dipole moments 
are zero.
In our weak lensing studies, we have to estimate accurately
moments
up to the fourth order. The shape of an object is constructed out of
the second moments $q_{ij}$, i.e.;
$$
q_{ij}=\int {\rm d}^2\theta\, W(\theta)\,\theta_i\theta_j I(\fat\theta),
$$ 
where we choose $W(\theta)=1/(2\pi r_{\rm 
g}^2)\exp(-|\theta|^2/(2r_{\rm g}^2))$ with $r_{\rm g}=3$ image pixels
henceforth. The following two fourth-order moments are important
for correcting object shapes for PSF effects in weak lensing studies
\citep[see][]{ksb95}:
$$
X^{\rm sh} = \int {\rm d}^2\theta\left[
                     \begin{array}{cc}
		     2W\theta^2+2W'\eta_1^2 &  2W'\eta_1\eta_2\\
		     2W'\eta_1\eta_2 &  2W\theta^2+2W'\eta_2^2\\
		     \end{array}\right] I(\fat\theta)\\
$$
and
$$
X^{\rm sm} = \int {\rm d}^2\theta\left[
                     \begin{array}{cc}
		     \eta_3+2W''\eta_1^2 & W''\eta_1\eta_2\\
		     W''\eta_1\eta_2 & \eta_3+W''\eta_2^2
		     \end{array}\right]  I(\fat\theta),\\
$$
where $\eta_1:=(\theta_1^2-\theta_2^2)$, $\eta_2:=2\theta_1\theta_2$,
$\eta_3:=W+2W'\theta^2$ and
$W'$ denotes the derivative with respect to $\theta^2$.  As the
quantities under consideration are linear in the light distribution
$I(\fat\theta)$, the measurement in the mosaic can be predicted from the
measurements in the individual images (if a linear image co-addition is
performed).  If the mosaic is constructed by a straight mean of $N$
input images, the value $q^{\rm mosaic}_{ij}$, measured in
the co-added image has to be equal to:
$$
q^{\rm mean}_{ij}=\frac 1N\sum_{k=1}^N q^k_{ij},
$$ where the $q^k_{ij}$ are the moments in the individual images.  A
systematic deviation of the quantity $q_{ij}^{\rm diff}:=q^{\rm
mosaic}_{ij}-q^{\rm mean}_{ij}$ (and in corresponding expressions for
$X^{\rm sh}$ and $X^{\rm sm}$) from zero is hence a good indicator
that the astrometric solution should be improved.  Results of this
test for a deep $R$-band mosaic constructed out of 30 exposures with
sub-arcsecond seeing are shown in Figs. \ref{fig:qdiff},
\ref{fig:xshdiff} and \ref{fig:xsmdiff}. The agreement between the
predictions and the measurements is in general very good, only the
results of $X^{\rm sh}_{\alpha\beta}$ may indicate a small systematic
offset in the co-added image. We conclude that the alignment accuracy
of {\tt Astrometrix} is sufficient for our weak lensing studies.  For
our WFI@2.2m $R$-band observations {\tt Astrometrix} formally
estimates an internal astrometric accuracy of $1/10-1/5$ of a WFI
pixel $(0\myarcsec02-0\myarcsec04)$ over the whole field-of-view.
This result is consistent with the accuracy obtained by \citet{ars01}
for WFI@2.2m observations. However, this accuracy strongly depends on
filter, instrument and the observing strategy, i.e. the number of
bright stars that can be extracted, the kind of distortions and the
dither pattern.  For instance, \citet{rar04} obtain an alignment
accuracy of about $1/3$ of a pixel for WFI@2.2m $U$-band observations
and \citet{mrb03} $2/5$ of a pixel for $B$, $V$, $R$, $I$ observations with the
CFH12K@CFHT instrument.  Finally we compare the size of the PSF in
the individual images and the mosaic. \figref{fluxrad} shows that also
the size of the PSF in the mosaic can very well be predicted by
measurements in the individual exposures entering the co-addition.
\subsection{Photometry}
\label{subsec:photometric}
Besides the astrometric alignment, special care has to be taken
in the photometric calibration. First, we usually want to co-add
data obtained under different photometric conditions (for instance
if a target is observed during different nights of an observing run).
Second, the individual images can have intrinsically different
zero points\footnote{The sky-background equalisation described in 
\sectionref{subsec:gainequalisation} should remove zero point differences
within an exposure but we do not rely on this assumption here.}.
It is possible to arrive at an {\sl internally consistent} photometric
system for all exposures by comparing instrumental magnitudes of
overlap sources. The following procedure of a relative photometric
calibration is very similar to that described in \citet{kkg98}.
It is implemented in the LDAC {\tt relphotom} programme.
Having a full astrometric solution and hence an accurate table of
paired objects of all SCIENCE images at hand, a relative photometric solution is
straightforward. Given two overlapping 
images $k$ and $j$, consider all $i=1...N$ common objects and calculate the mean 
deviation of magnitudes $K$ and $J$
\begin{equation}
M_{k,j}:=\frac{\Sigma_{i}W_{i}(K_{i}-J_{i})}{\Sigma_{i}W_{i}}\;,   
\end{equation}
with $W_{i}=(\sigma^{2}_{K}+\sigma^{2}_{J})^{-1}$, where the
${\sigma}$ are the measurement errors of the corresponding magnitude
estimates. Objects deviating in $K_{i}-J_{i}$ more than a user-defined
threshold are excluded from the following statistics. The relative
zero points $\mathit{ZP}_{l}$ for all $N$ overlapping images are estimated by a
$\chi^{2}$ minimisation with respect to $\mathit{ZP}_{k}$,
\begin{equation}
\chi^{2} = \sum_{k,j}^{N}\left[M_{k,j}-(\mathit{ZP}_{k}-\mathit{ZP}_{j})\right]^{2}\;.
\end{equation}
In addition we demand that $\sum_k \mathit{ZP}_k=0$ to make the system
non-degenerate. We test the robustness of our
photometric algorithms with simulations and on real data in the next
sections.  We note however that our relative photometric calibration
depends on two important assumptions that may not be fulfilled:
\begin{itemize}
\item We assume that the relative zero points are constant on the scale
of each image. As reported by \citet{kgo04a}, zero point variations of
up to $\approx 0.06$ mag within a WFI@2.2m image can occur due to
non-uniform illuminations (e.g. scattered light or reflections
inside the instrument). \citet{msj01} show that these deviations are
not corrected for by standard flat-field techniques (such as the
application of SKYFLATs or SUPERFLATs). Indeed, they argue that the
aim of a flat sky-background is incompatible with the aim of
homogeneous photometry on the scale of an image. Similar effects
have been observed with the CFH12K@CFHT instrument
\citep[see][]{mac04}.
In other words, the main reason for the observed variations of the photometric
zero point is the mixing of multiplicative (flat-field) and additive effects 
(scattered light, reflections) in the flat-fields. Simple division by such a 
flat then leads to a seemingly flat sky-background, but introduces the 
observed variations in the photometric zero point. A solution would be to
disentangle the multiplicative and additive component. At the moment it 
is not clear how this could be achieved in the general case.
\item We neglect that objects of different spectral type can
be affected differently by varying photometric conditions
(e.g. absorption by clouds). At this stage we have not investigated
this effect.    
\end{itemize}
For the moment we do not take into account these effects in our
photometric calibration.

For a full photometric calibration of our SCIENCE images we need to
include the results of our standard star calibration
(\sectionref{sec:standards}) in the estimation of our relative zero
points. We are currently implementing and testing this.
\subsection{Simulations for investigating the robustness of the
photometric calibration}
\label{sec:photomsimulations}
\begin{figure}[ht]
  \includegraphics[width=1.0\hsize]{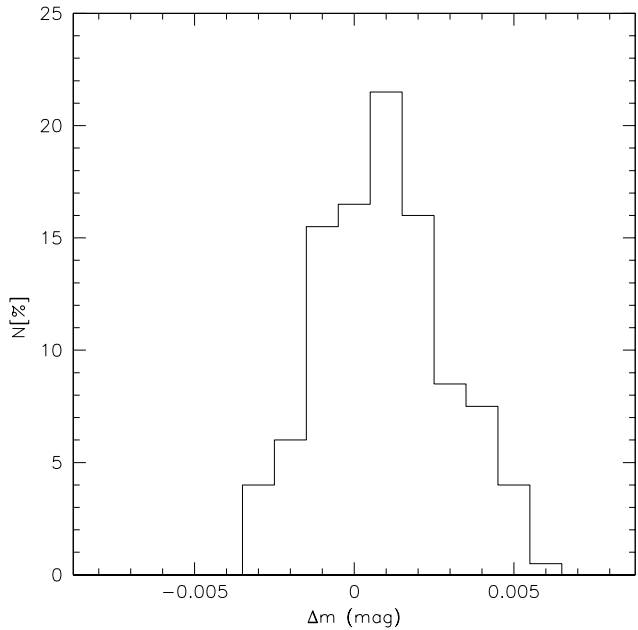}
  \caption{\label{fig:relphotom}\small{The figure
shows the error distribution of relative photometric
zero point recovery in Monte Carlo simulations. The $x$-axis displays
the difference $\Delta m$ (in magnitudes) between input and recovered relative 
zero point, the $y$-axis the recovery probability. Our algorithm recovers
the input values formally with $\Delta m=0\pm 0.002$.
See the text for more details on the simulations.}}
\end{figure}

\begin{figure*}[ht]
  \centerline{\includegraphics[width=0.47\hsize]{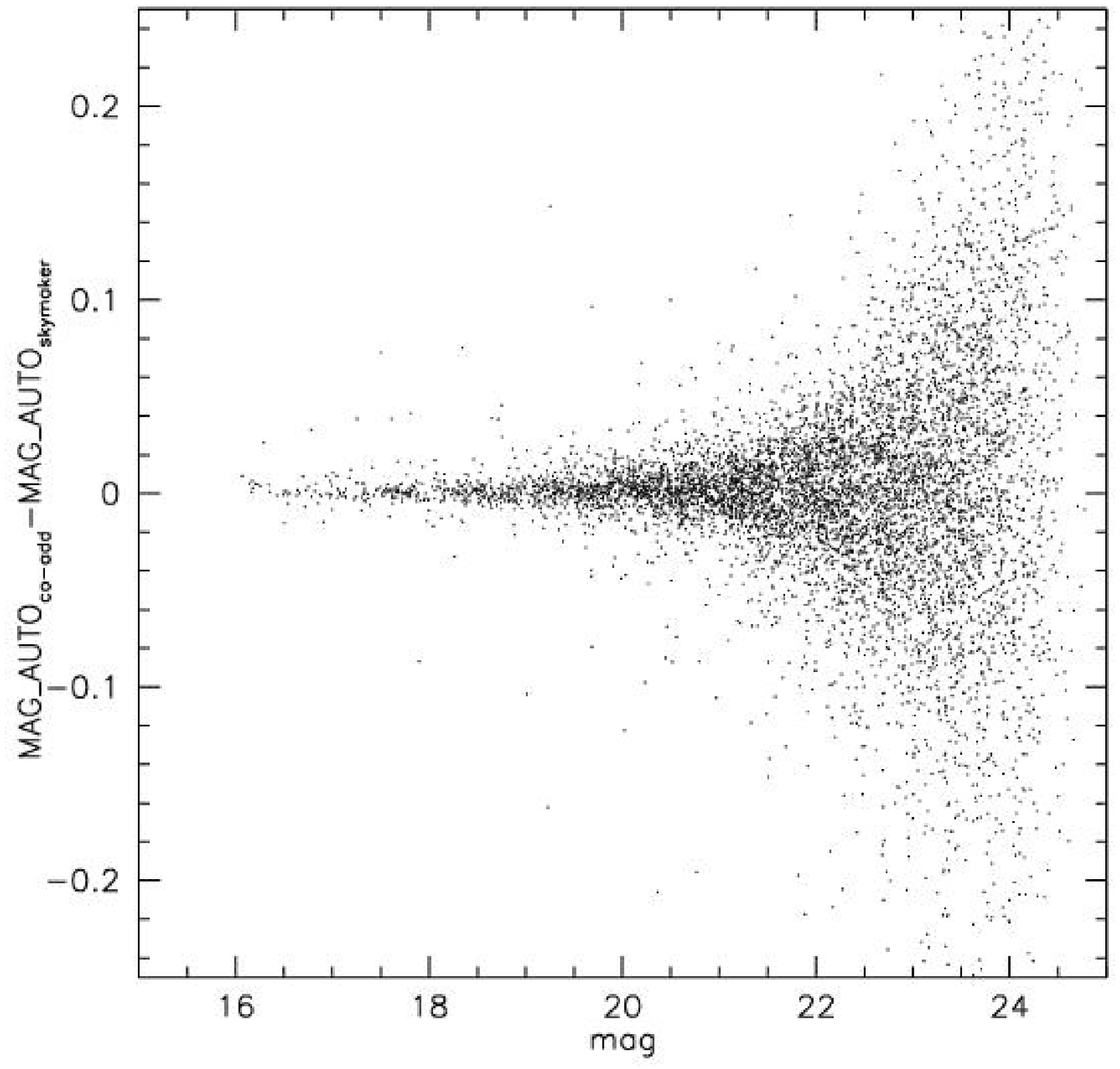}
  \includegraphics[width=0.47\hsize]{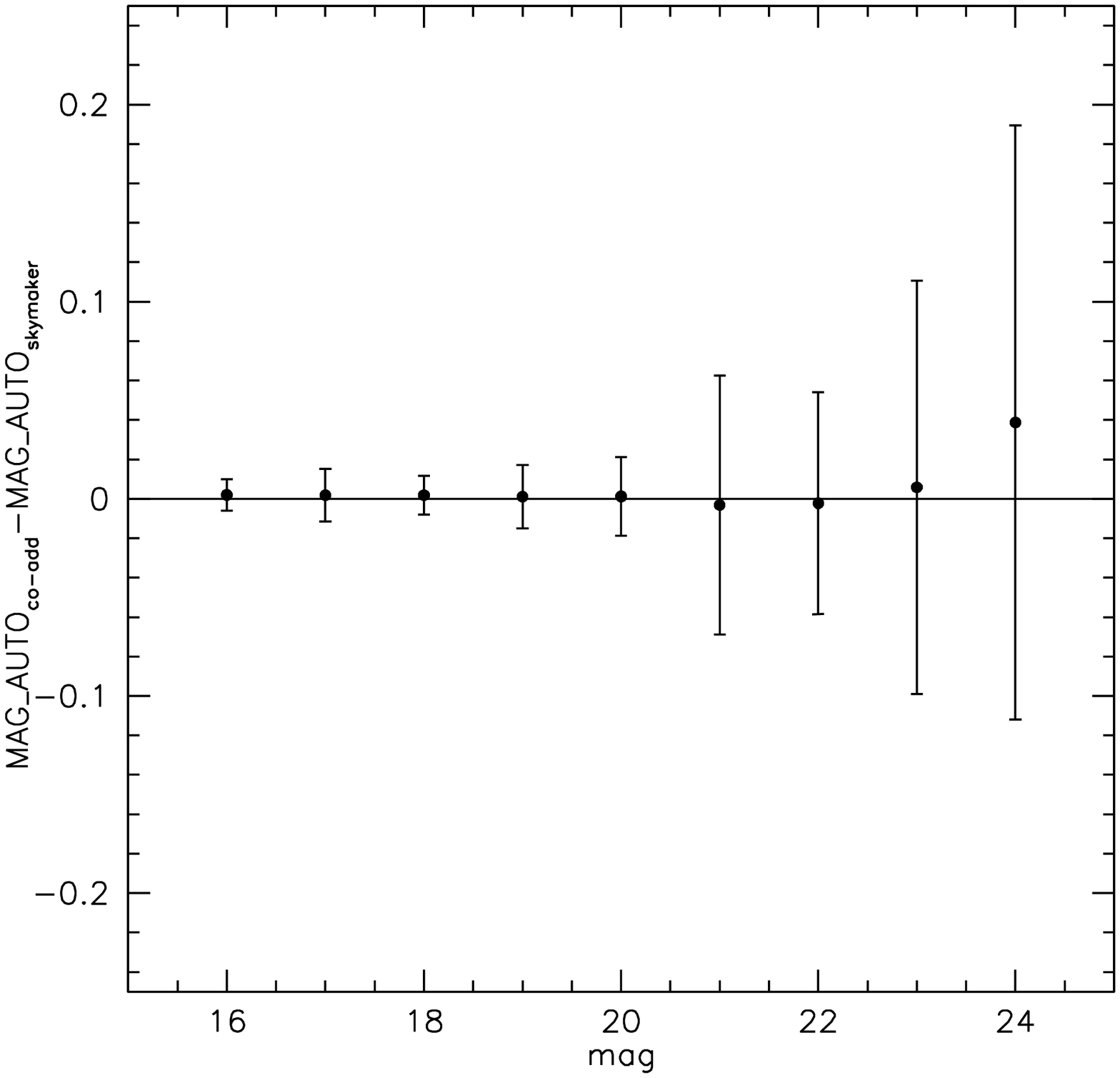}}
  \caption{\label{fig:photombias}\small{Investigation of possible flux
  measurement biases in our mosaics: We simulated 25 dithered WFI
  exposures with skymaker and processed them through our pipeline
  in exactly the same manner as real data. In addition we created
  one large skymaker mosaic not undergoing any pipeline processing.
  The figure shows the difference of magnitude estimates
  in these two images, i.e. it mainly reveals possible biases due to
  inaccurate sky subtraction in the mosaic. The left panel shows the
  raw differences, the right panel bins with a width of 1 magnitude.
  We notice that our processing does not introduce systematic biases
  up to the faintest magnitudes. See the text for more details.}}
\end{figure*}

The simulations described in this section aim at quantifying
the following issues of our photometric calibration:
\begin{itemize}
\item How accurately can we recover relative photometric 
zero points between individual images?
\item We subtract the sky-background from our images before
they are finally co-added (see \sectionref{sec:co-addition}
for details on the procedure). Does this procedure introduce
any systematic bias in the flux measurements of faint objects?
\end{itemize}
To this end we simulated WFI exposures with Emmanuel Bertin's
{\tt skymaker}\footnote{the TERAPIX tool for the simulation of
astronomical images} \citep[see][]{ewb01}
and processed them through our pipeline in exactly the same
way as real data. In this way we include all possible uncertainties
from measurements, possible pixelisation effects etc. in our
simulations. The simulations have the following characteristics:
\begin{itemize}
\item We created 25 exposures of 8 chips in the layout
of the WFI@2.2m telescope (i.e. a pixel size of $0\myarcsec238$ and a
chip size of $2\mathrm{k}\times 4\mathrm{k}$). Each exposure mimics a 500 s
integration in an $R$ broad-band filter. The images do not
contain any flat-field or fringe residuals but resemble
perfectly pre-reduced data.
\item Each exposure is put randomly within a 
$2'\times 2'$ dither box. No geometric
distortions have been introduced.
\item Each exposure has a random image seeing between $0\myarcsec9$ and
$1\myarcsec1$ (no PSF anisotropy) 
and a random sky $R$-band background between 20.0
and 20.5 mag/arcsec$^2$.
\item Each image is assigned a random magnitude zero point between
$24.9\pm0.3$, i.e. there is no correlation of zero point
with exposure, extinction or gain.
\end{itemize}
The images were processed through our pipeline starting at the
astrometric calibration up to the final mosaic.  In addition we
simulated with {\tt skymaker} a single $10\mathrm{k} \times 10\mathrm{k}$ exposure with a
formal integration time of $25\times 500$ s (skymaker mosaic in the
following) for flux comparisons with the co-added mosaic. The
accuracy of the zero point estimation is shown in
\figref{fig:relphotom}.  We see that our procedure recovers the input
values with a formal uncertainty of 0.002 mag.  In
\figref{fig:photombias} we show the comparison of flux measurements
from the co-added mosaic with the skymaker mosaic.  No systematic
deviations in the magnitudes up to the faintest objects are visible.
We conclude that our procedures accurately recover relative zero point
deviations between images and that our mosaics do not suffer from
systematics in estimating magnitudes.  Errors connected to the
algorithms are, at least for WFI@2.2m, small compared to systematic
errors (variations of about 0.05 mag within an image) due to possible
CCD illumination problems (see above).
\subsection{Photometric measurements in NGC 300}
\label{subsec:NGC300}
\begin{figure}
\centering
   \includegraphics[width=3.5in]{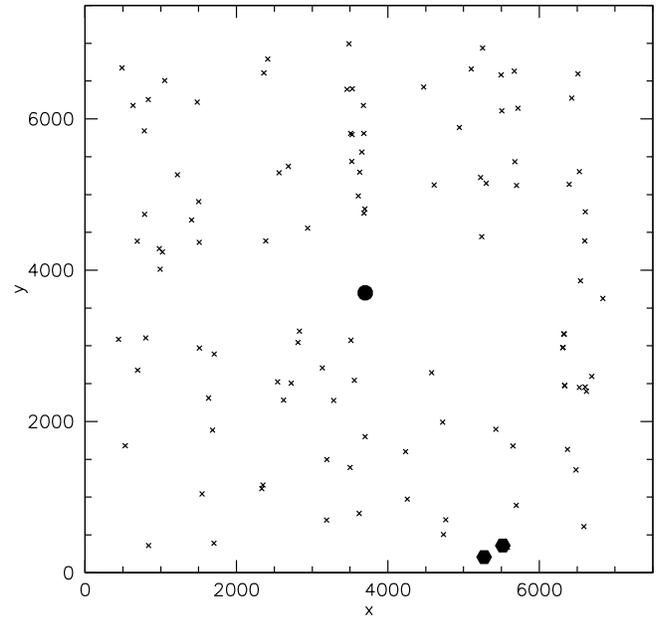}
   \caption{Position of the chosen standard stars in the co-added
     $V$-band image. Crosses denote the positions of the secondary
     standards from \citet{pgu02}, the big dot the centre of NGC 300
     and the filled hexagons two photoelectric standards from
     \citet{gra81}.}
\label{overview} 
\end{figure}

To finally quantify the quality of our current photometric
calibration, we compare flux measurements in the field of NGC 300
\citep[see][]{ses03} with previously published secondary standards 
by \cite{pgu02}.
PSF photometry was carried out on co-added $B$ (exposure time:
39\,600 sec) and $V$ (exposure time: 37\,440 sec) WFI@2.2m images of NGC 300
by using the DAOPHOT task \citep*{ste87} implemented in IRAF Version
2.12.2. We used about 100 out of 390 secondary photometric
standard stars that were published by \cite{pgu02} to uniformly
cover our field-of-view. Areas, in which crowding became significant or
where estimations of the sky-background became difficult (galaxy
centre, spiral arms, saturated stars), were
excluded. \figref{overview}
shows the spatial distribution of the chosen standard stars in the
co-added image. The galaxy-centre
of NGC 300 coincides with the field centre. The photometric standards
published in \cite{pgu02} were observed in four fields with the
Warsaw 1.3\,m telescope at the Las Campanas Observatory.

Using the data we solved the following transformation where we neglect
the airmass-term which is constant over the field-of-view:
\begin{eqnarray}
b&=&B + b_{1} \cdot (B-V)+b_{2}\\
v&=&V + v_{1} \cdot (B-V)+v_{2}.
\end{eqnarray}  
\begin{figure*}
\centering
   \includegraphics[width=0.47\hsize]{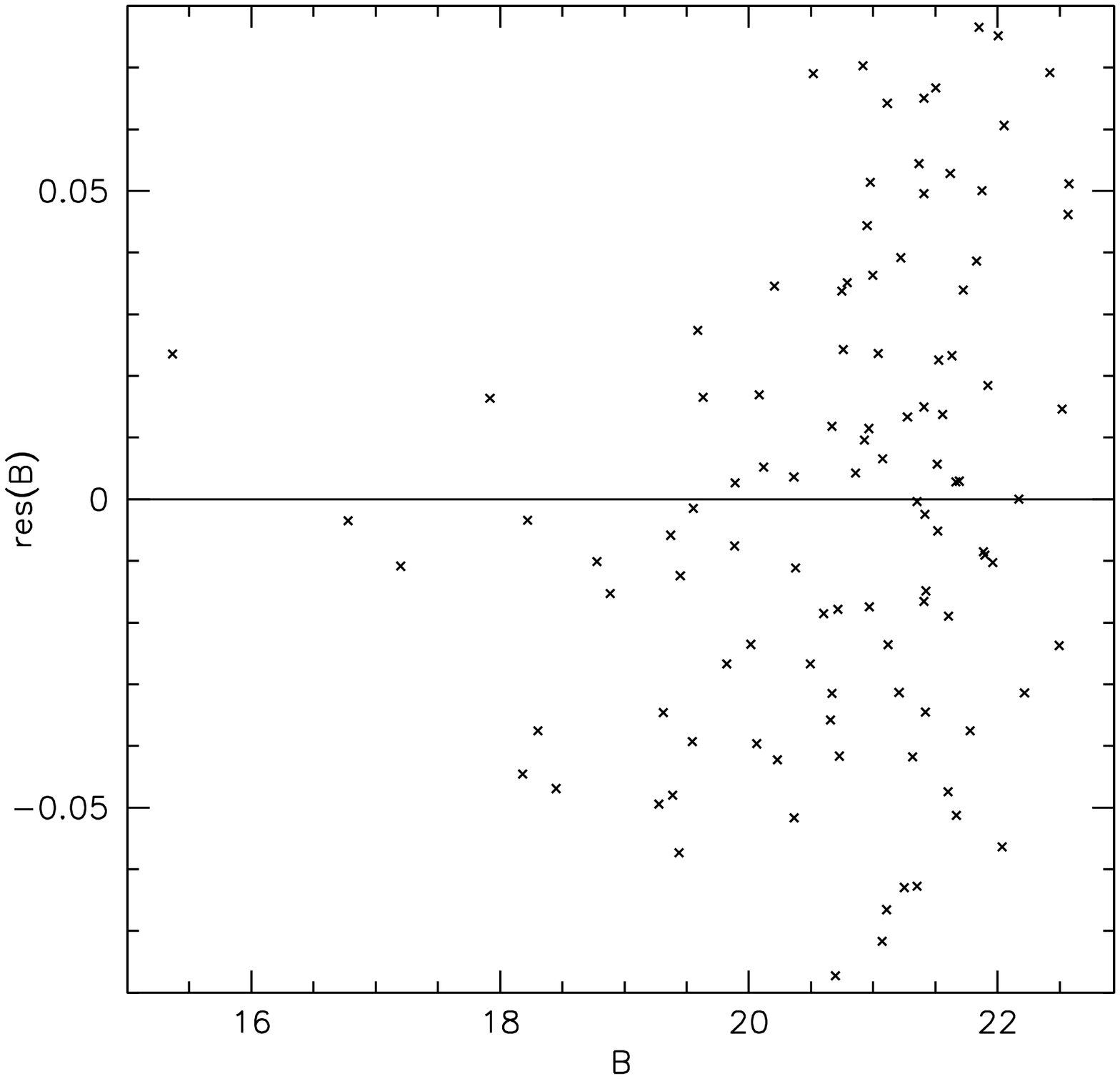}
   \includegraphics[width=0.47\hsize]{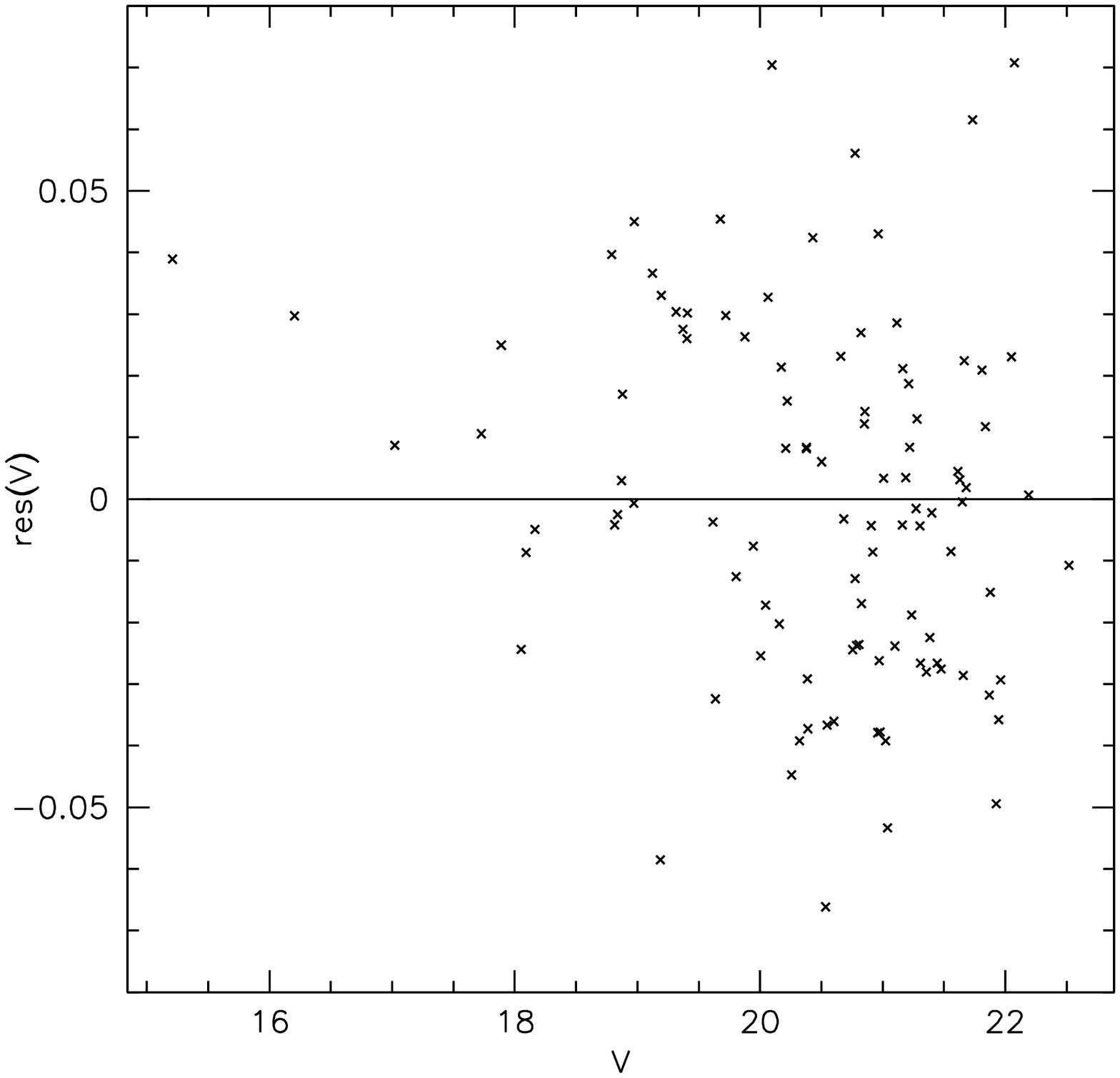}
   \caption{Residuals vs. magnitude plots of 107 chosen standard
     stars. The left panel shows the results for the $B$-band, whereas 
     the right panel shows those for the $V$-band}
\label{fig:scatter} 
\end{figure*}

Capital letters represent standard star magnitudes in $B$ and $V$
respectively whereas lower case letters indicate the measured
instrumental magnitudes. Values of $b_{[1,2]}$ and $v_{[1,2]}$ have been
determined by solving these equations for the standard stars using the
Levenberg-Marquardt algorithm implemented in the IRAF routine
INLFIT. This leads to values of $b_{1}=-0.28 \pm 0.007$,
$b_{2}=0.69 \pm 0.006 $, $v_{1}=0.078 \pm 0.005$ and
$v_{2}=0.89 \pm 0.005$. In \figref{fig:scatter} we plot the
residuals of this fit as a function of $B$-band (left-hand side) and
$V$-band (right-hand side) magnitude and see that there is (to a first
approximation) no significant tendency for the residuals to correlate
with magnitude\footnote{this especially holds for stars fainter
than mag=19}. Calculating the standard deviation $\sigma$ for
these fits we get $\sigma(B)=0.037$ and $\sigma(V)=0.029$ with
maximum residuals of about 0.07 magnitudes.

This means the errors and/or uncertainties introduced by a global
solution for the majority of stars are smaller than 0.04 magnitudes
and are therefore negligible for most photometric studies
taking into account that they are of the same order as the
uncertainties for some of the secondary photometric standards.

We also tested the global solution in relation to the different
chips. For this, we fitted the global solution separately to areas
where mostly images of a particular chip overlap and checked the
standard deviation of these fits. The results are presented in
\tableref{tab:standardcheck} and it becomes obvious that the standard
deviation for a single chip differs from the standard deviation of the
global solution only by a constant smaller than 0.005. Due to this,
our global solution can be regarded as sufficient and therefore shows
that our photometric calibration is capable of removing chip-to-chip
variations in a more than satisfying manner. This has also to be seen
in the context that the observations of the very extended NGC 300
galaxy were spread over 34 nights in six months, encompassing very
different sky conditions.

Finally we have performed a check of our transformation with two
photoelectric standards published by \cite{gra81} which are not
saturated in our images. Applying the global solution to these
photoelectric standards leads to residuals of $-0.036$ and $-0.069$
in $V$ and $-0.055$ and $-0.022$ in $B$. 
Considering the fact that on the one hand the
photometric errors for these stars are about 0.05 magnitudes, and on
the other hand some of our fitted standards had residuals up to 0.07
magnitudes, these residuals are in good agreement with the previous
results.

We point out that the checks performed so far give us
confidence that to date the relative photometry of the pipeline permits us to
reduce and to co-add images to a photometric accuracy of about 0.05
magnitudes (the average between the standard deviations of the fits to
secondary photometric standards of Pietrzynski et al., \citeyear{pgu02}, and
photoelectric standard stars from Graham, \citeyear{gra81}). This is in
agreement with the expected errors due to the above mentioned illumination
issues.

\begin{table}
\caption{\label{tab:standardcheck} Standard deviation of the global
solution applied to each chip. The relation between chip number and 
actual position on the mosaic can be seen in \figref{fig:WFIAREA}.}
\begin{center}
\begin{tabular}{ccc}
\hline
\hline
Chip No. & $\sigma(B)$ & $\sigma(V)$ \\
\hline
1& 0.035 & 0.024 \\
2& 0.040 & 0.034 \\
3& 0.039 & 0.033 \\
4& 0.040 & 0.031 \\
5& 0.030 & 0.026 \\
6& 0.036 & 0.033 \\
7& 0.039 & 0.033 \\
8& 0.039 & 0.027 \\ 
\hline 
1-8 & 0.037 & 0.028 \\
\hline
\end{tabular}
\end{center}
\end{table}
\subsection{{\sl Set} quality characteristics}
\label{subsec:setcharacteristics}
\begin{figure*}
  \centerline{\includegraphics[width=0.95\hsize]{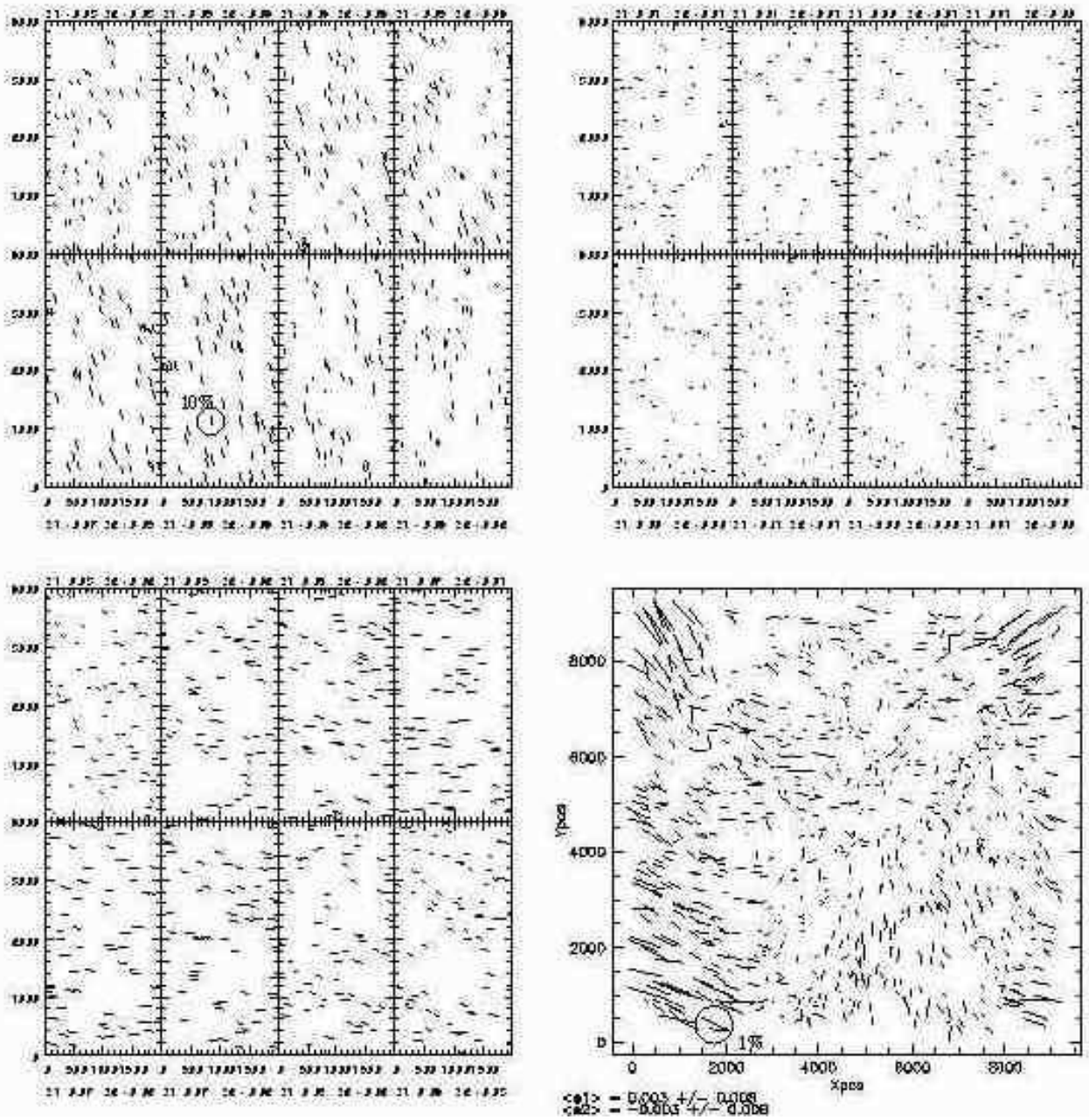}}
  \caption{\label{fig:psf_overview}\small{PSF properties of WFI@2.2m
  fields: For each image within a set
  we characterise PSF properties (PSF size and anisotropy) 
  over the complete field-of-view of
  a WFI. To this end we run the KSB algorithm [a weak lensing analysis
  technique for accurate shape measurements of astronomical sources; see
  \citet{ksb95}] over selected stars of all images. The upper left
  panel shows PSF anisotropies for an intra-focal, the upper right for
  a focal and the lower left for an extra-focal exposure. The sticks
  represent amplitude (given as $\sqrt{e_1^2 + e_2^2}$) and
  orientation $\phi=0.5\arctan (e2/e1)$ of the PSF anisotropy. The
  chosen scale for the stick length is the same for these three plots in order
  to show the increase in the anisotropies with respect to the focused
  exposure. The mean stellar ellipticities are 0.066, 0.09 and 0.059,
  respectively. The lower right panel depicts typical PSF anisotropies of a
  WFI@2.2m $R$-band mosaic ($\sim57$ exposures with $\sim500$ s exposure time 
  each). Note that the largest PSF anisotropies in the mosaic are as small as 
  $\approx 0.01$. Compared to the other three PSF plots, a different scale for 
  the stick length was used in order to clearly show the
  anisotropies. We note that WFI@2.2m has excellent PSF properties if
  data are obtained under favourable focus conditions.}}
\end{figure*}

\begin{figure*}[ht]
\begin{center}
  \begin{minipage}[t]{0.49\textwidth}
    \psfig{figure=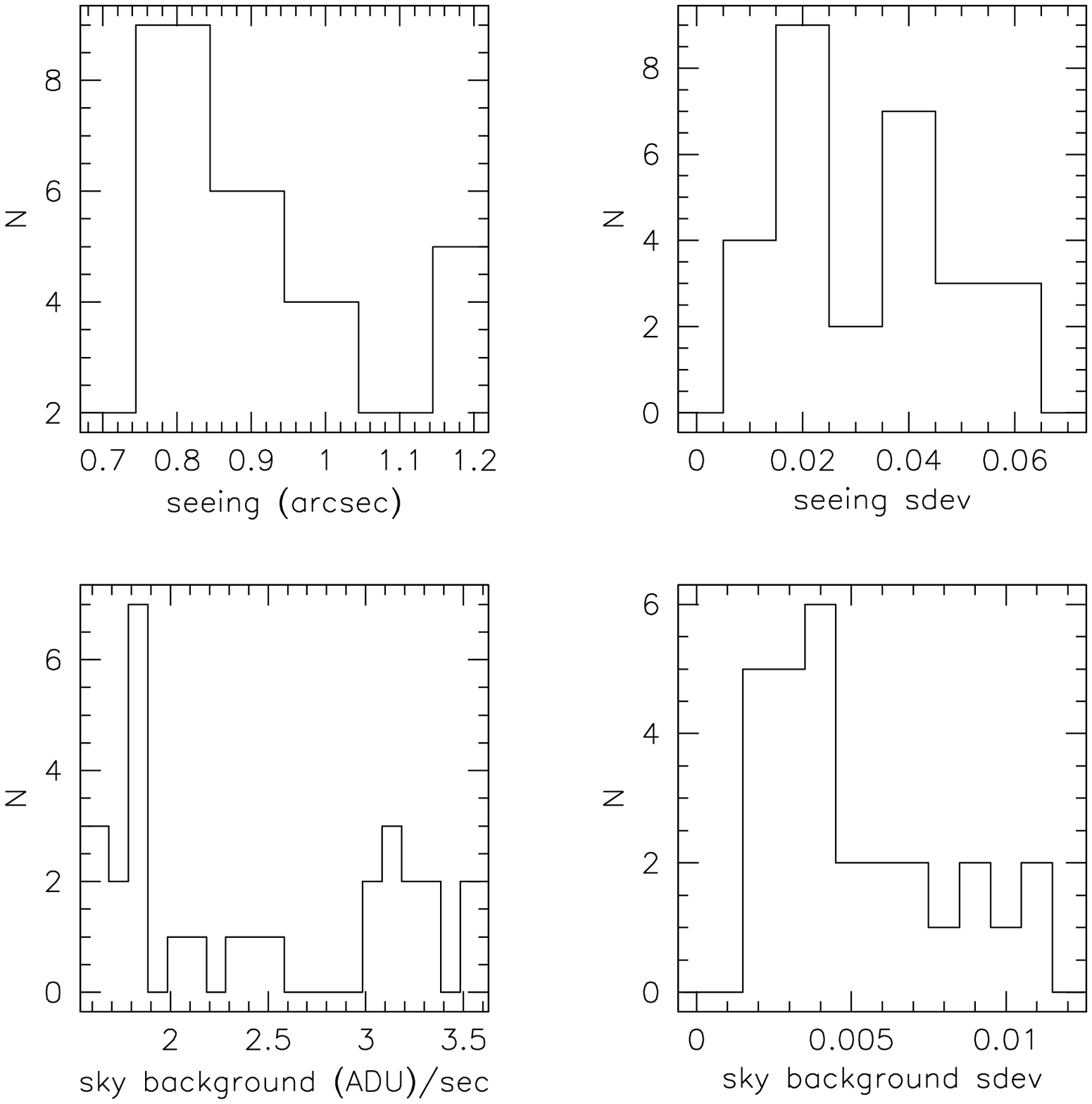,width=\hsize,angle=0}
  \end{minipage}
  \begin{minipage}[t]{0.49\textwidth}
    \psfig{figure=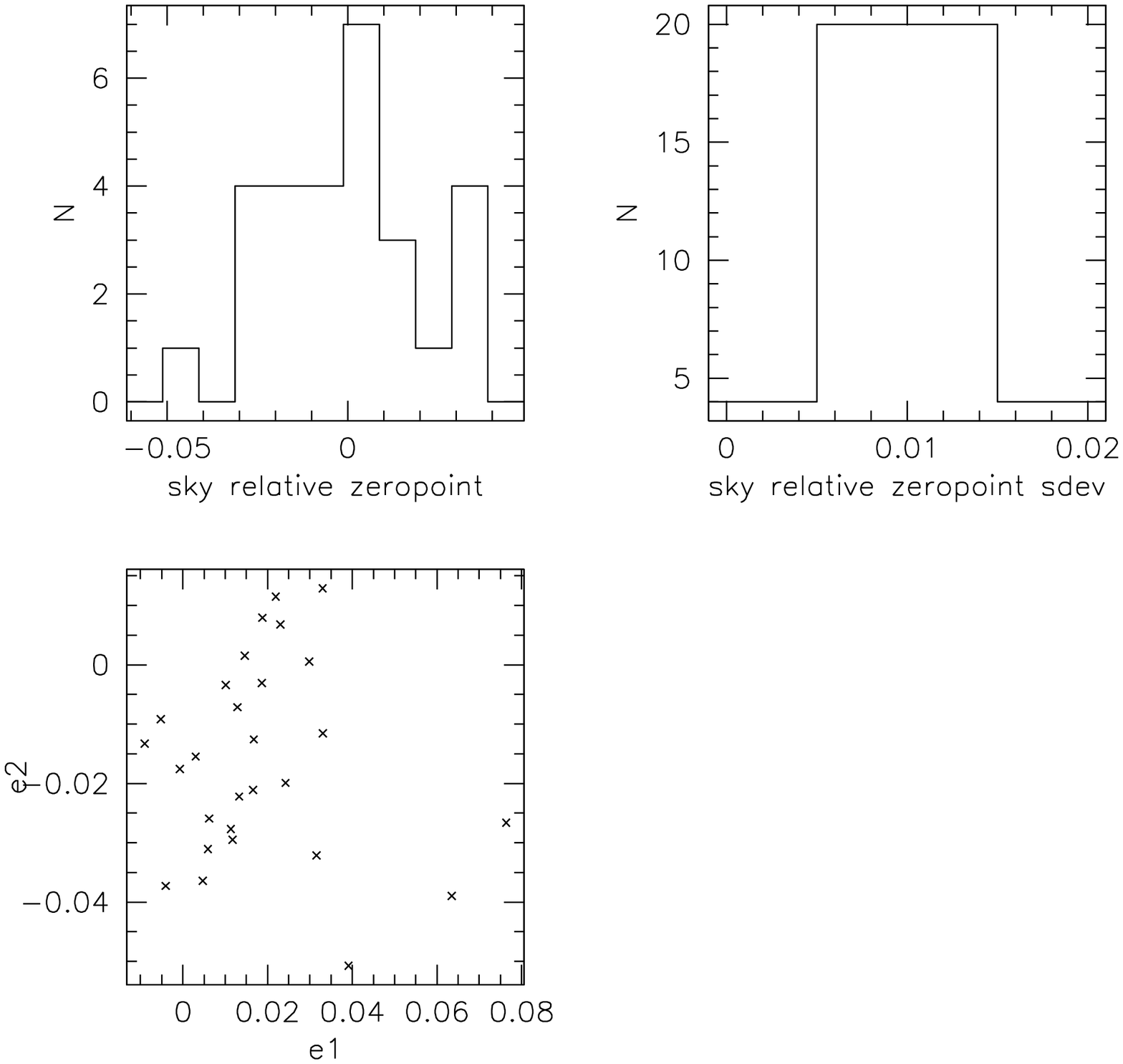,width=\hsize,angle=0}
  \end{minipage}
\end{center}
\caption{Shown are quality control plots for a set of 28 $R$ band
exposures from WFI@2.2m. They primarily allow us to select the images
that enter the final co-addition process. The two upper left plots
show the seeing distribution in the set, where the seeing value of an
exposure is estimated from the mean seeing in the individual chips.
The distribution of seeing variations within an exposure is also shown
(second left plot). The two lower left plots show the night sky
amplitude (normalised to an exposure time of one second) during the
observations. They allow us to identify exposures taken during twilight
or unfavourable moon phases. In the upper right we have a quick check
on the photometric conditions through the distribution of relative
zero points (see \sectionref{subsec:photometric}). The standard
deviations of relative zero points show that our sky-background equalisation (see
\sectionref{subsec:gainequalisation}) adjusts zero points typically to
0.01-0.03 mag within a WFI@2.2m mosaic. In the lower right we see the
PSF ellipticity distribution (see \figref{fig:psf_overview}). Here each
point represents the mean ellipticity value from all stars within an
exposure.}
\label{fig:setchar}
\end{figure*}

The object catalogues generated for the astrometric and photometric
calibration give us a good overview of the data quality at this
stage. They allow us to investigate in more detail the PSF properties
of individual exposures which is crucial in weak lensing studies. Also
the night sky conditions (transparency, sky brightness) can be studied
in detail and images that should be excluded from the final
co-addition can be identified. \figref{fig:psf_overview} discusses in
detail PSF properties of WFI@2.2m and \figref{fig:setchar} shows an
overview of our currently implemented quality control.

\section{Weighting and Flagging}
\label{sec:weightingflagging}
With astrometric and photometric information for all images at hand we
can finally co-add them and produce a deep mosaic. This section
motivates our choices for the implemented co-addition procedure which is
very close to that of the EIS Wide Survey described in \citet{nbc99}.

\begin{figure}[ht]
  \centerline{\includegraphics[width=0.9\hsize,angle=-90]{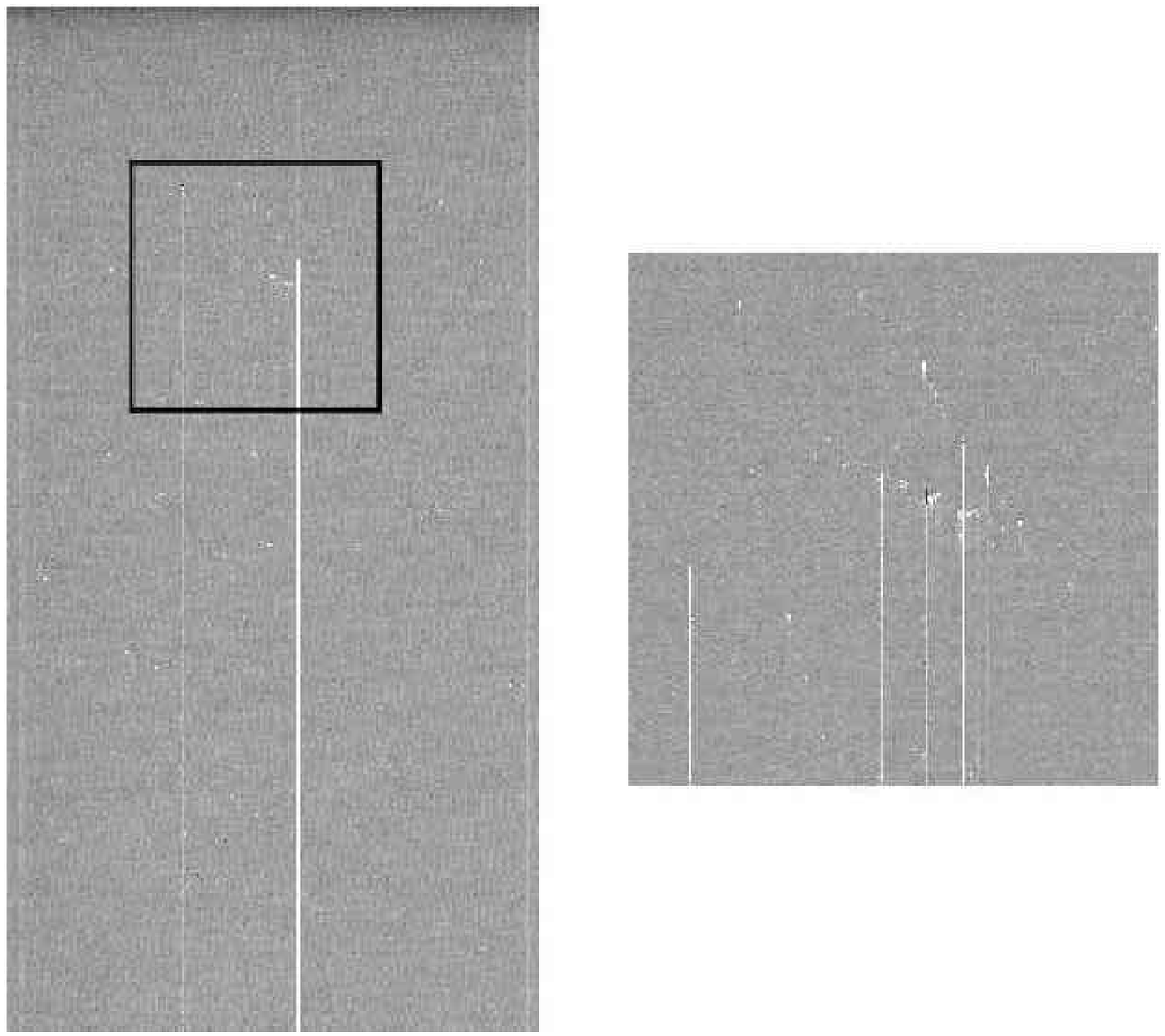}}
  \caption{\label{fig:dark} Shown is a masDARK image and a zoomed in
region. In a stack of several long-exposed DARK frames hot and cold
pixels, which often come in groups and affect complete rows/columns,
show up with high significance and are easily identified by
applying a pixel value threshold to the masDARK.}
\end{figure}

\begin{figure*}[ht]
  \includegraphics[width=0.47\hsize,angle=-90]{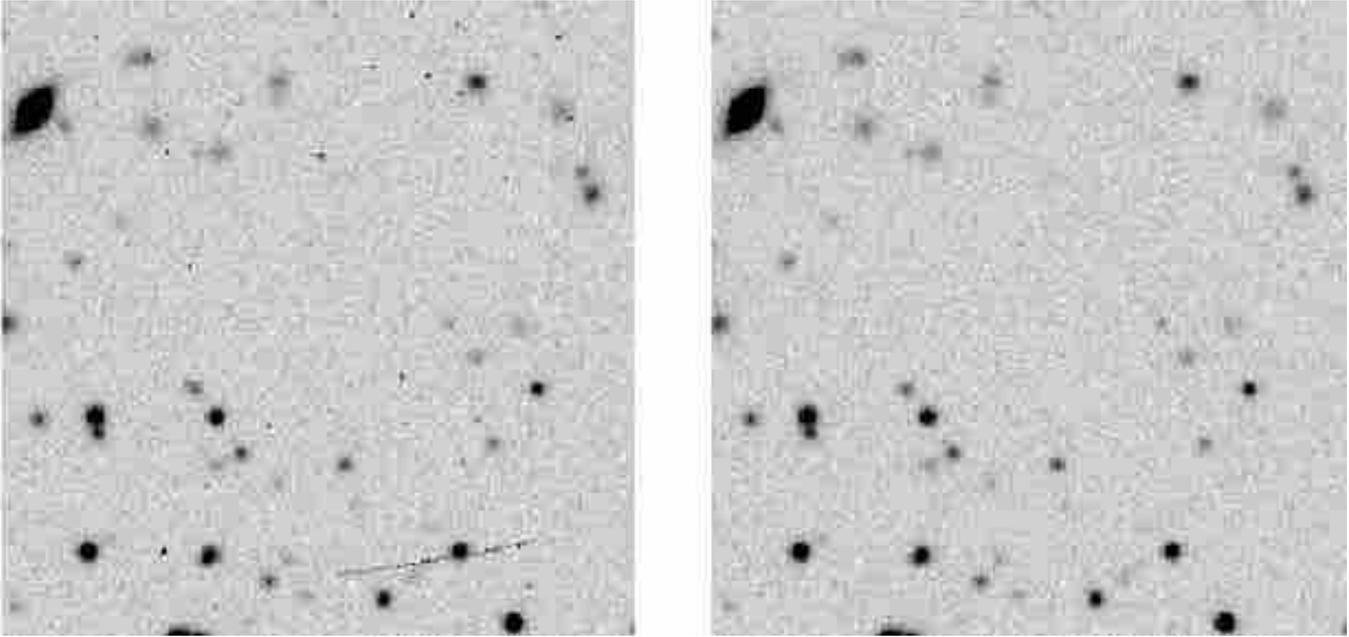}
  \caption{\label{fig:cosmicrem} The left panel shows 25 co-added
WFI@2.2m frames where no cosmic ray cleaning has been performed.  The
right panel shows the result with the cleaning procedure turned on.
We find that for WFI@2.2m frames {\tt SExtractor} (in connection with
a neural network filter generated with {\tt EYE}) performs a very
efficient detection of small-scale cosmic ray features.}
\end{figure*}

As long-exposed SCIENCE frames are dominated by Gaussian sky noise the
optimal result (in terms of noise properties in the co-added image) is
obtained by a weighted mean co-addition of all input images. However, this is not
straightforward as many image pixels carry non-Gaussian noise
properties (such as bad columns, vignetted image regions, cosmics)
or other defects that we would like to exclude from the co-addition
process (such as satellite tracks or extended stellar
reflections). To perform a weighted mean co-addition all
{\sl bad} image pixels have to be known beforehand and assigned a zero
weight in the co-addition process. We note that identifying defects
before the co-addition has several advantages over rejecting them
during the stacking process:
\begin{itemize}
\item Rejection methods during the co-addition, such as
median stacking or sigma-clip rejection require a minimum
number of 5-10 input images for robust statistics. With prior
knowledge of the defects good results can already be obtained with
a small number of images.
\item Co-additions based on median or sigma-clip rejection
are problematic when observations under varying seeing
conditions have to be combined. In this case
object profiles in the co-added image might be affected by 
the rejection algorithm.
\item A practical advantage is that a weighted mean without
rejections is a strictly linear process. Hence, new images can 
directly be added to an already existing mosaic and not all
input images need to be co-added again.
\end{itemize}
We identify bad pixels in the individual exposures
as follows:
\begin{enumerate}
\item Every chip contains {\sl hot} and {\sl cold} pixels, i. e.
pixels that always have a high/low charge value. They are typically
stable over the period of an observing run and are most
effectively identified in the masDARK frames (see
\figref{fig:dark}). If DARK frames are not at hand, SKYFLATs
and/or SUPERFLATs often provide a sufficient bad pixel map.
\item Saturated image pixels are identified by applying a pixel
value threshold to the SCIENCE frames.
\item Pixels affected by cosmic rays are not identified as easily as
their appearance is different in each SCIENCE frame. We detect them
with {\tt SExtractor} with a special filter for cosmic ray detection
generated with {\tt EYE} \citep[][]{ber01,nbc99}. The filter we actually use
was specifically designed for cosmic ray removal on EMMI@NTT data in the
framework of the EIS Wide project. It is difficult to say for which
other instruments it can be used without modification or to quantify
its actual performance. The visual impression from our co-added
WFI@2.2m frames is that the detection efficiency for small-scale
cosmic ray features is very high and gives a sufficient image cleaning
for this camera (see \figref{fig:cosmicrem}).
\item Other extended defects such as satellite tracks or
reflections from bright stars are masked by hand. They have been
identified in the manual pass through the data during the {\sl Run}
processing (see \sectionref{sec:manualpass}).
\end{enumerate}
In this way we can generate a {FLAG} map for each image. It is an
integer FITS image containing zeros for every good pixel and values
characterising the different defects otherwise. We use these FLAG maps
in the creation of WEIGHT maps (see below) and to mark standard stars
whose magnitude measurement is affected by a defect (see
\sectionref{sec:standards}).\footnote{The FLAG and WEIGHT maps
described in this section are actually created on the {\sl Run}
level before the standard field processing. 
Because of their very close relation to the image co-addition
we shifted their description to the {\sl Set} processing.}

\begin{figure*}
\begin{center}
  \begin{minipage}[t]{0.4\textwidth}
    \psfig{figure=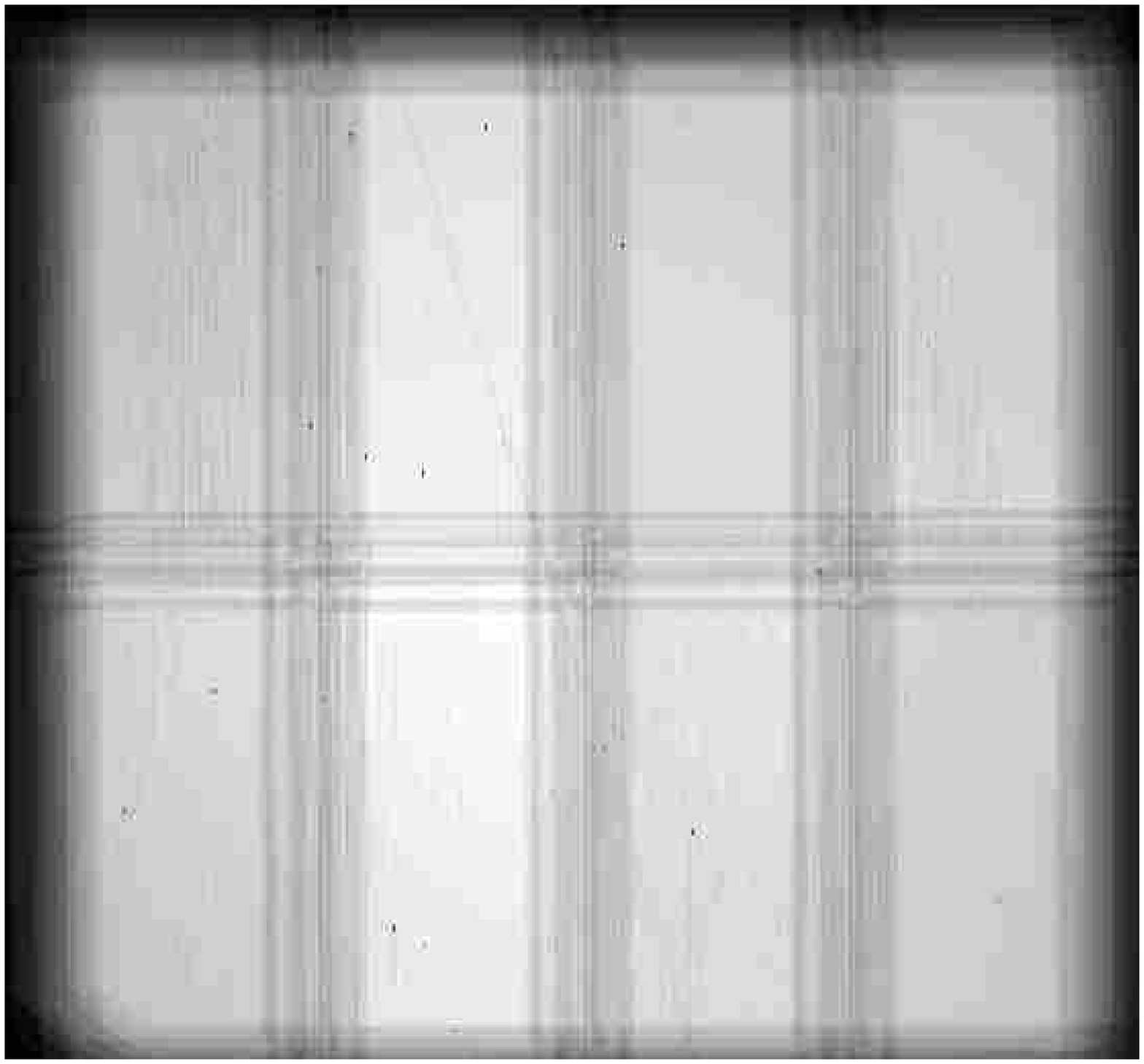,height=0.28\textheight,angle=0}
  \end{minipage}
  \begin{minipage}[t]{0.4\textwidth}
    \psfig{figure=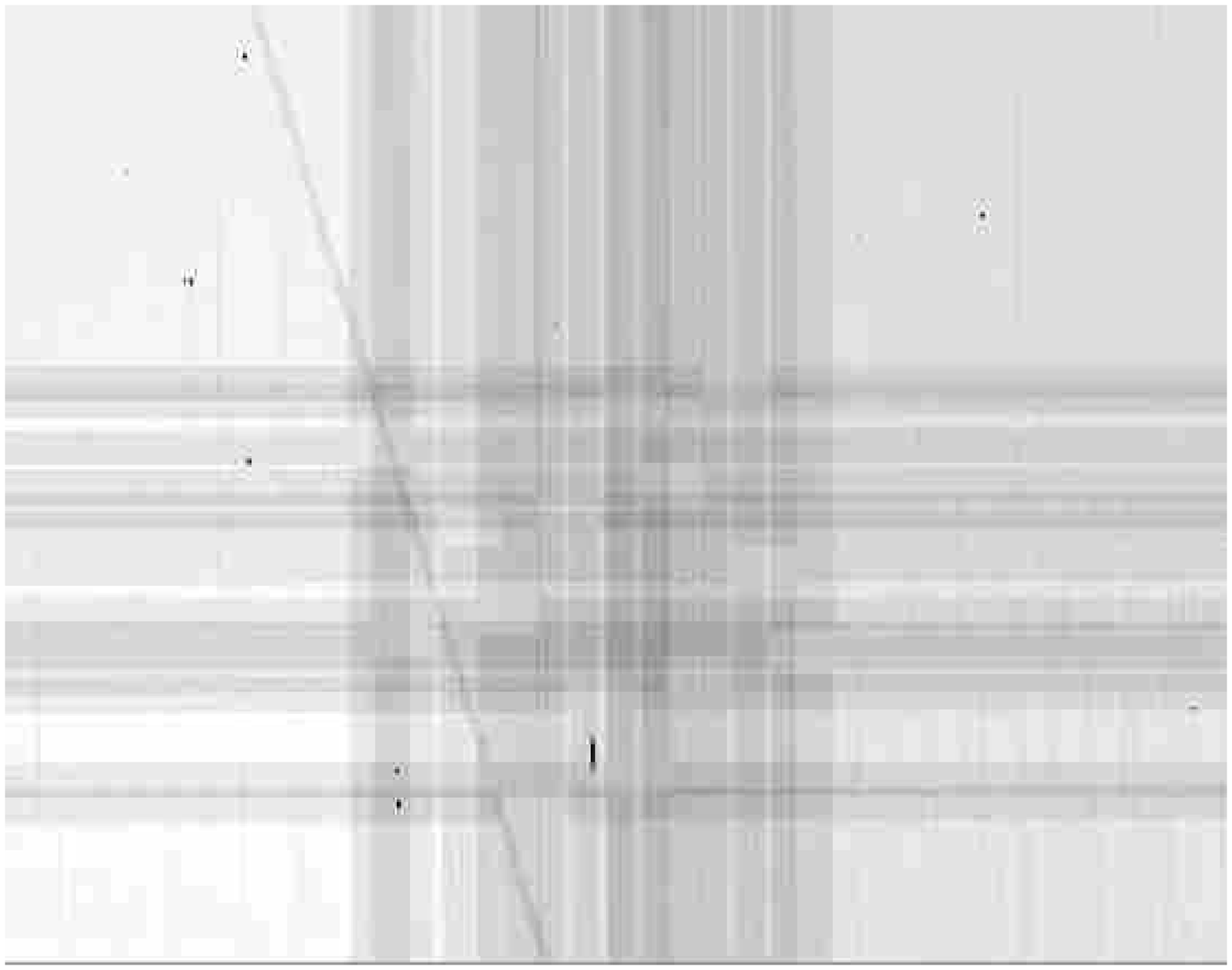,height=0.28\textheight,angle=0}
  \end{minipage}
\end{center}
\caption{Shown is the WEIGHT map (left panel) characterising the noise
properties of a co-added science image and a zoom-in to its centre
(right panel). The lighter the colour, the higher the relative weight
(the lower the noise) of the pixel. The darker regions at the
positions where different chips meet have about half the weight of
well-covered regions.  Different weights between regions where the
same number of input images have contributed show intrinsic sensitivity
variations. We note that the noise structure is quite
complex and cannot be taken into account appropriately without the
WEIGHT map. These maps are used by {\tt SExtractor} in the object detection
phase and for noise calculations. Having the map at hand one also does
not need to cut the outermost regions of the coadded images where the
noise is considerably higher than in the inner regions. See also
\figref{fig:weightcat}.}
\label{fig:weight}
\end{figure*}

\begin{figure*}[ht]
  \centerline{\includegraphics[width=0.49\hsize,angle=-90]{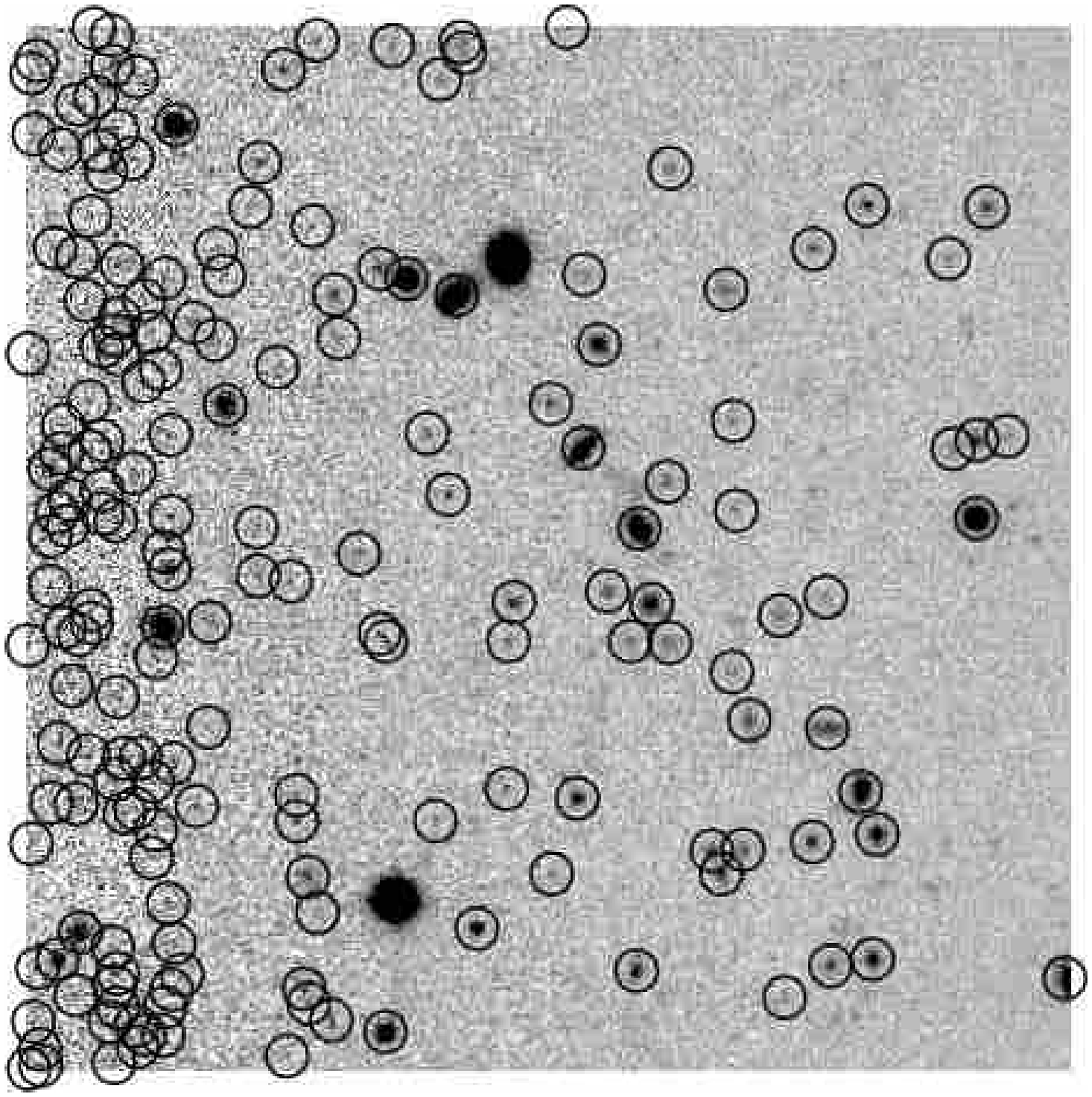}
  \includegraphics[width=0.48\hsize,angle=-90]{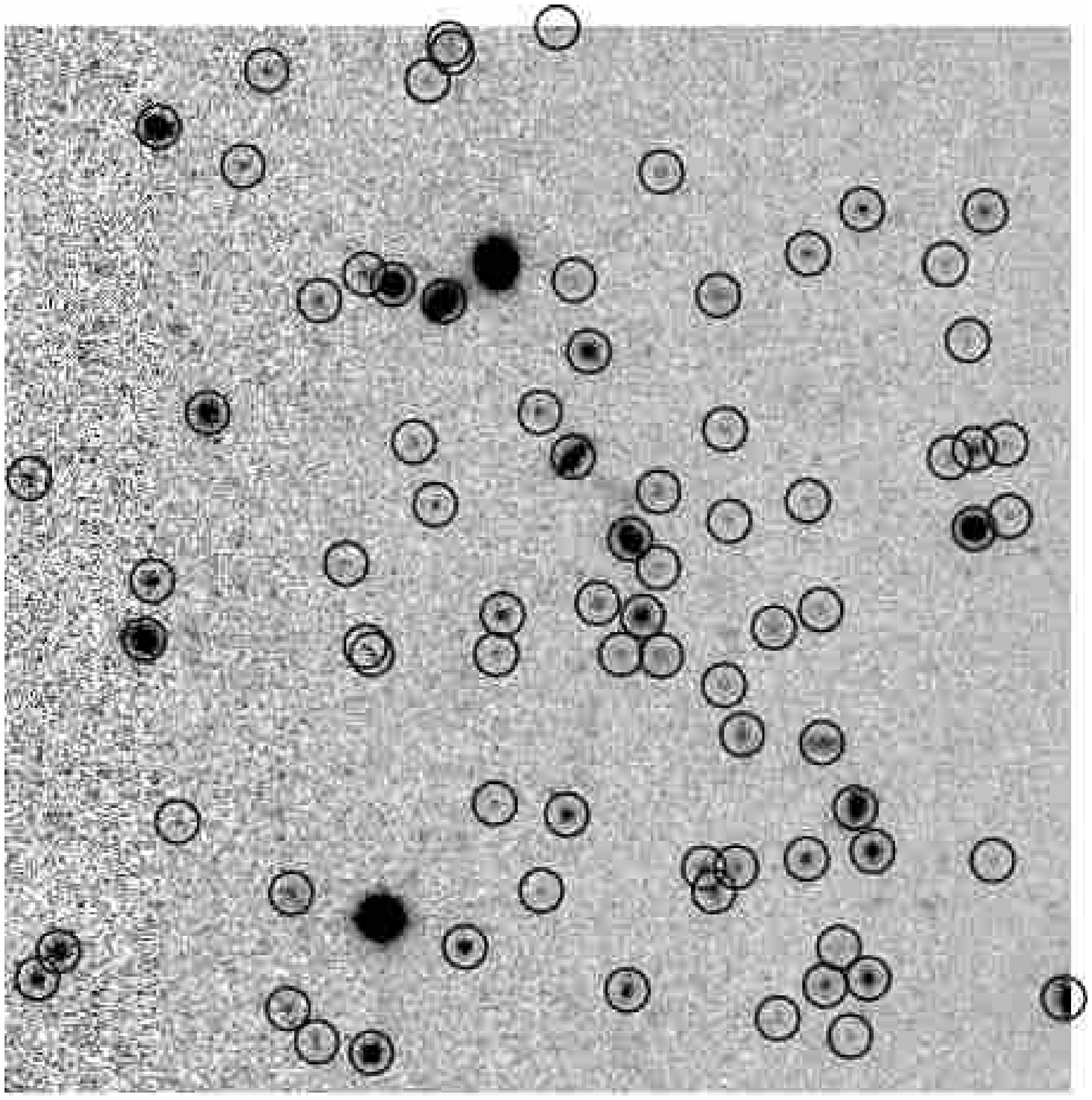}}
  \caption{\label{fig:weightcat}\small{{\tt SExtractor} source
  catalogue obtained from the same mosaic (with an area of
  $95\myarcsecnodot\times 95\myarcsecnodot$) without (left) and with
  (right) a WEIGHT image as an additional {\tt SExtractor} argument.
  The number of spurious detections in regions with higher noise is
  obvious when no WEIGHT image is used as an additional input.}}
\end{figure*}

Besides to the actual image co-addition algorithm we need to pay
special attention to the noise properties in the final co-added
image. The gaps in multi-chip cameras and the different intrinsic
gains of the various detectors lead to complex noise properties in a
co-added image consisting of dithered exposures. These noise
variations need to be taken into account properly when estimating a
threshold for object detection or for estimating errors based on pixel
noise (such as object fluxes for instance). For a full
characterisation of the noise, another image besides the final
co-added image, the {WEIGHT} map, describing the relative noise
properties for each image pixel, is created during the co-addition
process. We arrive at this WEIGHT map in the following way:
\begin{enumerate}
\item We produce a WEIGHT map for every input image to be co-added.
The starting point for these individual WEIGHTs is the masFLAT that
is rescaled to a mode of unity. It provides information on the
relative sensitivity (and hence noise) variations within an image. This basic
map is modified by the FLAG map by setting the value of defect pixels 
to zero.
\item These individual WEIGHTS are co-added together with the science
images. The co-added WEIGHT provides a full characterisation of the
relative noise properties later. The exact co-addition procedure is
described in \sectionref{sec:co-addition}. \figref{fig:weight} shows
an example for a co-added WEIGHT map and \figref{fig:weightcat} its merits for
object detection.
\end{enumerate} 

FLAG and WEIGHT maps are produced with the TERAPIX tool
{\tt weightwatcher} \citep[see][]{ber98}.
\section{Image co-addition}
\label{sec:co-addition}
Before the final co-addition, all images are brought to the same
background level by subtracting the night sky. To accurately model the
sky-background and to avoid biased estimates close to large
astronomical sources we apply the following two-stage process:
\begin{enumerate}
\item  We run {\tt SExtractor} and detect all large-scale objects having 
at least 50 adjacent pixels with 1.5$\sigma$
over the sky-background (these values are for WFI@2.2m and can vary
according to the data set). All image pixels belonging to 
detected objects are replaced with the image mode.
\item From this object-subtracted image we create a {\tt SExtractor}
BACKGROUND check-image (BACK\_SIZE=90 for WFI@2.2m images) and
subtract it from the original SCIENCE image.
\end{enumerate}
We showed in \sectionref{sec:photomsimulations} that this sky
subtraction does not introduce a significant bias in the object
photometry on the co-added images.

For the co-addition we can use Richard Hook's {\tt EIS Drizzle} which
is implemented within IRAF or {\tt SWarp} (E. Bertin, TERAPIX). Both
software packages perform a weighted mean co-addition on a sub-pixel level 
taking into account the full astrometric solution and allowing
a variable output pixel size. {\tt EIS Drizzle} was
developed within the EIS Wide project \citep[][]{nbc99} 
to co-add quickly large
volumes of data with low demands on CPU and virtual memory. It
uses strongly simplified calculations during the pixel resampling
process and allows only strictly linear co-additions (see below). In
contrast, {\tt SWarp} offers a large variety of pixel resampling and
co-addition algorithms. 

{\tt SWarp} first undistorts and resamples all input SCIENCE and WEIGHT images
according to the astrometric solution. The user can choose
between several sophisticated kernels for the pixel remapping. We use
by default the LANCZOS3 kernel \citep[see][for details on {\tt SWarp's}
resampling kernels]{ber02}. The final co-addition of the resampled
images is performed in a second pass. Having all resampled input
images belonging to a given output pixel at hand simultaneously, {\tt SWarp}
can calculate the final result in a variety of ways, such as median,
mean or the weighted mean which is our method of choice.  {\tt EIS
Drizzle's} implementation is significantly different. It uses a method
known as forward mapping. It performs the complete co-addition in one
pass by putting the input images consecutively on the output
grid. Hence, {\tt EIS Drizzle} is limited to linear co-addition methods.
Applying the astrometric solution, an individual input pixel is mapped
somewhere on top of several neighbouring output pixels, and is
distorted and rotated. Its flux is then distributed accordingly
amongst these output pixels.\footnote{The user can vary the fraction of
the input pixel entering this procedure (drizzle PIXFRAC parameter).
We always use the original pixel size in our co-additions
(PIXFRAC=1).} The {\tt EIS Drizzle} approach strongly simplifies the
calculation of the flux distribution, in the sense that only
non-integer shifts are taken into account, whereas rotations and
distortions of the mapped pixels are neglected. This corresponds to
the {\sl Turbo Kernel} in the current {\tt MultiDrizzle} implementation
\citep[][]{kfh02}. See also \citet{frh02} for a more detailed
description of the drizzling approach. We discuss the differences of
the pixel resampling kernels of {\tt SWarp} and {\tt EIS Drizzle} in
more detail at the end of this section.

Choosing a weighted mean co-addition for {\tt SWarp}, both software packages
calculate the value of an output pixel in the same way.  Four factors
contribute to the final result in our weighted mean co-addition. Given
are the value $I_i$, representing the part of an input pixel that goes
to a specified output pixel $I_{\rm out}$ in the co-added
image. $W_{i}$ represents the associated value in the WEIGHT map.
$I_{i}$ is scaled with factors $f_{i}$ to the consistent
photometric zero point and normalised to a fixed exposure time of 1 s. 
This scaling reads
\begin{equation}
f_{i}=10^{-0.4\,\mathit{ZP}_{i}}/t_{i},
\end{equation}
where $t_i$ is the exposure time and $ZP_i$ the relative photometric
zero point. All images are in addition weighted according to their sky
noise. This weight scale is given by
\begin{equation}
w_i=\frac 1{\sigma_{{\rm sky},i}^2 f_i^2}.
\end{equation}
We take into account that the noise also scales with the flux
scale $f_i$. The values $I_{\rm out}$ and $W_{\rm out}$ in a stack of $N$
exposures then read
\begin{eqnarray}
I_{\rm out} = \frac{\sum_{i=1}^{N}I_if_iW_iw_i}{\sum_{i=1}^{N}W_iw_i} & , 
\;\;\;\;\; & W_{\rm out}=\sum_{i=1}^{N}W_iw_i\;.
\end{eqnarray}
Besides the co-added SCIENCE and WEIGHT mosaics {\tt EIS Drizzle}
can produce a CONTEXT map which allows the identification of all
input frames that contribute to a given pixel in the co-added images.

As standard sky projection of the mosaic we use the TAN
projection \citep[see][for further information on sky
projections]{grc02} with an orientation such that North is up and East
to the left. A reference coordinate can be specified for the
co-addition, associated with an integer reference pixel.  Thus, if
multi-colour information is available for a particular {\sl Set}, the
mosaics in the different bands can be registered with sub-pixel
accuracy if required.

\begin{figure}[ht]
  \includegraphics[width=1.0\hsize]{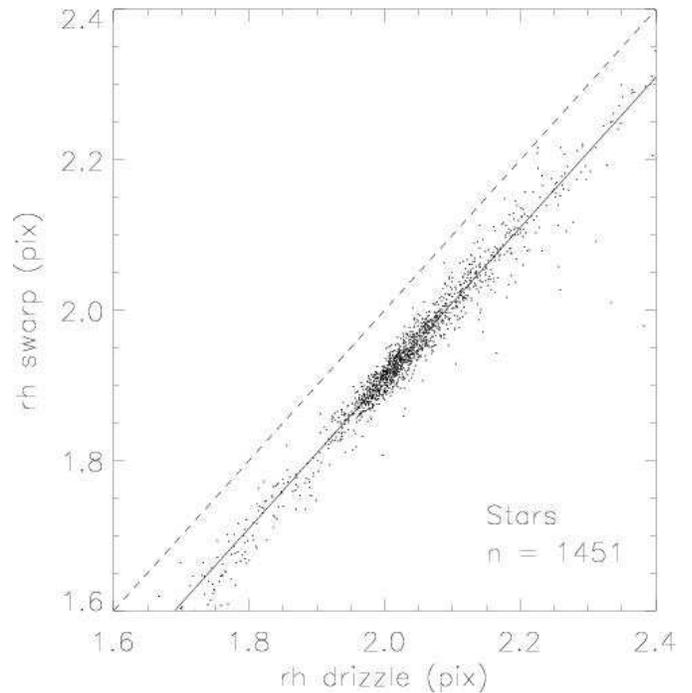}
  \caption{\label{fig:radius}\small{Comparison of the image seeing
  between two mosaics created with {\tt SWarp} and {\tt EIS Drizzle}. Shown are
  the half-light radii for unsaturated stars in the two
  co-additions. The swarped image has an image seeing that is $~0.09$
  pixels smaller than the one for the drizzled image. The mean values
  for the image seeing are $0\myarcsec95$ for the drizzled image and
  $0\myarcsec91$ for the swarped one.  The same astrometric solution was
  used for both co-additions. Thus {\tt EIS Drizzle} slightly increases the
  size of the PSF, an effect of its simplified kernel. In
  \figref{fig:combinedpsf} it is shown that the PSF anisotropies are
  identical for both co-addition strategies.}}
\end{figure}

\begin{figure*}[ht]
  \centerline{\includegraphics[width=0.47\hsize]{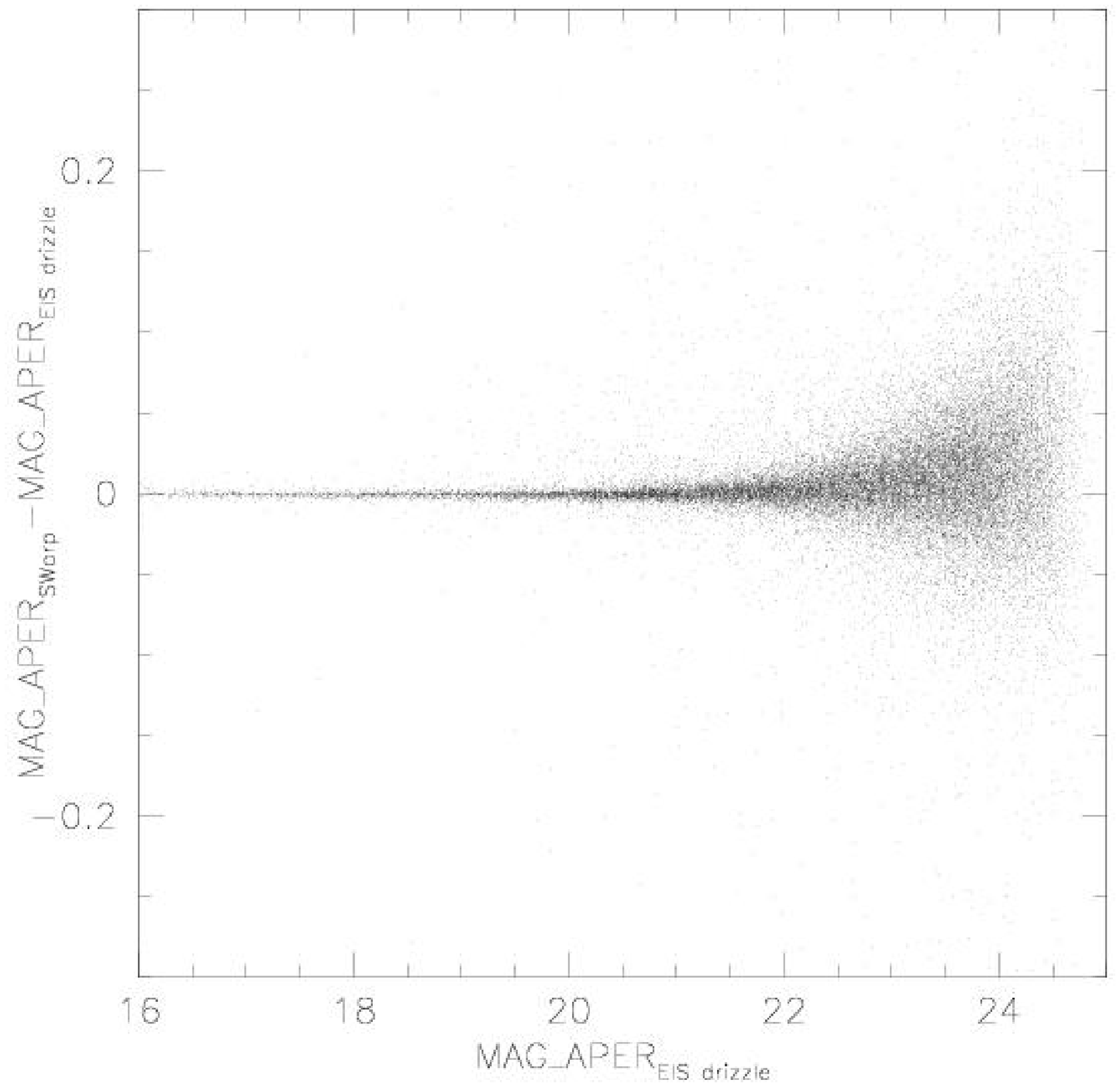}
  \includegraphics[width=0.47\hsize]{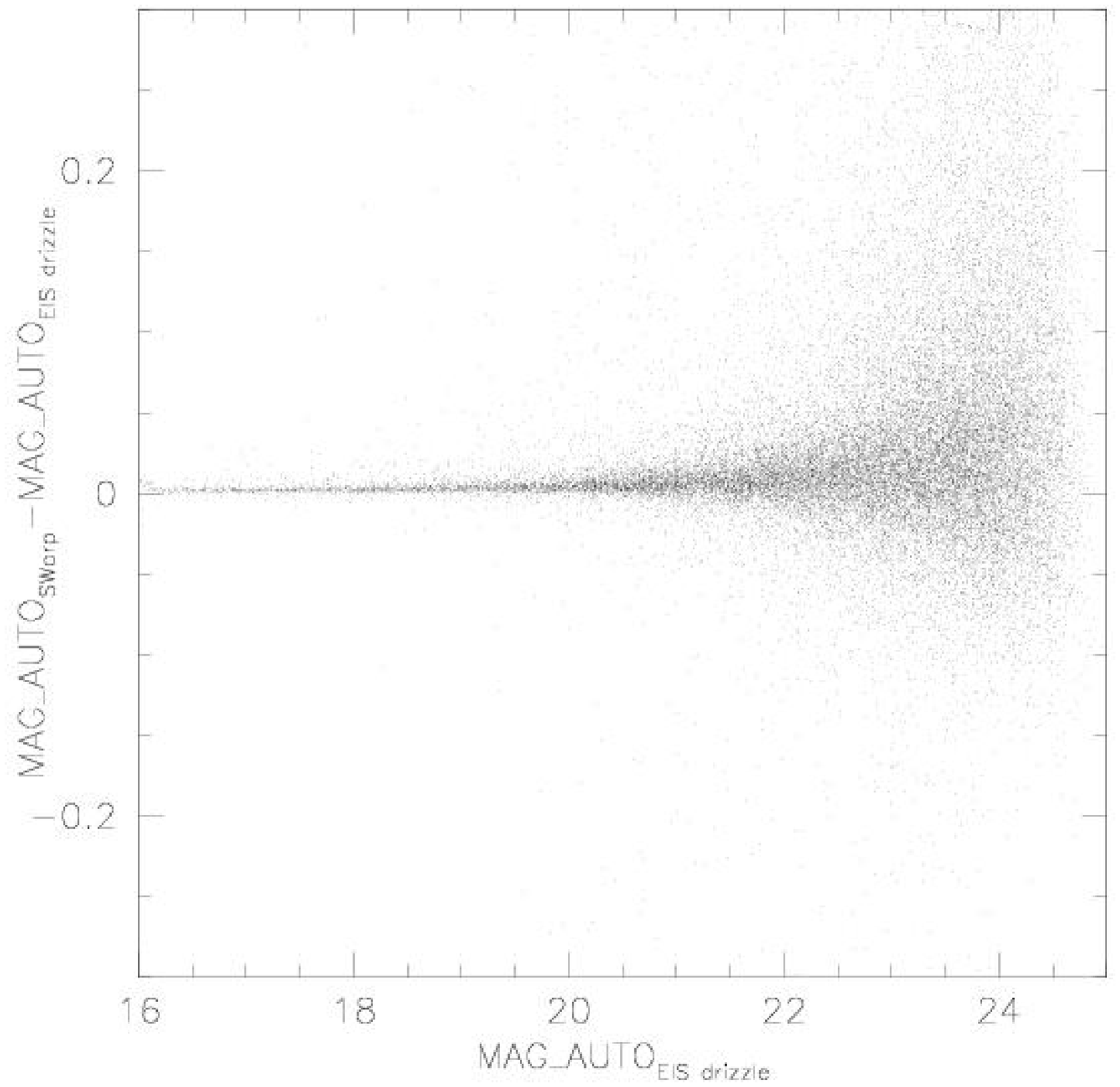}}
  \caption{\label{fig:magdrizswarpcomp}\small{We show comparisons of
magnitude estimates from an {\tt EIS Drizzle} and a {\tt SWarp} co-addition from
the same set. Both measurements are in very good agreement if the flux
is measured within a fixed aperture (here with a diameter of
$3\myarcsec5$). However, for {\tt SExtractor's} MAG\_AUTO estimate (Kron like
total magnitudes), the aperture used for faint sources is larger
in the drizzled images than in the {\tt SWarp} mosaic. The reason is
the more correlated noise in the {\tt EIS Drizzle} approach.}}
\end{figure*}

\begin{figure*}[ht]
  \includegraphics[width=1.0\hsize]{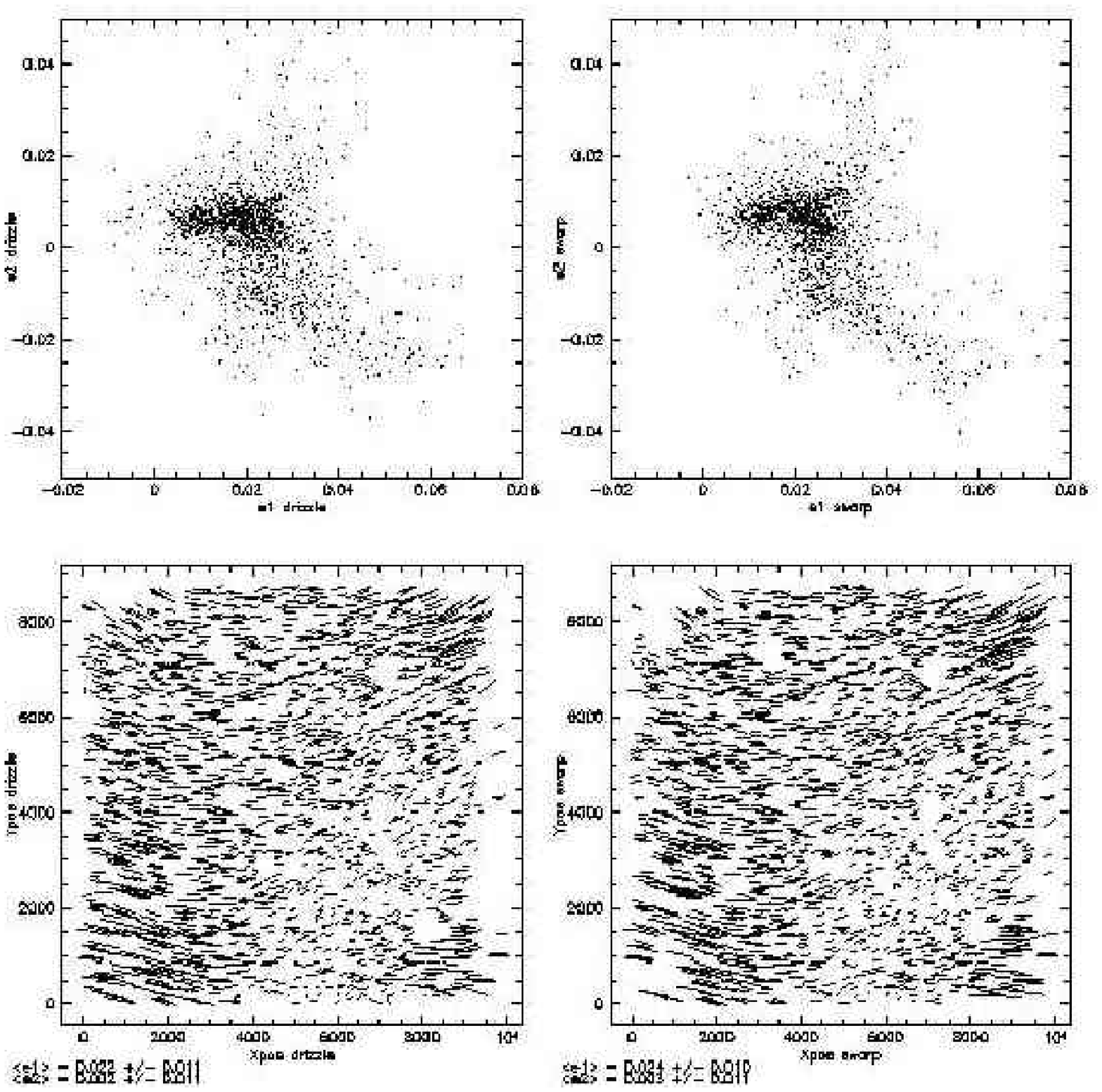}
  \caption{\label{fig:combinedpsf}\small{PSF anisotropies in a drizzled (left
  column) and a swarped (right column) mosaic of the same data set. The 
  patterns are virtually identical. Yet {\tt EIS Drizzle} marginally
  increases the size of the PSF, as was shown in the left panel of
  Fig. \ref{fig:radius}. The mean PSF anisotropy for 
  this particular mosaic amounts to mere $0.022\pm 0.015$.}}
\end{figure*}

In the following we show a comparison of photometric properties
from objects extracted from the same set co-added with {\tt EIS Drizzle}
and {\tt SWarp}. We expect that the simpler resampling kernel from 
{\tt EIS Drizzle} leads to stronger noise correlations and image blurring 
than the more sophisticated {\tt SWarp} approach. This is illustrated in
\figref{fig:radius} and \figref{fig:magdrizswarpcomp}. The {\tt EIS Drizzle}
co-addition has a slightly larger image seeing and {\tt SExtractor} chooses
a larger {\sl optimal} radius for its MAG\_AUTO parameter. In
contrast, {\tt EIS Drizzle} has no significant impact on the PSF
anisotropies (see \figref{fig:combinedpsf}). Also more sophisticated
weak lensing analyses (such as cluster mass reconstructions) 
revealed that results based on object shape
measurements do not differ significantly between the
two co-additions. 
\section{First Data quality assessments of co-added {\sl Sets}}
\label{sec:co-addquality}
The implementation of a thorough, automatic quality check on our 
co-added mosaics is still in its infancy. In this section we 
describe basic tests on extracted object catalogues that are done
without any user interaction.
\subsection{Galaxy counts}
\begin{figure}
\centering
\resizebox{\hsize}{!}{\includegraphics[width=\textwidth,clip]{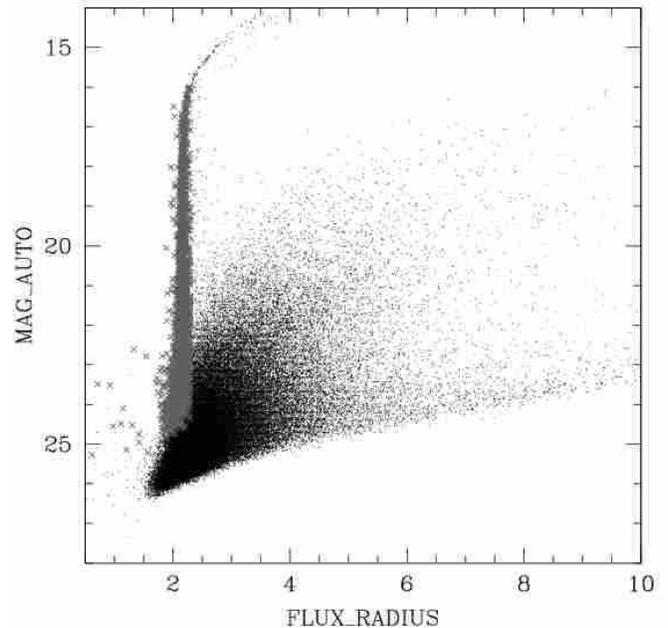}}
\caption{Magnitude vs. half-light radius plot. Grey crosses: stars;
appear as a vertical branch in the plot and are selected by the
CLASS\_STAR parameter and half-light radius.  Dots: extended sources.
}
\label{fig:magrh}
\end{figure}

\begin{figure}
\centering
\resizebox{\hsize}{!}{\includegraphics[width=\textwidth,clip]{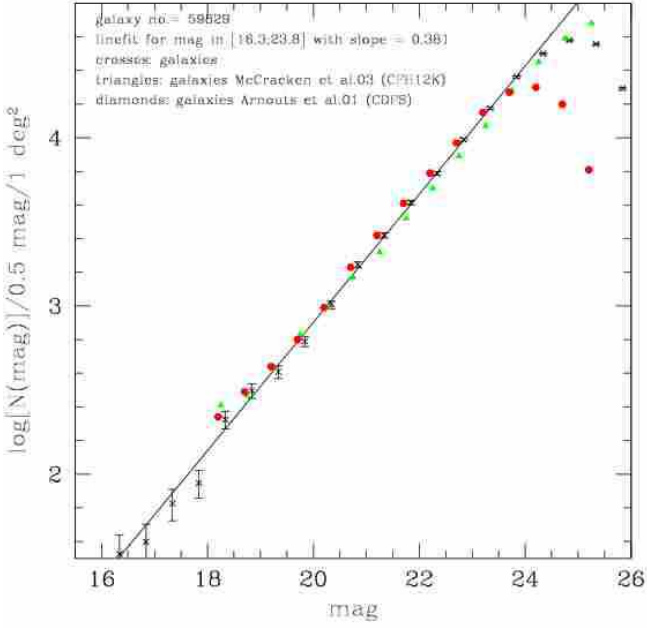}}
\caption{The figure displays the logarithmic galaxy counts in
$0.5\,{\rm mag}$ bins per one square degree from our reduction of
the Chandra Deep Field South \citep[CDFS; see][]{gfk04}.  
The error bars are due to Poisson noise; the line fit is
an error-weighted linear regression in a magnitude range between the
saturation and the limiting magnitude (here: $R\in[16.3;23.8]$).  For
the normalisation of the area we take into account that each detected object
occupies an area in which fainter objects cannot be detected.  As a
comparison to our galaxy counts we also plot the galaxy number counts
from the CFH12K-VIRMOS deep field \citep[][]{mrb03}. For the CDFS we also 
plot the number counts from \citet{avb01}.}
\label{fig:magdistribution}
\end{figure}

The programme {\tt SExtractor} is used to create
a raw catalogue of all objects that consist of at least 3 contiguous
pixels (DETECT\_MINAREA=3) with a flux $2\sigma$ above the flux of the
sky-background (DETECT\_THRESH=2). The source extraction is done on
a filtered image; we use a normalised Gaussian filter with a full
width half maximum of 4.0 pixels (FILTER\_NAME=gauss\_4.0\_7x7.conv).
This conservative threshold is chosen in order to minimise the number
of spurious detections.  In the following, we use the {\tt SExtractor}
parameters MAG\_AUTO for magnitudes and FLUX\_RADIUS for the
half-light radius.  To create a galaxy catalogue, the {\tt SExtractor}
parameter CLASS\_STAR in combination with the half-light radius is
used; saturated objects are rejected.  We define every object which
has CLASS\_STAR less than 0.95 as a galaxy.  A simple check of this
selection is a magnitude over half-light radius plot.  All stars have
the same half-light radius and therefore show up as a vertical branch
in this plot, see \figref{fig:magrh}.
%
%
\\ We count the number of galaxies
in $0.5\,{\rm mag}$ wide bins per one square degree.  To
normalise the area to one square degree, we take into account that
each object occupies an area in which fainter objects can not be
detected.  For this correction we use the {\tt SExtractor} parameter ISO0,
which is the isophotal area above the analysis threshold.  We note
that in the case of empty fields this effect is almost negligible.  An
error-weighted linear regression to the logarithmic galaxy counts is
performed and the slope, $\dd \log\,N/\dd\, {\rm mag}$ determined.  We
routinely compare our galaxy counts with those of \citet{mrb03}, see 
\figref{fig:magdistribution}.  With this comparison, a rough test of the
magnitude zero point and the limiting magnitude can be performed.
%
%

\subsection{Clustering of extended sources}
\begin{figure*}
\centering
\resizebox{\hsize}{!}{\includegraphics[width=0.44\textwidth]{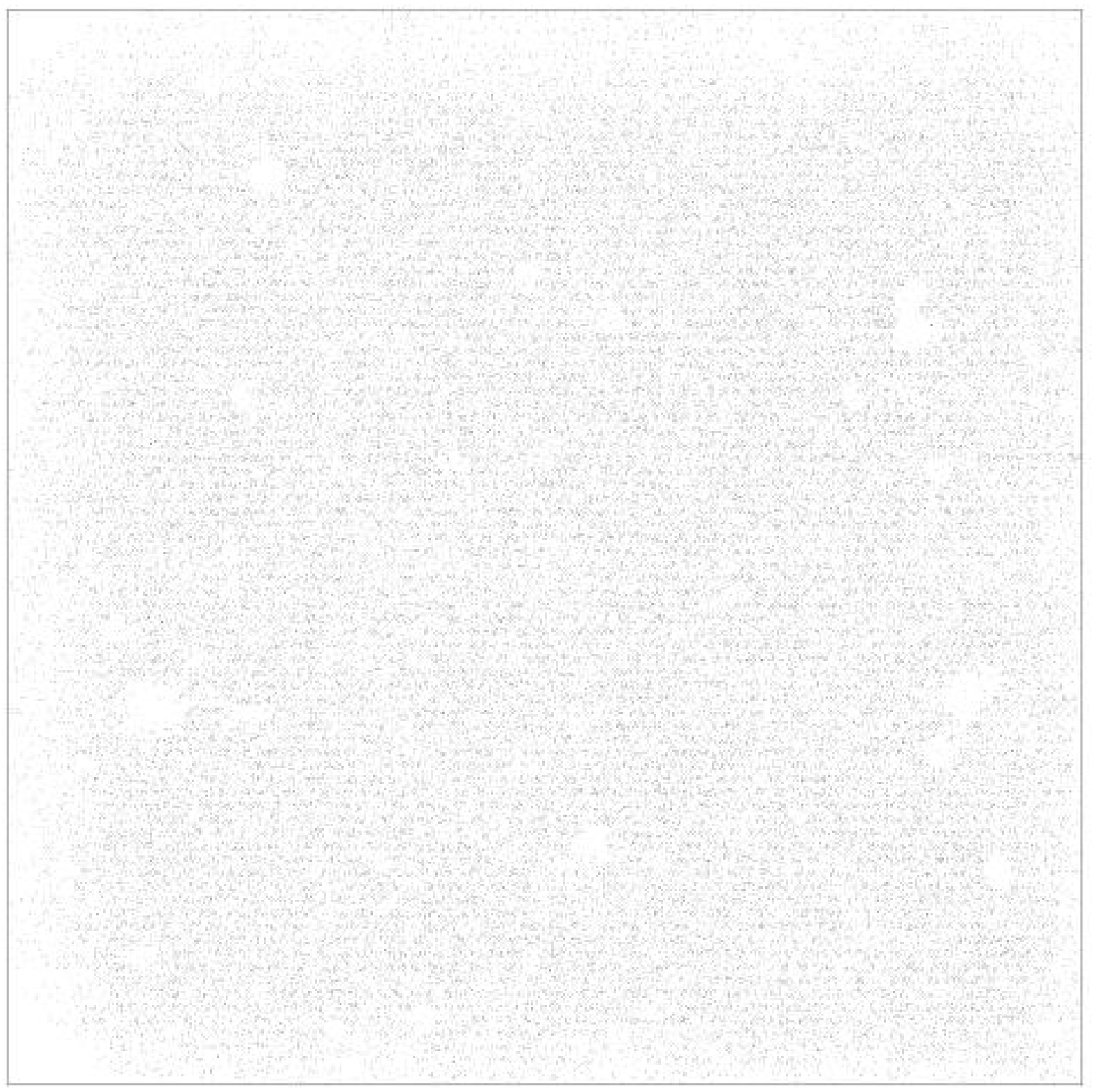}
\includegraphics[width=0.438\textwidth]{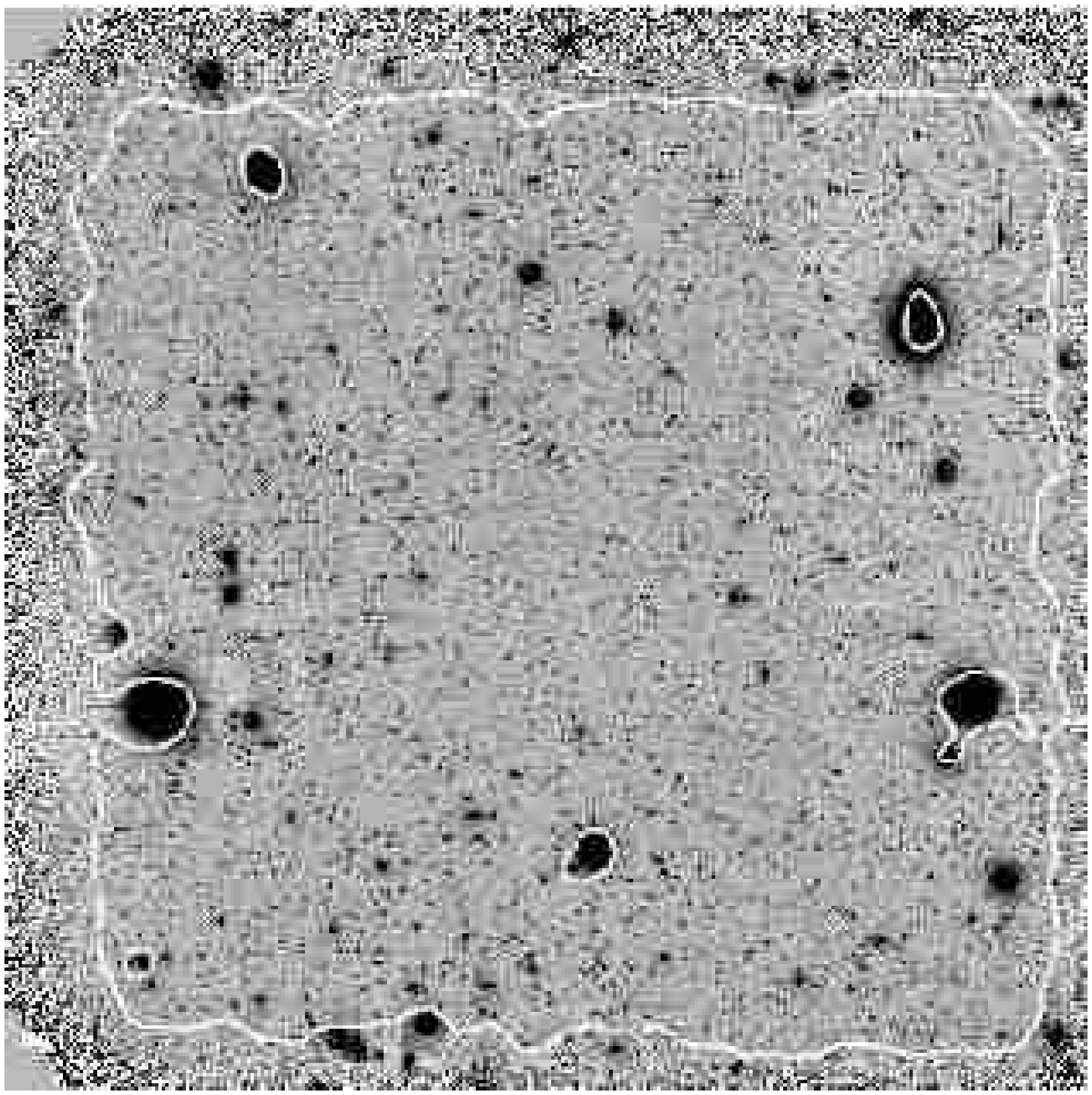}}
\caption{Obscuration mask for the CDFS.
Left panel: Distribution of detected sources.
Right panel: FITS image of the CDFS. White contours indicate masked
region. In deep fields, our automatic obscuration mask based on object density
variations reliably marks large-scale astronomical sources and noisy
borders which would significantly influence the area of
our correlation function analysis. This approach turned out to be
sufficient for quality check purposes.}
\label{fig:corrmask}
\end{figure*}

A further test for the quality of the co-added image is the clustering
of sources.  For that purpose we use the two-point angular correlation
function, $\omega(\theta)$, where $\omega(\theta)\delta\theta$ is the
excess probability of finding a pair separated by an angle between
$\theta$ and $\theta+\delta\theta$.  We estimate this quantity by
creating a large number of random catalogues (by default 40 mock
catalogues are created) and count the pairs within the data catalogue,
$DD$, within the random catalogue, $RR$ and between the data and
random catalogues, $DR$.  The estimator for $\omega(\theta)$, proposed
by \citet{las93}, is \be \omega(\theta)=\frac{DD-2DR+RR}{RR}.  \ee The
random fields must have the same geometry as the data field.
Therefore we calculate an obscuration mask out of the number density
of extracted sources as follows.  A mesh with $512\times 512$ mesh
cells is placed on top of a data field and the number of objects in
each cell is counted.  Then the cell count matrix is smoothed with a
Gaussian kernel of about $512/60$ cells FWHM, and all matrix elements
lower than a given fraction (we use 75\% as default but varying this
parameter between 50\% and 80\% does not change the results
significantly) of the mean number of galaxies inside a cell are
defined as a masked region.  The borders of a field, bright stars, and
part of their halos are masked automatically by this method (see
\figref{fig:corrmask}).

To maximise computational speed, we perform our calculations by
creating an index tree for galaxy position as explained in
\citet{zhp04}.  The error bars in our check-plots for each angular bin
are simply estimated by Poisson noise and are therefore a lower limit
to the uncertainty in $\omega(\theta)$.  

\begin{figure}
\centering
\resizebox{\hsize}{!}{\includegraphics[width=\textwidth,clip]{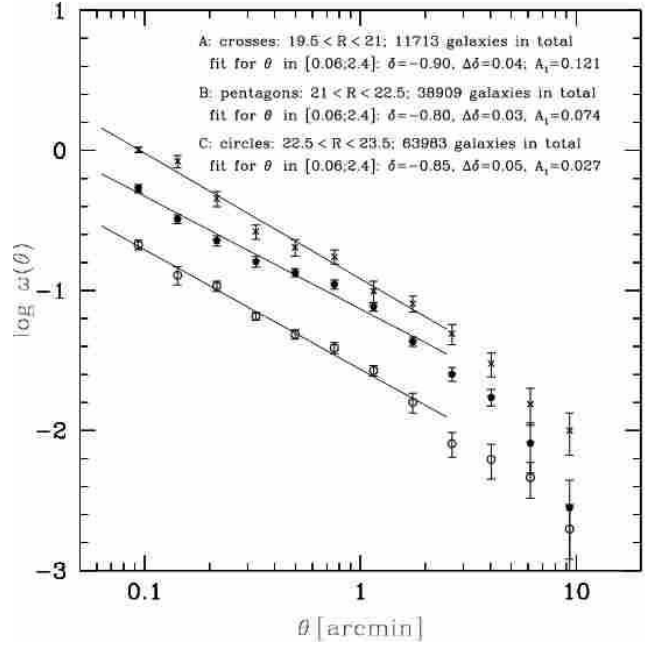}}
\caption{The two-point angular correlation function of galaxies for
different magnitude bins, for 11 WFI-fields in the DPS.  We show the
logarithm of the amplitude, $\log\,\omega(\theta)$, as a function of
the angular separation $\theta$ in arcmin.  Here the error bar in
each bin is due to the field-to-field variance.  We perform a simple
error-weighted linear regression in the angular interval
$\theta\in[0\myarcsec06;2\myarcminnodot]$ (where the integral constraint is
negligible) to determine the slope $\delta$ and the amplitude
$A_1=\omega(\theta=1^\prime)$.}

\label{fig:correlation}
\end{figure}

As an example we present the two-point angular correlation function of
galaxies in our $R$-band reductions for 11 WFI-fields in the 
EIS Deep Public Survey (EIS DPS). 
For three different magnitude bins we fitted a power law,
$\omega(\theta)=A\,\theta^\delta-C$, to the data.  The variable $C$
is the so-called integral constraint \citep[see for
instance][]{rsm93}, which only becomes important for large fields at
larger scales ($\theta>2^\prime$).  We therefore neglect the integral
constraint and perform the fit for small scales ($\theta<2^\prime$).
The results are shown in \figref{fig:correlation}.
%
%

\begin{figure}
\centering
\resizebox{\hsize}{!}{\includegraphics[width=\textwidth,clip]{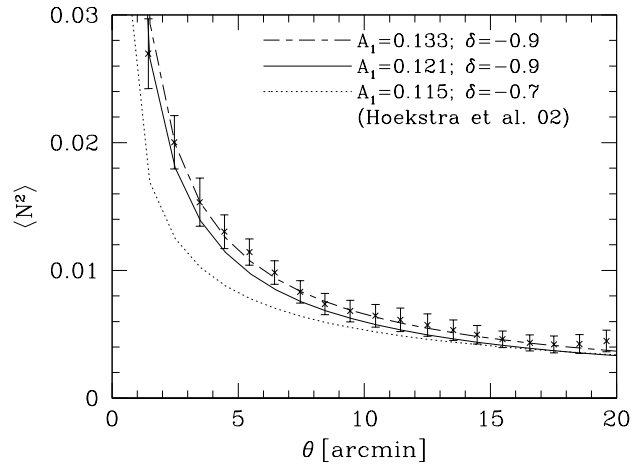}}
\caption{The aperture number count dispersion $\langle\textit{N}_{\rm
obs}^2\rangle(\theta)$ as a function of angular scale $\theta$ in
arcmin for 11 WFI-fields in the magnitude interval $R\in[19.5;21]$.
The error bars on the function $\langle\textit{N}^2\rangle(\theta)$
are due to the field-to-field variance of the 11 fields.  Note that
the points are correlated.  The lines display the function
$\langle\textit{N}_{\rm fit}^2\rangle(\theta)$ assuming a power law
for the angular two-point correlation function of the form
$\omega(\theta)=A_1\,\theta^\delta$ for different parameters
$A_1=\omega(\theta=1^\prime)$ and $\delta$.  Solid line:
$\langle\textit{N}_{\rm fit}^2\rangle(\theta)$ calculated for the fit
parameter obtained from the angular correlation function (see 
\figref{fig:correlation}); dotted line: comparison with
$\langle\textit{N}^2\rangle(\theta)$ calculated for the fit parameters
determined by \citet{hvg02}; dashed line: best fit to
$\langle\textit{N}^2\rangle(\theta)$.  }
\label{fig:NN}
\end{figure}

The results from our clustering analysis can be cross-checked
by considering
the aperture number count dispersion
${\langle\textit{N}^2\rangle(\theta)}$. It
is directly related to the
angular correlation function $\omega(\theta)$ by \be
\langle\textit{N}^2\rangle(\theta)=\int\dd
\vartheta\,\frac{\vartheta}{\theta^2}\,\omega(\vartheta)\,T_+\left(\frac{\vartheta}{\theta^2}\right),
\ee where the function $T_+$ is defined as eq. (35) in \citet{svm02}
and it has the nice property of being
independent of the integral constraint. As an example we use the
measured $\omega(\theta)$ for the magnitude interval $R\in[19.5;21]$
from the 11 WFI-fields of the last section to calculate
$\langle\textit{N}_{\rm obs}^2\rangle(\theta)$, see \figref{fig:NN}.  
To compare the slope $\delta$ and amplitude $A_1$ of the
fitted power law to the measured $\omega(\theta)$ of the previous
section (\figref{fig:correlation}), $\langle\textit{N}_{\rm
fit}^2\rangle(\theta)$ is calculated for the power law,
$\omega(\theta)=A_1\,\theta^\delta$.  The results show that the slope
$\delta$ obtained from the power law fit of the angular correlation
function is correct.  The amplitude $A_1$, however, seems, with
$A_1=0.133$, to be a bit larger; this small discrepancy can easily
arise due to the fact that $\langle\textit{N}_{\rm
obs}^2\rangle(\theta)$ is calculated using $\omega(\theta)$ over the
entire $\theta$-range and that the $\omega$-fit is slightly influenced
by the integral constraint.  As a comparison, \figref{fig:NN}
displays the function $\langle\textit{N}^2\rangle(\theta)$ for the fit
parameters determined by \citet{hvg02} for $R_{\rm C}$-band data of
the same magnitude range.  Note that this can only be a rough
comparison, because of different source extraction algorithms and
slightly different $R$-band filter.
%
%
%
\section{Conclusions and Outlook}
\label{sec:conclusions}
We have presented our image processing methods for multi-chip cameras
that we developed in the course of the GaBoDS project. A significant
fraction of GaBoDS is a virtual survey in which observational data were
collected from the ESO Science archive within an ASTROVIRTEL
program\footnote{ASTROVIRTEL Cycle 2: Gravitational lensing studies in
randomly distributed, high galactic latitude fields; P.I. Erben}
primarily for weak gravitational lensing studies. This allowed us to
test and to apply our procedures on data sets which were acquired for
a large variety of scientific programmes (e.g. deep field
observations, search for moving objects, Cepheid studies in nearby
galaxies) and obtained in many different ways (very compact to very
wide dither patterns, {\sl Sets} observed within a single night or
over several years). Our experiences regarding the processing of
these data sets from WFI@2.2m are the following:
\begin{itemize}
\item The techniques described perform very well on empty field
observations with WFI@2.2m. The excellent optics of this instrument
allow an accurate astrometric alignment of images which is crucial for
weak lensing studies \citep[][\citeyear{ses04}]{ses03}.  Also very large data
sets, obtained over several years, can be processed. For example, we
collected and reduced all WFI@2.2m observations from the CDFS, which
consist of more than 100 individual exposures in each of the 
$U$, $B$, $V$ and $R$ bands \citep[][]{gfk04,mib04}.
\item Data sets from crowded fields or from large-scale objects, whose
extent is comparable to the field-of-view, can be processed with good
results but require substantial manual intervention \citep[see][for
processing details of the field around NGC 300]{ses03}.
\item Our absolute photometric calibration is currently accurate to
about 0.05 mag as discussed in \sectionref{subsec:photometric} and
\sectionref{subsec:NGC300}. Also the comparison of our standard star
calibrations with independent measurements in
\sectionref{sec:standards} shows errors of the same order. However, it
seems that non-uniform illuminations that probably contribute most to
the error budget, do not strongly depend on wavelength and hence
errors on colours are smaller \citep[see][]{kgo04a}. Our reduction
of the CDFS data gives good results in photometric redshift studies
of different groups \citep[][]{mib04,gss04,hbe04}.
\end{itemize}
As was described in \sectionref{sec:concepts} our processing system
has already been used with a larger variety of single- and multi-chip
cameras. However, more sophisticated tests on the astrometric and
photometric accuracies, as presented here for WFI@2.2m, have not been
done except for the single-chip camera FORS1@VLT \citep[][]{bep04} and
FORS2@VLT with a two-chip camera \citep[][]{emc03}.

We are currently performing a thorough comparison of released fields
from the EIS DPS with our own reductions of this survey. This will give
us additional insights in the quality and the properties 
of our reductions and allows us to identify useful and necessary
quality tests for the co-added mosaics. This work will be presented
in Dietrich et al. (in prep.).  
\begin{acknowledgements}
We thank Emmanuel Bertin and Oliver Czoske for comments and
suggestions on the manuscript, and are deeply grateful to Yannick
Mellier and Ludovic van Waerbeke for their support of the GaBoDS
project and their long-standing collaboration.  
The MPA and the MPE in Garching, the TERAPIX data centre at
IAP in Paris and the Astrophysics department of the University of
Innsbruck kindly gave us access to various computer platforms. We are
very thankful to the people who helped us to improve our image
processing system by using our tools, adjusting them to new
instruments, making suggestions for their improvement and by reporting
and/or fixing bugs. Many of us learned data reduction techniques
while working within the ESO Imaging Survey Team at ESO.

This work was supported by the German Ministry for Science and
Education (BMBF) through the DLR under the project 50 OR 0106 and
through DESY under the project 05AE2PDA/8, and by the Deutsche
Forschungsgemeinschaft (DFG) under the project SCHN 342/3--1. The
support given by ASTROVIRTEL, a Project funded by the European
Commission under FP5 Contract No. HPRI-CT-1999-00081, is acknowledged.
\end{acknowledgements}
\bibliographystyle{aa}
\bibliography{pipeline}
\appendix
\section{Pipeline Image header}
\label{sec:imageheader}
Our pipeline replaces the original FITS headers of all individual
CCDs by a new one containing only a minimum set of keywords.
In this way we unify the headers for all instruments and avoid 
inconsistencies especially in the astrometric calibration. An example
header for WFI@2.2m is:
\begin{verbatim}
SIMPLE  =                    T  /
BITPIX  =                   16  /
NAXIS   =                    2  /
NAXIS1  =                 2142  /
NAXIS2  =                 4128  /
BSCALE  =                   1.  /
BZERO   =               32768.  /
CTYPE1  =           'RA---TAN'  /
CTYPE2  =           'DEC--TAN'  /
CRPIX1  =                -416.  /
CRPIX2  =                -224.  /
CD1_1   =            -6.61E-05  /
CD2_2   =             6.61E-05  /
CD1_2   =                   0.  /
CD2_1   =                   0.  /
CRVAL1  =            12.505009  /
CRVAL2  =            -52.15978  /
RADECSYS=                'FK5'  /
FILTER  =   'BB#Rc/162_ESO844'  /
AIRMASS =             1.113885  /
EXPTIME =             599.9176  /
EQUINOX =                2000.  /
IMAGEID =                    3  /
GABODSID=                 1401  /
EISID   =                   10  /
OBJECT  =         'BPM16274_3'  /
ZP      =                 -1.0  /
COEFF   =                 -1.0  /
DUMMY0  =                    0  /
DUMMY1  =                    0  /
DUMMY2  =                    0  /
DUMMY3  =                    0  /
DUMMY4  =                    0  /
.
.
END                              
\end{verbatim}
The IMAGEID, GABODSID and EISID are unique identifiers for the chip
position within the mosaic, the night of the observation and the image
respectively. ZP and COEFF will finally contain magnitude zero point
and extinction coefficient.  At the end we introduce 20 DUMMY keywords
allowing the user to transfer important information from the original
headers. For an explanation of the rest of the keywords see
\citep{hfg01, grc02}.
\section{Released software}
We will release all modules of our processing system that are based on
publicly available code. Notes on their retrieval will be given in the
final, accepted version of this paper.
\end{document}